\documentclass{article}

\usepackage{amssymb}
\usepackage[utf8]{inputenc}
\usepackage{textcomp}
\usepackage{hyperref}
\usepackage{graphicx}
\usepackage{amsmath}
\usepackage[version=4]{mhchem}
\usepackage{siunitx}
\usepackage{longtable,tabularx}
\usepackage{comment}
\usepackage{graphicx}
\usepackage{epstopdf, epsfig}
\usepackage{xcolor}
\usepackage{ulem}
\usepackage{mathtools}
\usepackage[numbers]{natbib}
\usepackage{graphicx}
\usepackage{epstopdf, epsfig}
\usepackage{bm}
\usepackage{subcaption}
\usepackage{float}
\usepackage{multirow}
\usepackage{booktabs}

\usepackage{color}
\definecolor{nCol}{rgb}{0.2, 0.7, 0}
\definecolor{lCol}{rgb}{0.0, 0.0, 0}

\newcommand\Rey{\mbox{\textit{Re}}}  
\newcommand\Pran{\mbox{\textit{Pr}}} 
\newcommand\xmach{\mbox{\textit{\textcolor{lCol}{M}}}}  

\DeclareMathOperator*{\argmin}{arg\,min}

\usepackage[utf8]{inputenc}

\title{Towards high-accuracy deep learning inference of compressible turbulent flows over aerofoils}

\author{Li-Wei Chen, Nils Thuerey \\ Technical University of Munich, D-85748 Garching
\\ liwei.chen@tum.de}
\date{03 September 2021}

\begin{document}

\maketitle

\begin{abstract}
The present study investigates the accurate inference of Reynolds-averaged Navier-Stokes solutions for the two-dimensional compressible flow over aerofoils with a deep neural network.
Our approach yields networks that learn to generate precise flow fields for varying body-fitted, structured grids by providing them with an encoding of the corresponding mapping to a canonical space for the solutions.
We apply the deep neural network model to a benchmark case of incompressible flow at randomly given angles of attack and Reynolds numbers 
and achieve an improvement of more than an order of magnitude compared to previous work.
Further, for transonic flow cases, the deep neural network model accurately predicts complex flow behavior at high Reynolds numbers, such as shock wave/boundary layer interaction, and quantitative distributions like pressure coefficient, skin friction coefficient as well as wake total pressure profiles downstream of aerofoils. The proposed deep learning method significantly speeds up the predictions of flow fields and shows promise for enabling fast aerodynamic designs. 
\end{abstract}

\section*{Nomenclature}


{\renewcommand\arraystretch{1.0}
\noindent\begin{longtable*}{@{}l @{\quad=\quad} l@{}}
$a$ &  the speed of sound \\
$C_p$& pressure coefficient \\
$c$   & chord length \\
\textcolor{lCol}{$E$} & \textcolor{lCol}{total energy} \\
($i$,$j$)  & the array indices of a structured grid \\
\textcolor{lCol}{$\mathrm{i}$, $\mathrm{j}$ and $\mathrm{k}$}  & \textcolor{lCol}{indices for Einstein summation notation} \\
$J$ & the transformation Jacobian \\
$\mathcal{G}$ & geometric information\\
$L_1$ & L1 loss \\
\xmach & Mach number \\
\textcolor{lCol}{\Pran{}} & \textcolor{lCol}{Prandtl number} \\
$p$ & pressure \\
\textcolor{lCol}{$q_{x_{\mathrm{i}}}$} & \textcolor{lCol}{heat flux} \\
pad & the number of padding on the boundary \\
\Rey{} & Reynolds number \\
$S$ & the number of pixels \\
$\mathbf{s}$ & coordinates of an aerofoil profile \\
SD & standard deviation \\
str & stride for convolutional layers \\
$T$ & the number of samples \\
$\mathbf{T}$ & coordinates transformation matrix \\
$u$ & \emph{x} component of velocity\\
$v$ & \emph{y} component of velocity\\
$\mathbf{w}$ & weights of a neural network \\
($x$,$y$) & the coordinates in the physical space  \\
$\mathbf{x}$ & inputs of a neural network \\
$\mathbf{y}$ & outputs of a neural network \\
\textcolor{lCol}{$y_n^{+}$} & \textcolor{lCol}{the non-dimensional wall distance} \\
$\alpha$ & the angle of attack (AoA) \\
\textcolor{lCol}{$\gamma$} & \textcolor{lCol}{the ratio of specific heats} \\
\textcolor{lCol}{$\delta_{\mathrm{i}\mathrm{j}}$} & \textcolor{lCol}{Kronecker delta} \\
\textcolor{lCol}{$\Theta$} & \textcolor{lCol}{temperature} \\
\textcolor{lCol}{$\mu$} & \textcolor{lCol}{dynamic viscosity} \\
$\rho$ & density\\
$\tau_w$ & skin friction \\
\textcolor{lCol}{$\tau_{x_{\mathrm{i}} x_{\mathrm{j}}}$ } & \textcolor{lCol}{shear stress} \\
($\xi$,$\eta$) & the curvilinear coordinates in the computational space \\

\multicolumn{2}{@{}l}{Subscripts}\\
$\infty$ & free-stream condition\\
max	& maximum value \\
min & minimum value \\
train & training set \\
test  & test set \\
\textcolor{lCol}{$T$} & \textcolor{lCol}{turbulent quantities} \\
$t$     & values at a stagnation point \\
0   & reference points \\

\multicolumn{2}{@{}l}{Symbols}\\
\textcolor{lCol}{$\breve{(\cdot )}$} & \textcolor{lCol}{values predicted by the neural network} \\

\textcolor{lCol}{$\hat{(\cdot )}$} &
\textcolor{lCol}{non-dimensional forms of transformation matrix components} \\

\textcolor{lCol}{$\tilde{(\cdot )}$} &
\textcolor{lCol}{normalized values for the neural network}
\end{longtable*}}

\section{Introduction}

The recent years have witnessed great advances for applications of deep learning to fluid mechanics \cite{ling_kurzawski_templeton_2016, Lusch_Brunton2018embeddingNature, Duraisamy2019annuReview, fukami_fukagata_taira_2019, Holl2020Learning, Kochkov2101784118}.
Owing to the significance in both fundamental research and industrial practice, 
deep learning methods to infer solutions of Navier-Stokes equations have attracted considerable interest.
For many approaches, convolutional neural networks (CNNs) provide an important building block, e.g., for
low-cost approximations of steady flows \cite{Guo2016ConvolutionalNN}.
Aiming for the goal of accurate predictions of flow solutions, state-of-the-art deep learning methods and architectures have been developed for the inference of Reynolds-averaged Navier-Stokes solutions as well as steady laminar flow fields \cite{thuerey2018deep, Duru2022105312, chen2019unet, Eichinger2019stationary}. 
More recently, trained deep neural network models that infer flow fields have also been used as surrogate models to optimize shapes \citep{chen_cakal_hu_thuerey_2021}. 
\textcolor{lCol}{There are also noteworthy attempts to build accurate reduced order models based on classical linear operators, e.g. proper orthogonal decomposition methods \citep{Li_Zhang2016performancePOD, WuZhangPengWang2019, cao2019, Liu_Zhang2022dataAssimilation},
and nonlinear alternatives \citep{Hesthaven2018NonIntrusiveROM, fukami_fukagata_taira_2019, LiKouZhang2021}, which have shown promising results.}

Although progress has been made, there are notable challenges for deep learning to achieve the high accuracy that real-world engineering applications demand.
Most importantly, such applications require flexible domain discretizations that adapt to the embedded geometries, and are able to resolve fine features in flow fields, such as thin boundary layers at high Reynolds numbers and shock waves in compressible flows.
Specifically, in most of the previous studies, geometric information is encoded with signed distance function \citep{Bhatnagar_2019,chen_cakal_hu_thuerey_2021} or binary masks  \citep{thuerey2018deep,Leer2021fast}, and flow field data has to be interpolated onto an auxiliary Cartesian grid for the neural networks. As these grids are uniformly spaced, the resolution of the auxiliary grid has to be redundantly fine to prevent undesirable blurring of important flow structures. 
Other recent works resort to fully connected networks to resolve aerofoil geometries of transonic flows \citep{disun2021}.

A potential solution is the use of graph neural networks (GNNs) which directly encode and employ information about the discretization, i.e. mesh points and edges \citep{deAvila_icml2020_6414, pfaff2021learning}. 
However, three reasons lead us to seek other ways. First, a GNN is essentially unstructured. As we are particularly interested in aerodynamics, we realise that structured meshes are widely used not only in industry, e.g. aerospace engineering, but also in fundamental research areas such as the direct numerical simulation based on high-order finite-difference method \citep{Sandberg2015lpt, Jacobs2017opensbli}. 
Second, many current deep learning techniques are specifically designed for structured-grid-based convolutional neural networks \citep{Holl2020Learning, um2020sol, List_Chen_Thuerey2021}. 
And third, GNNs still face many open challenges such as questions regarding the theory of graph representations, demanding hardware requirements, non-trivial parallelism and distributed solving \citep{li2021graphtheta}.
Taking these considerations into account, we instead develop an inference method that uses information about the discretization of 
\textcolor{black}{the physical domain}
for a given structured curvilinear mesh that adapts to geometries.



With a finite-volume or finite-difference method based on a smooth, structured grid, it is possible to apply a local univalent transformation to the physical space coordinates (\emph{x}, \emph{y}) to obtain the curvilinear coordinates ($\xi$, $\eta$) with curved coordinate lines, i.e. structured body-fitted grids \citep{Gordon1973constructionCurvilinear, Krist_Biedron_Rumsey1998manual, blazek2001computational, WangSW2002, Sandberg2015lpt, Jacobs2017opensbli}. 
\textcolor{black}{For a given mesh}, the transformation matrix uniquely maps the physical space to the computational space. As the latter represents a Cartesian coordinate system, we use it as a canonical representation of the desired solutions with different discretizations. Due to its Cartesian nature, this space makes the use of efficient, regular convolutional architectures for the neural networks possible.
Early theoretical work has addressed the equivalence of image representations in networks \citep{lenc2015understanding} and showed promising applications with warped images in polar and spherical coordinate systems \citep{esteves2018polar, cohen2018spherical}. However, studies on coordinate transformations with arbitrary grids and the associated deep neural network architectures are relatively scarce. In this regard, the present work can be seen as a step  towards a generalized handling of geometric encodings for neural networks that deal with arbitrary warped images.

We recall that the combination of structured-grid solvers and deep learning has naturally led to architectures related to curvilinear coordinates \citep{Obiols_Sales_2020, tsunoda2021accuracy}. 
In these studies, encouraging results have been achieved, e.g., via CFDNet \citep{Obiols_Sales_2020}. However, the coordinate transformation is not fully incorporated into the neural network, which is designed for providing an approximate initial guess to accelerate the main computational fluid dynamics (CFD) solver. 
Recently, elliptic coordinate transformations were taken into account in mapping irregular 2D domain to rectangular ones, such that physics constrained CNN models can be trained to learn solutions of parametric PDEs \citep{Gao2021PhyGeoNet}. In contrast, the present study investigates the effect of different generic forms of coordinate transformations for deep learning, and evaluates their accuracy and robustness in detail for aerodynamic flows.


Unlike other deep learning frameworks that rely on the coupling with CFD solvers \citep{deAvila_icml2020_6414, Obiols_Sales_2020, tsunoda2021accuracy}, the present method is a pure convolutional neural network architecture trained in a fully supervised manner. The network directly makes high-fidelity predictions about flow fields, and hence works as a flexible surrogate model that produces a full flow field \citep{chen_cakal_hu_thuerey_2021}. Once trained, it can be used in different optimization tasks with multiple objectives and can be seamlessly integrated with deep learning algorithms, such as differentiable physics-based methods \citep{Holl2020Learning,List_Chen_Thuerey2021}.

Moreover, compared to others \citep{thuerey2018deep,Bhatnagar_2019,Gao2021PhyGeoNet}, the present deep neural network model accurately predicts complex flow behavior at high Reynolds numbers and transonic speeds, such as shock wave/boundary layer interaction, and quantitative distributions like
pressure coefficient, skin friction coefficient as well as wake total pressure profiles downstream of aerofoils. This is of particular importance in practical engineering designs for aerofoils and compressor/turbine blades \citep{MichelassiChen2015dns}. To the best of our knowledge, no previous studies exist that achieve a comparable prediction accuracy and robustness using a pure neural network model. 

The purpose of the present work is to demonstrate the capabilities of deep learning techniques for accurate predictions of compressible flow, and for achieving an improved understanding 
of the fundamental phenomena involved in such flows. \textcolor{lCol}{Given the importance of sectional aerofoil data in the preliminary designs in the aerospace industry, we focus on the inference of two-dimensional statistically steady flows over aerofoils.} The methods developed in the present study open up an avenue for the use of deep learning techniques in the inference of precise CFD solutions using structured body-fitted grids.
This paper is organized as follows. The mathematical formulation and numerical
method are briefly presented in \S \ref{sec:methodology}. 
The detailed experiments and results are then given in \S \ref{sec:results_and_discussion}, the performance of deep neural network models is reviewed in \S \ref{sec:performance} and concluding remarks are in \S \ref{sec:concluding_remarks}.

\section{Methodology}\label{sec:methodology}
\subsection{Inference task for deep neural networks}
\noindent 
The goal of the present work is to study the inference of Reynolds-averaged Navier-Stokes (RANS) solutions for the compressible flow over different aerofoil geometries in two dimensions with a deep neural network (DNN).
\textcolor{lCol}{
To non-dimensionalize the RANS equations, we use the freestream
variables including the density $\rho_{\infty}$, speed of sound $a_{\infty}$, 
and the chord length of the aerofoil $c$ as characteristic quantities.
Then, the non-dimensional form can be expressed in tensor notation as
\begin{equation}\label{eq:RANS-1}
\frac{\partial \rho u_\mathrm{i}}{\partial x_\mathrm{i}}=0
\end{equation}
\begin{equation}\label{eq:RANS-2}
\frac{\partial \rho u_\mathrm{i} u_\mathrm{j}}{\partial x_\mathrm{j}}=-\frac{\partial p}{\partial x_\mathrm{i}} + \frac{\partial \tau_{x_{\mathrm{i}} x_{\mathrm{j}}}}{\partial x_\mathrm{j}}
\end{equation}
\begin{equation}\label{eq:RANS-3}
\frac{\partial (\rho E + p)u_\mathrm{i}}{\partial x_\mathrm{i}} = \frac{\partial (-q_\mathrm{i} +u_\mathrm{j} \tau_{x_{\mathrm{i}} x_{\mathrm{j}}} )}{\partial x_\mathrm{i}}
\end{equation}
}
\textcolor{lCol}{
where the shear stress (with Stokes’ hypothesis) and heat flux terms are defined as
\begin{equation*}
\tau_{x_{\mathrm{i}} x_{\mathrm{j}}}=(\mu+\mu_T)\frac{\xmach_{\infty}}{\Rey_{\infty}}[(\frac{\partial u_\mathrm{i}}{\partial x_\mathrm{j}}+\frac{\partial u_\mathrm{j}}{\partial x_\mathrm{i}})-\frac{2}{3}\frac{\partial u_{\mathrm{k}}}{\partial x_{\mathrm{k}}}\delta_{\mathrm{i}\mathrm{j}}]
\end{equation*} 
and 
\begin{equation*}
q_{x_{\mathrm{i}}}=-(\frac{\mu}{\Pran}+\frac{\mu_T}{\Pran_T})\frac{\xmach_{\infty}}{\Rey_{\infty}(\gamma-1)}\frac{\partial \Theta}{\partial x_{\mathrm{i}}}.
\end{equation*}
Here, Reynolds number is defined as $\Rey_{\infty}=\rho_{\infty}\sqrt{u_{\infty}^2+v_{\infty}^2}c/\mu_{\infty}$; $\gamma$ is the ratio of specific heats, 1.4 for air; the laminar viscosity $\mu$ is obtained by Sutherland's law (the function of temperature), and the turbulent viscosity $\mu_T$ is determined by turbulence models; laminar and turbulent Prandtl numbers are constants, i.e.
$\Pran=0.72$, $\Pran_T=0.9$.
The relation between pressure $p$ and total energy $E$ is given by
\begin{equation*}
p=(\gamma-1)\bigl[\rho E-\frac{1}{2}\rho u_{\mathrm{i}}u_{\mathrm{i}}\bigr].
\end{equation*}
Note also that from the equation of state for a perfect gas, we have $p=\rho a^2/\gamma$ and temperature $\Theta=a^2$. According to Einstein summation convention, the indices i, j and k can range over the set $\{1,2\}$ in 2D, so ($x_1$, $x_2$) is equivalent to the traditional ($x$, $y$); and ($u_1$, $u_2$) is equivalent to ($u$, $v$).
}

\textcolor{lCol}{
The continuous flow domain over an aerofoil is discretized on a mesh with a finite number of degrees of freedom. 
On the aerofoil surface, the no-slip adiabatic wall boundary condition is imposed for the viscous flow case; the slip wall condition is used for the inviscid case \cite{Krist_Biedron_Rumsey1998manual}. At the farfield boundary, the free-stream condition is given as
\begin{equation}\label{eq:RANS-4}
\left.\begin{aligned}
\rho_{\infty} = 1.0 \\
u_{\infty} = \xmach_{\infty}\mathrm{cos}(\alpha_{\infty}) \\
v_{\infty} = \xmach_{\infty}\mathrm{sin}(\alpha_{\infty}) \\
a_{\infty} = 1.0 \mathrm{\ (or\ } p_{\infty}=\rho_{\infty} a^2_{\infty}/\gamma) 
\end{aligned}\right\} 
\end{equation}
which is also used for initialization.
For the sake of brevity, the discretized PDEs Eqs. \ref{eq:RANS-1}-\ref{eq:RANS-4} can be expressed in an implicit form:
\begin{equation}\label{eq:discreteForm}
    \mathcal{R}(\mathbf{y}, \xmach_{\infty}, \alpha_{\infty}, \Rey{}_{\infty}, \mathcal{G}) = 0.
\end{equation}
The two-dimensional steady CFD simulation aims at obtaining the spatial distribution of four independent field variables (i.e. $\mathbf{y}=$[$\rho$, $u$, $v$, $p$] or [$\rho$, $u$, $v$, $a$]) by solving equations \ref{eq:discreteForm} in an iterative manner for given geometric information $\mathcal{G}$ which includes the discretization of flow domain and the embedded aerofoil coordinates.}

\textcolor{lCol}{In the aerodynamic designs, we are interested the steady-state deterministic flowfield $\mathbf{y}$ at a given free-stream condition ($\mathbf{x}=$[$\xmach_{\infty}$, $\alpha_{\infty}$, $\Rey{}_{\infty}$]) and a given geometric information $\mathcal{G}$.}
The converged solution of a RANS simulation is physically an ensemble average, and it represents a deterministic solution. 
\textcolor{lCol}{
Writing the discretized PDE equations \ref{eq:discreteForm} in an explicit input-output form as
\begin{equation}
    \mathbf{y}=f(\mathbf{x},\mathcal{G}) ,
\end{equation}
%
the output $\mathbf{y}$ denotes the reference solution of a RANS simulation, which is expressed by flow field variables in either primitive (e.g. density, velocity, temperature) or conservative forms (e.g. density, momentum and total energy).
We define the learning task as
finding the function $\breve{f}$ that maps inputs $\mathbf{x}$ and $\mathcal{G}$ to an approximation of the output $\mathbf{\breve{y}}$.
We approximate the output $\mathbf{y}$ of the true function $f$ as
closely as possible with a deep neural network representation $\Breve{f}$ 
by adjusting its weights
$\mathbf{w}$ such that $\mathbf{y}\approx~\textcolor{lCol}{\Breve{\mathbf{y}}=\Breve{f}}(\mathbf{x},\mathcal{G};\mathbf{w})$.
The training objective becomes
\begin{equation}\label{eq:training_objective}
    \argmin_{\mathbf{w}} |f(\mathbf{x},\mathcal{G})-\breve{f}(\mathbf{x},\mathcal{G};\mathbf{w})|_1.
\end{equation}
This represents a fully supervised learning approach
based on the input-output pairs provided in terms of the training dataset.}


%

\subsection{\textcolor{lCol}{Discretization and geometric information}}
In most existing studies, a constant, uniformly distributed auxiliary grid is used to infer RANS solutions with deep learning methods \citep{thuerey2018deep, Bhatnagar_2019}. The auxiliary grid itself takes the form of a Cartesian grid, and hence the geometric input $\mathcal{G}$ takes the form of a binary mask or a signed distance function to encode the embedded aerofoil. However, an inherent drawback is that the discretization of the physical domain for inference can not change, and hence the representation of $\mathbf{y}$ is constant at training time.
As a consequence the discretization has to be redundantly fine to avoid the undesirable blurring of important flow structures.

We propose an alternative approach where the domain discretization varies and adapts to each input. Information about this discretization is encoded in $\mathcal{G}$.
We design the geometric encoding based on the following considerations:
First, we leverage the existing domain expertise in CFD to
obtain a suitable body-fitted structured mesh. These offer an optimal discretization to resolve important features in the flow field. Second, assuming that the structured mesh is smooth, we map the variables $\mathbf{y}$ from the physical space to a canonical space, often also called ``computational space", in which the neural networks operate.
Hence, no matter how the domain discretization in the form of the CFD mesh changes, the canonical space for the DNN 
remains the same. However, we provide the DNN with information about the discretization in physical space in terms of a geometric encoding, such that the neural network can infer the solution at the correct location.
Since we investigate a varying aerofoil profile $\mathbf{s}$, the corresponding discretization of flow domain (i.e. mesh) for a given $\mathbf{s}$ can be expressed as $\mathcal{G}=g(\mathbf{s})$. Here, function $g$ can be seen as the practice of mesh generation subjected to a given aerofoil $\mathbf{s}$.
To find a good representation of $\mathcal{G}$ for geometric encoding in the canonical space, we introduce three methods in \S \ref{sec:encodingGeoInfo}.
%
%
%
We discuss details of the deep neural network architecture in \S \ref{sec:NNarchitecture}.

\subsection{Encoding geometric information}\label{sec:encodingGeoInfo}

\noindent
For a given structured mesh, we perform a univalent transformation from the coordinates (\emph{x}, \emph{y}) in the physical space along indices \emph{i} and \emph{j} to the curvilinear coordinates ($\xi$, $\eta$) in the computational space \citep{Gordon1973constructionCurvilinear}, as follows:

\begin{equation}
    \begin{bmatrix}
    x\\
    y
    \end{bmatrix}_{i,j}=
    \begin{bmatrix}
    x_0\\
    y_0
    \end{bmatrix}_{i,1}+
\int    \begin{bmatrix}
    dx \\
    dy
    \end{bmatrix}.
\end{equation}
Here, $x_0$ and $y_0$ at ($i$, 1) are the reference coordinates at $j=1$ (or $\eta=0$), normally containing the coordinate information of a given aerofoil surface. Then, we have 
\textcolor{lCol}{
\begin{equation}
    \begin{bmatrix}
     d x \\
     d y
    \end{bmatrix}=
    \begin{bmatrix}
\partial_{\xi}x & \partial_{\eta}x\\
\partial_{\xi}y & \partial_{\eta}y
\end{bmatrix}
\begin{bmatrix}
d\xi \\
d\eta
\end{bmatrix}
\end{equation}}
and 
\textcolor{lCol}{
\begin{equation}
    \begin{bmatrix}
     d \xi \\
     d \eta
    \end{bmatrix}=
    \begin{bmatrix}
\partial_x{\xi}  & \partial_y{\xi}\\
\partial_x{\eta} & \partial_y{\eta}
\end{bmatrix}
\begin{bmatrix}
d x \\
d y
\end{bmatrix}.
\end{equation}}
Here, the definition of $\xi$ and $\eta$ is $\xi=(i-1)/(i_{\rm max}-1)$ and $\eta=(j-1)/(j_{\rm max}-1)$ with given integers $i$ from 1 to $i_{\rm max}$ and $j$ from 1 to $j_{\rm max}$, which can be extended by analytic continuation into a real-number domain $[0,1]\times[0,1]$. The partial derivatives are \textcolor{lCol}{represented} as:
\textcolor{lCol}{$\partial_{\xi}x = \frac{\partial x}{\partial \xi}$, $\partial_{\eta}x = \frac{\partial x}{\partial \eta}$, $\partial_{\xi}y = \frac{\partial y}{\partial \xi}$, and $\partial_{\eta}y = \frac{\partial y}{\partial \eta}$}. The partial derivatives of curvilinear coordinates, also known as metrics \citep{Rumsey1997}, are: \textcolor{lCol}{$\partial_x\xi = \frac{\partial \xi}{\partial x}$, $\partial_y\xi = \frac{\partial \xi}{\partial y}$, $\partial_x\eta = \frac{\partial \eta}{\partial x}$, and $\partial_y\eta = \frac{\partial \eta}{\partial y}$}, respectively.

Then, we define the transformation matrix to be:
\textcolor{lCol}{
\begin{equation}
\mathbf{T}=\begin{bmatrix}
\partial_x\xi & \partial_y\xi\\
\partial_x\eta & \partial_y\eta 
\end{bmatrix}=
\begin{bmatrix}
\partial_{\xi}x & \partial_{\eta}x\\
\partial_{\xi}y & \partial_{\eta}y
\end{bmatrix}^{-1}.
\end{equation}
}

\begin{figure}
    \centering
    \begin{subfigure}{.35\textwidth}
    \centering
    \includegraphics[width=\linewidth, angle=-90]{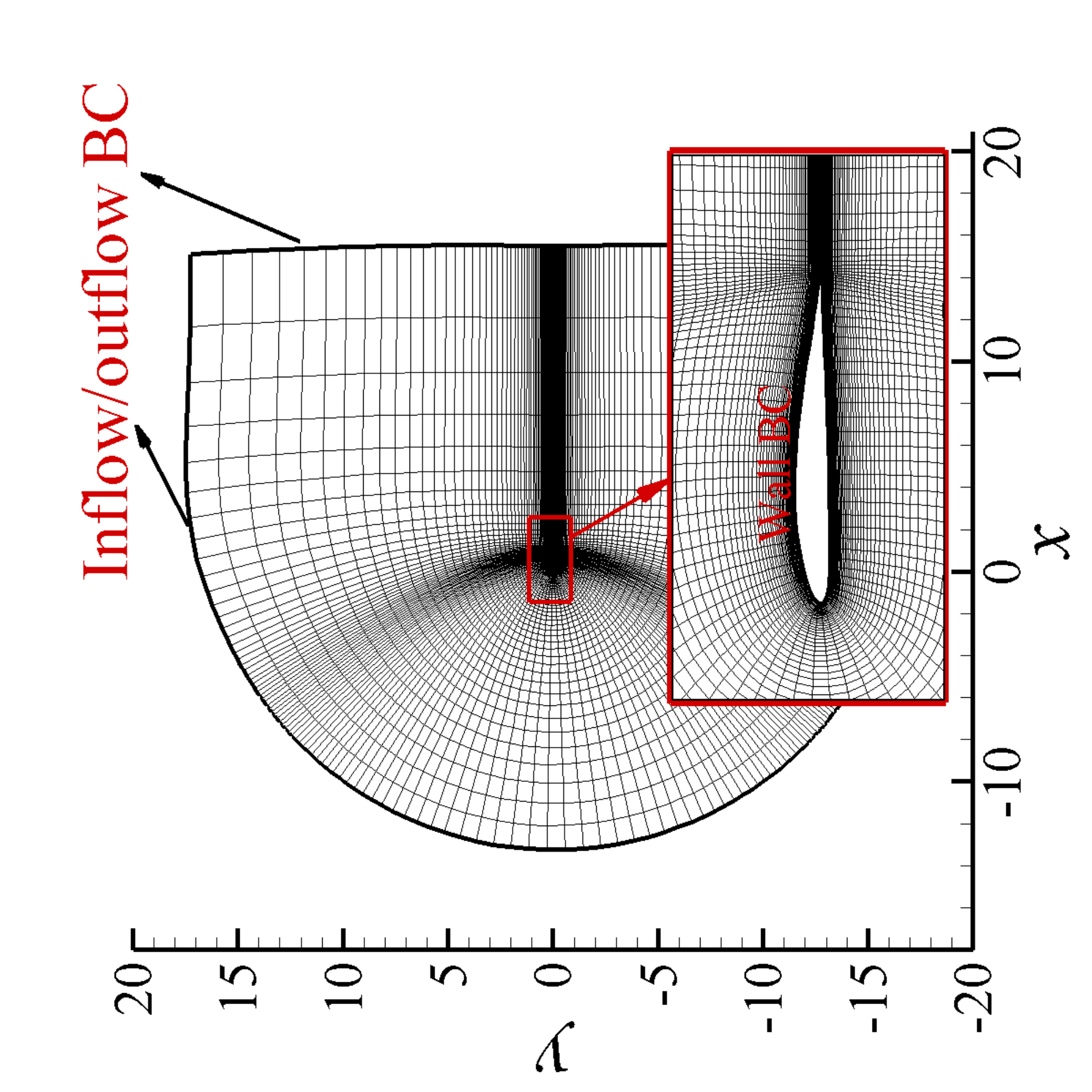}
    \caption{}
    \end{subfigure}
    \begin{subfigure}{.35\textwidth}
    \centering
    \includegraphics[width=\linewidth, angle=-90]{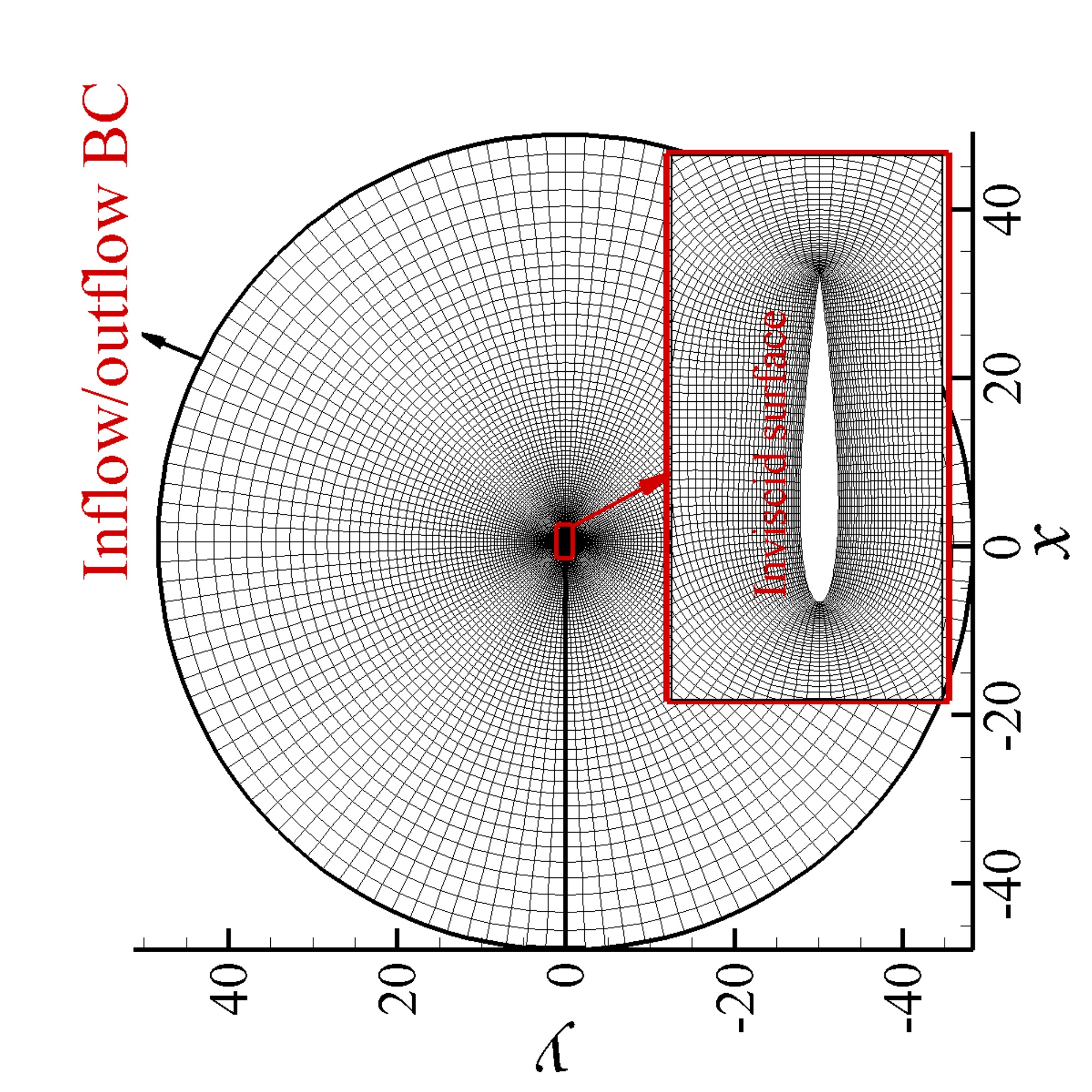}
    \caption{}
    \end{subfigure}
    \caption{a) A typical C-grid topology mesh (i.e. ``fx84w097'') in RANS simulations. 
    b) The O-grid topology mesh for ``naca0012'' in the Euler/inviscid simulation. The imposed boundary conditions are labeled.
    }
    \label{fig:cmesh_topology}
\end{figure}

\begin{figure}
    \centering
    \begin{subfigure}{.4\textwidth}
    \centering
    \includegraphics[width=\linewidth]{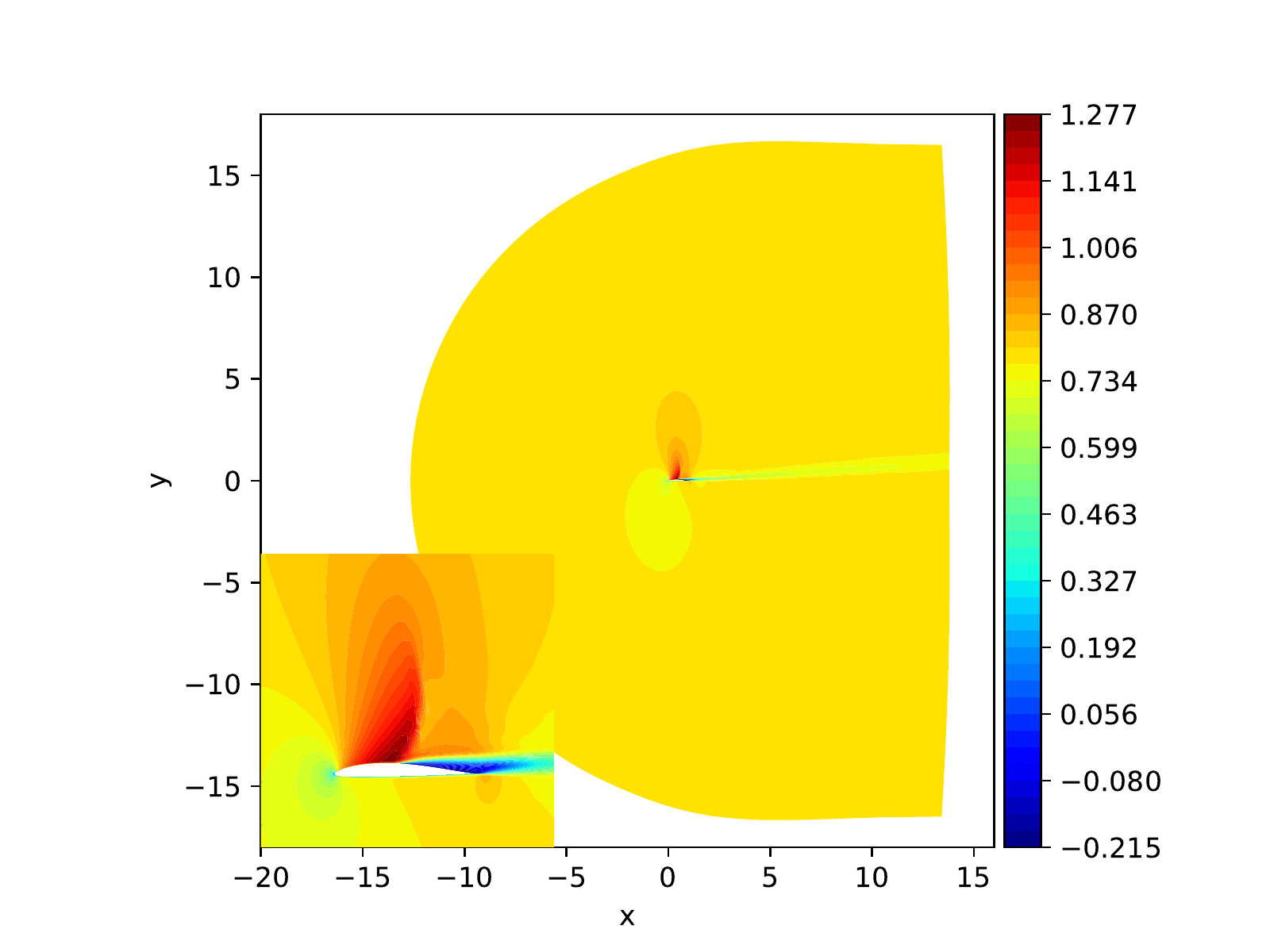}
    \caption{}
    \end{subfigure}
    \begin{subfigure}{.4\textwidth}
    \centering
    \includegraphics[width=\linewidth]{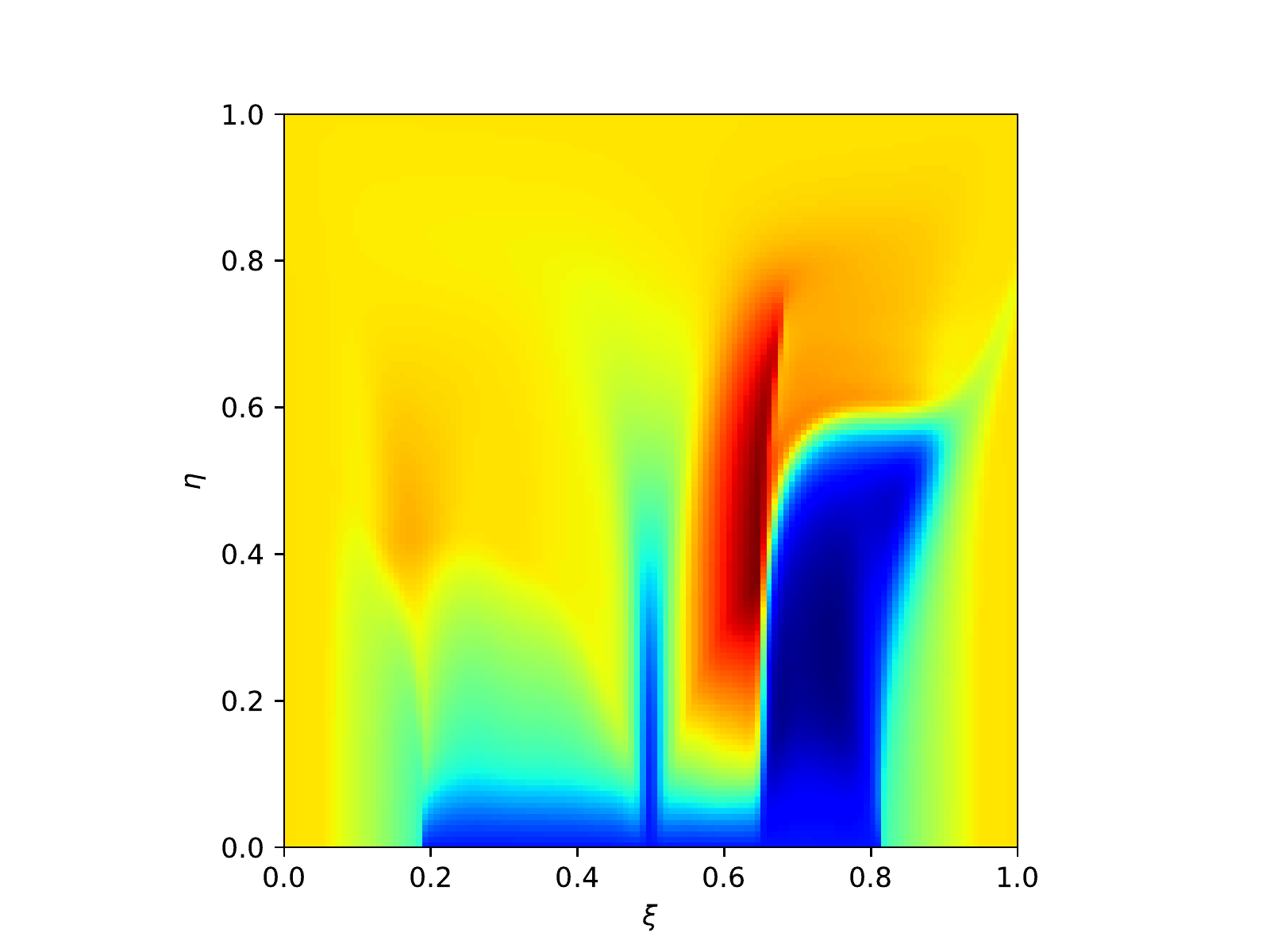}
    \caption{}
    \end{subfigure}
    \caption{a) The \emph{x}-component velocity field over aerofoil ``fx84w097'' in the physical space. b) The corresponding distribution in the curvilinear coordinate system ($\xi$, $\eta$).}
    \label{fig:variables_coordinates}
\end{figure}

\begin{figure}
\centering
\begin{subfigure}{.3\textwidth}
\centering
\includegraphics[width=\linewidth]{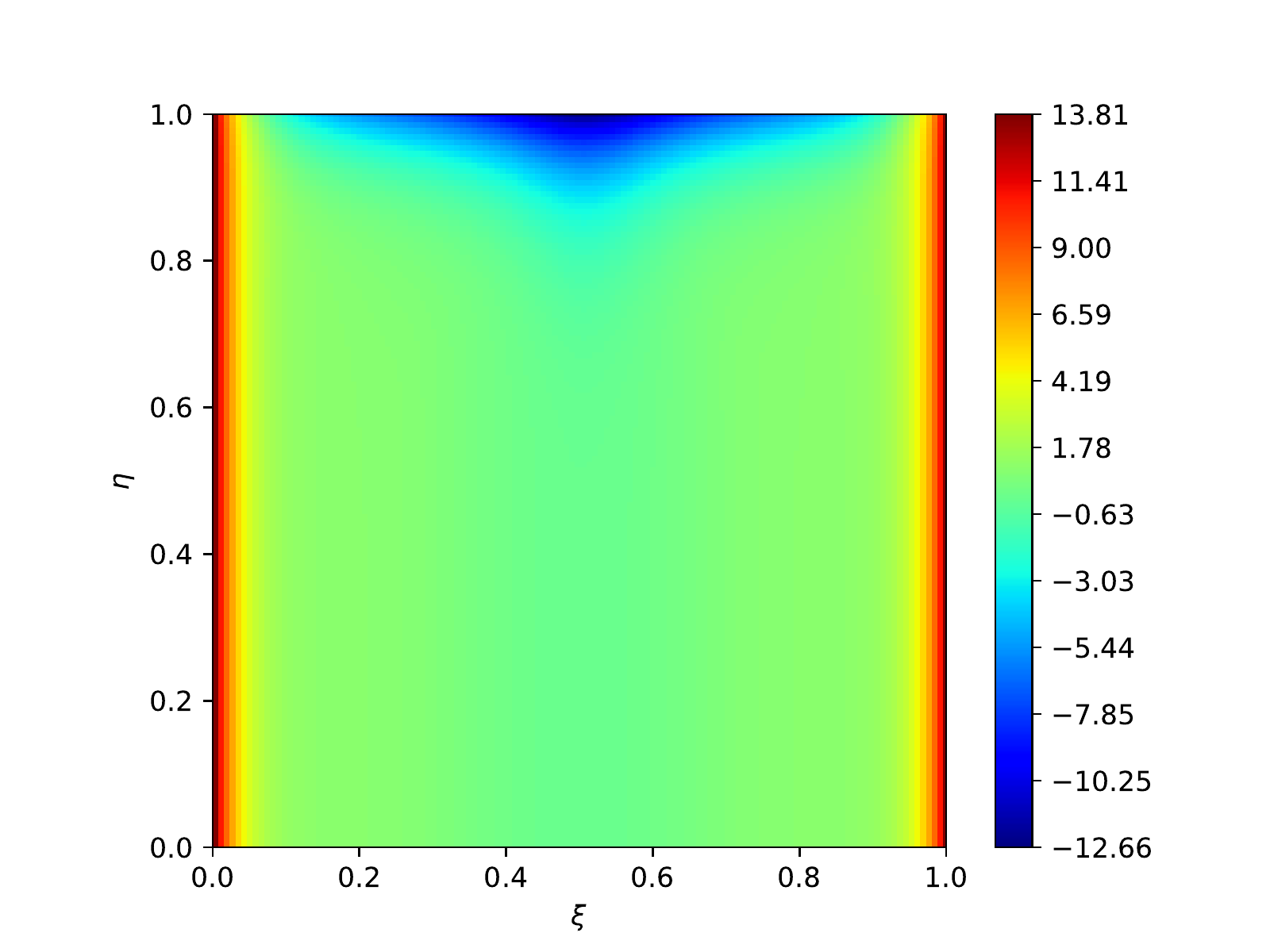}
\caption{}
\end{subfigure}
\begin{subfigure}{.3\textwidth}
\centering
\includegraphics[width=\linewidth]{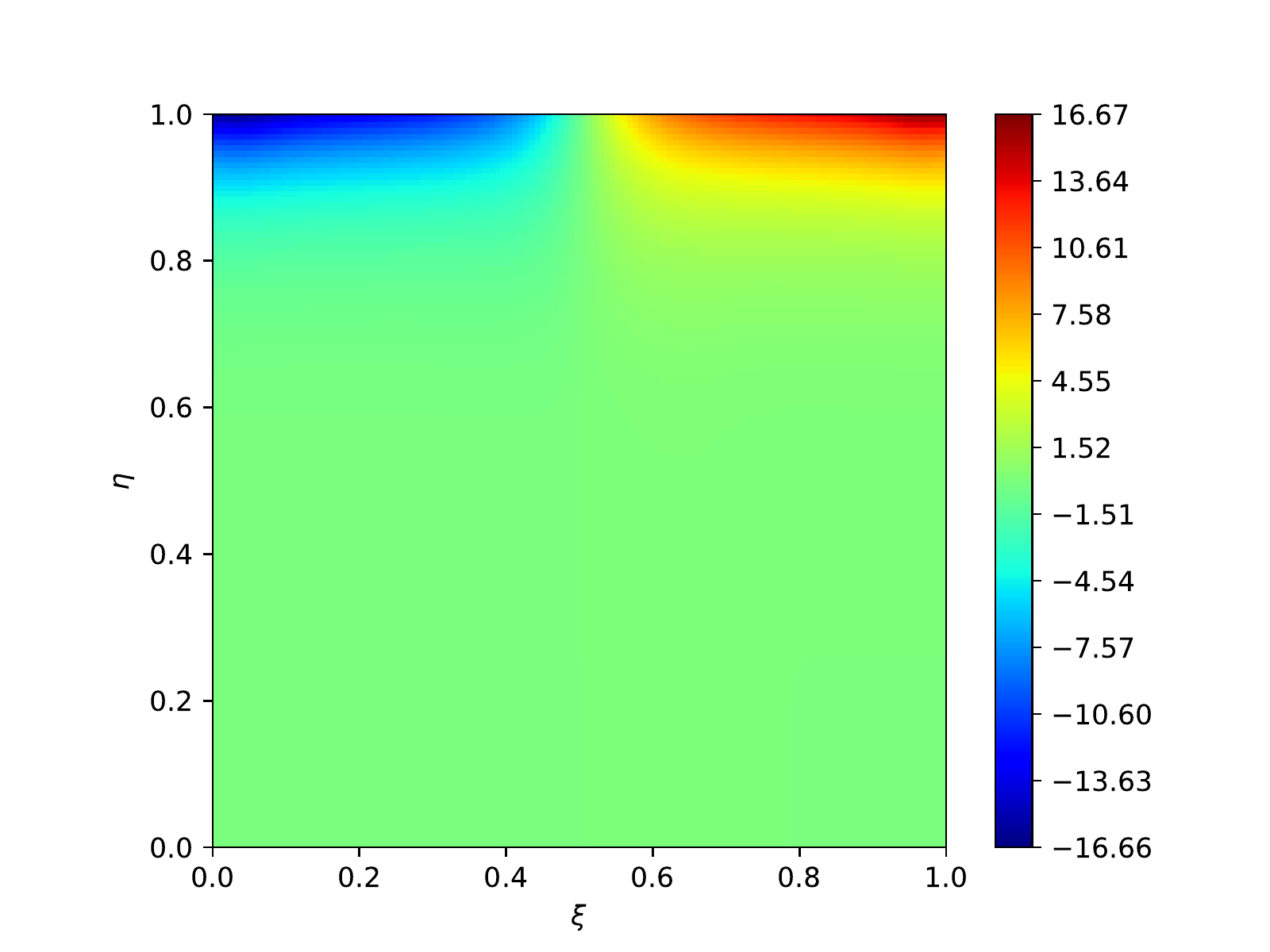} 
\caption{}
\end{subfigure}

\caption{The \emph{x} coordinate (a) and \emph{y} coordinate (b) of the mesh for aerofoil ``fx84w097'' in the curvilinear coordinate system ($\xi$, $\eta$).}
\label{fig:geo_info_method_a}
\end{figure}

This approach is fully in line with the discretization using structured grids, among which c-grid and o-grid topology meshes are most commonly used in numerical simulations of flows over aerofoils as shown in Fig. \ref{fig:cmesh_topology}. Through the coordinate transformation, variables in the physical space can be uniquely mapped into the computational space and back.
Hence, we define a 
unit cube of curvilinear coordinates as the canonical space in which the neural networks operate.
Figure \ref{fig:variables_coordinates}a shows the \emph{x}-component velocity field over the aerofoil ``fx84w097'' in the physical space. In terms of the curvilinear coordinates ($\xi$, $\eta$), which align with the array indices (\emph{i}, \emph{j}), the field becomes the image shown in Fig. \ref{fig:variables_coordinates}b. Similarly, the \emph{x} and \emph{y} coordinates of the mesh in the physical space can be visualised in the curvilinear coordinate system ($\xi$, $\eta$) as shown in Figs. \ref{fig:geo_info_method_a}a and \ref{fig:geo_info_method_a}b. 
The encoding method of directly using images of mesh coordinates $x$ and $y$ as geometric information in Fig. \ref{fig:geo_info_method_a} is denoted as ``Method A'', i.e.

\begin{equation*}
    \mathcal{G}_{_{i_{\rm max}\times j_{\rm max}\times2}}
    =\begin{bmatrix}
    x & y
    \end{bmatrix}.
\end{equation*}

\begin{figure}
\centering
\begin{subfigure}{.3\textwidth}
\centering
\includegraphics[width=\linewidth]{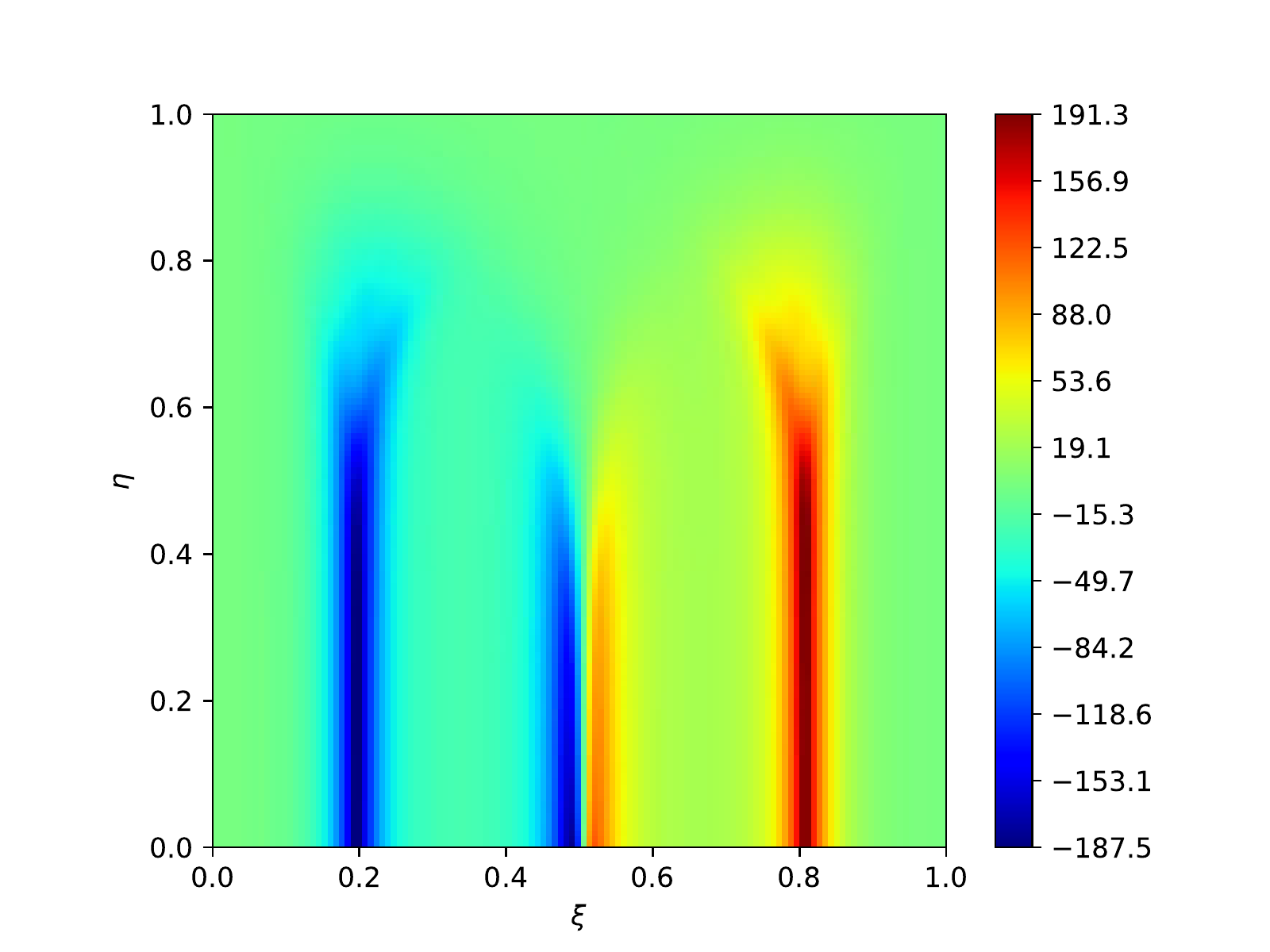}
\caption{$\xi_x$}
\end{subfigure}
\begin{subfigure}{.3\textwidth}
\centering
\includegraphics[width=\linewidth]{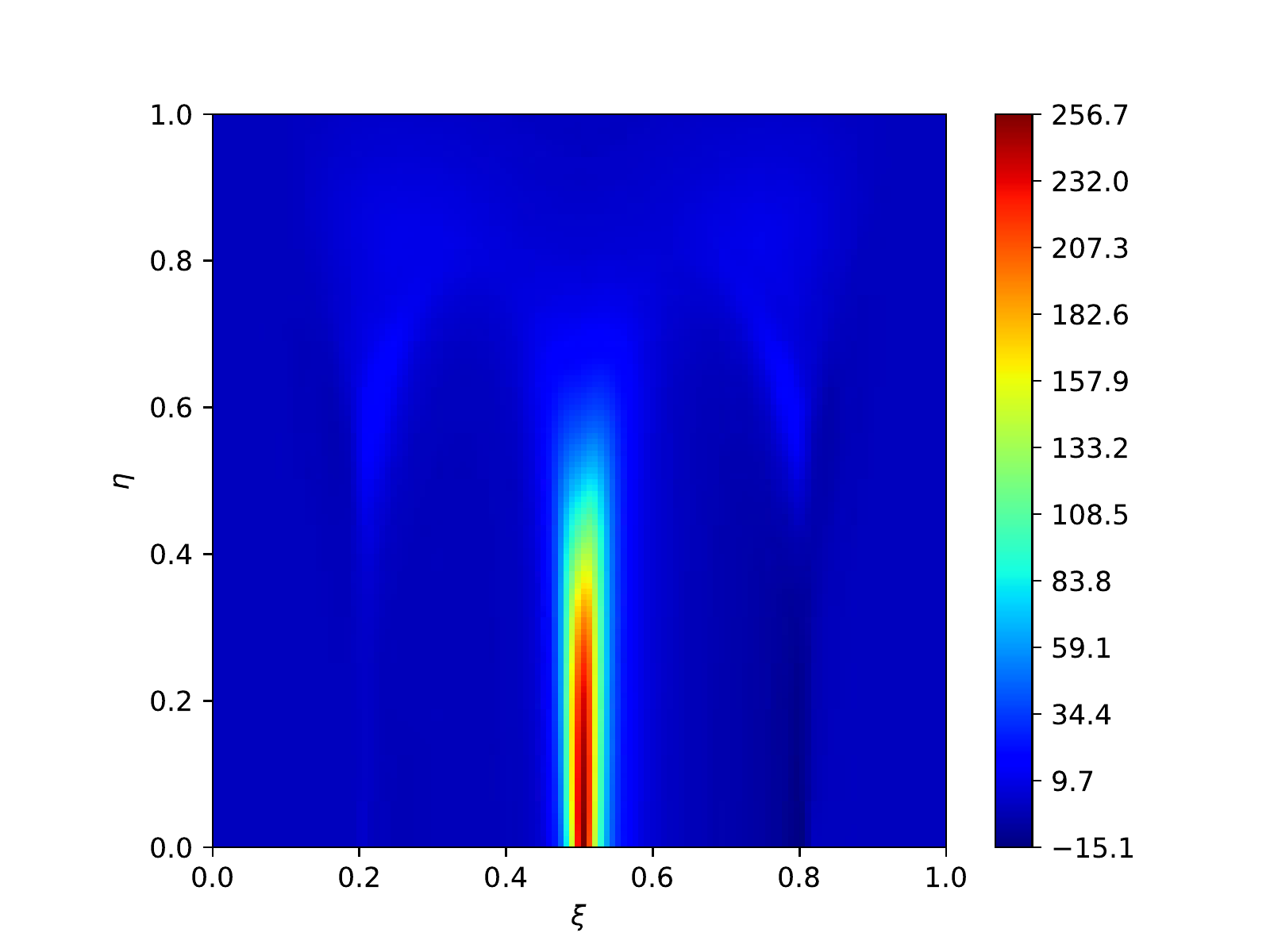}
\caption{$\xi_y$}
\end{subfigure}
\begin{subfigure}{.3\textwidth}
\centering
\includegraphics[width=\linewidth]{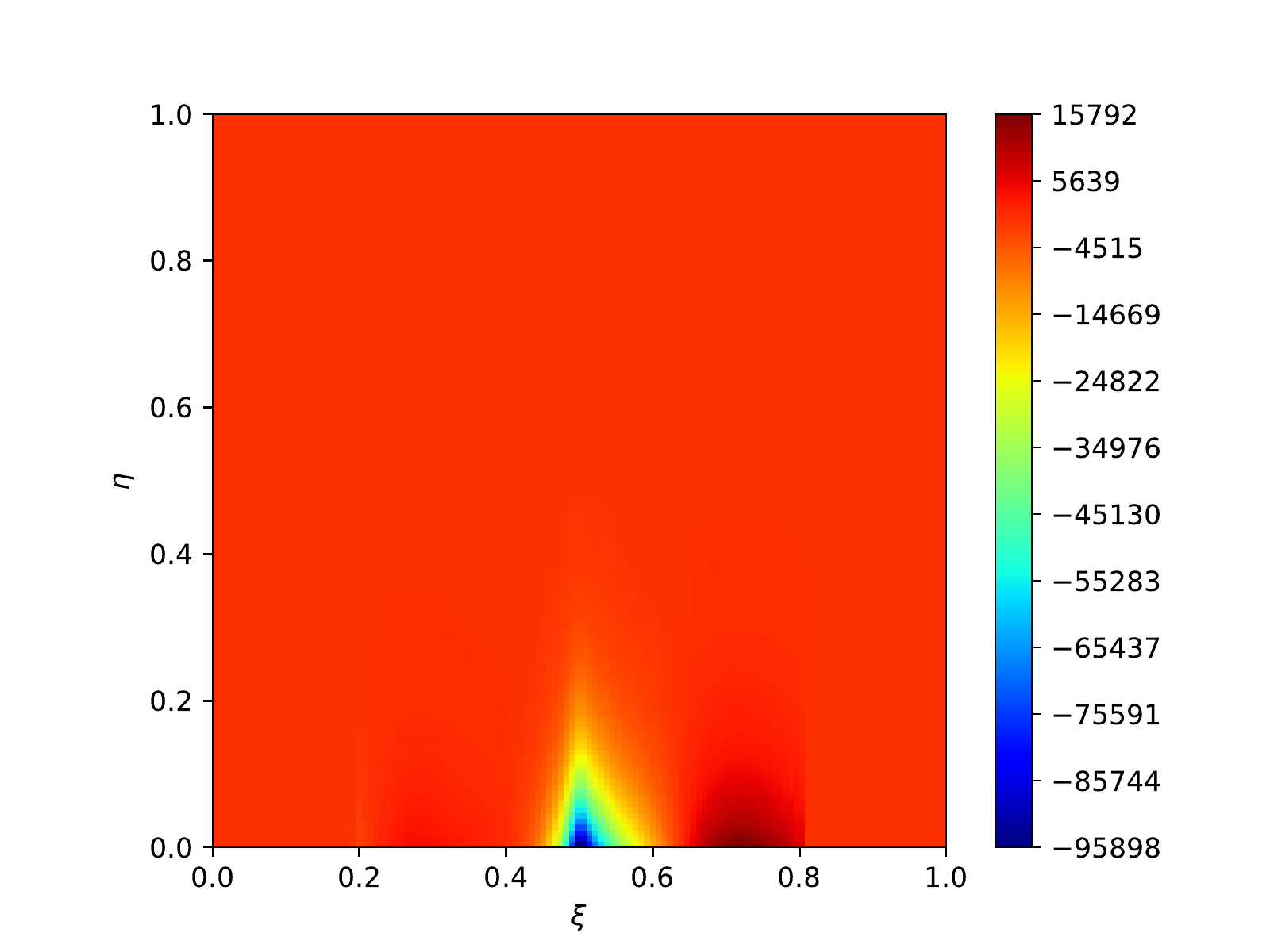}
\caption{$\eta_x$}
\end{subfigure}

\begin{subfigure}{.3\textwidth}
\centering
\includegraphics[width=\linewidth]{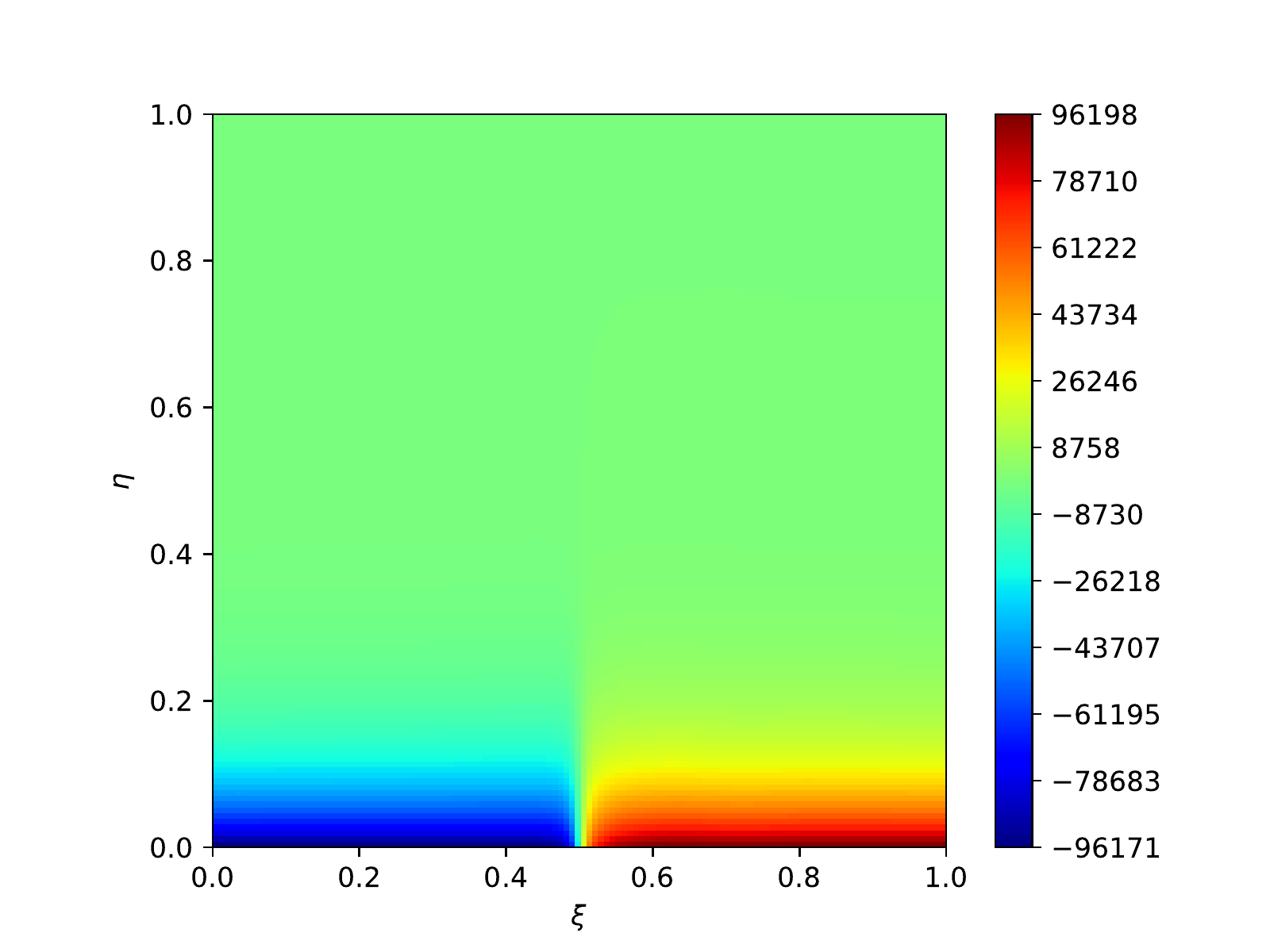}
\caption{$\eta_y$}
\end{subfigure}
\begin{subfigure}{.3\textwidth}
\centering
\includegraphics[width=\linewidth]{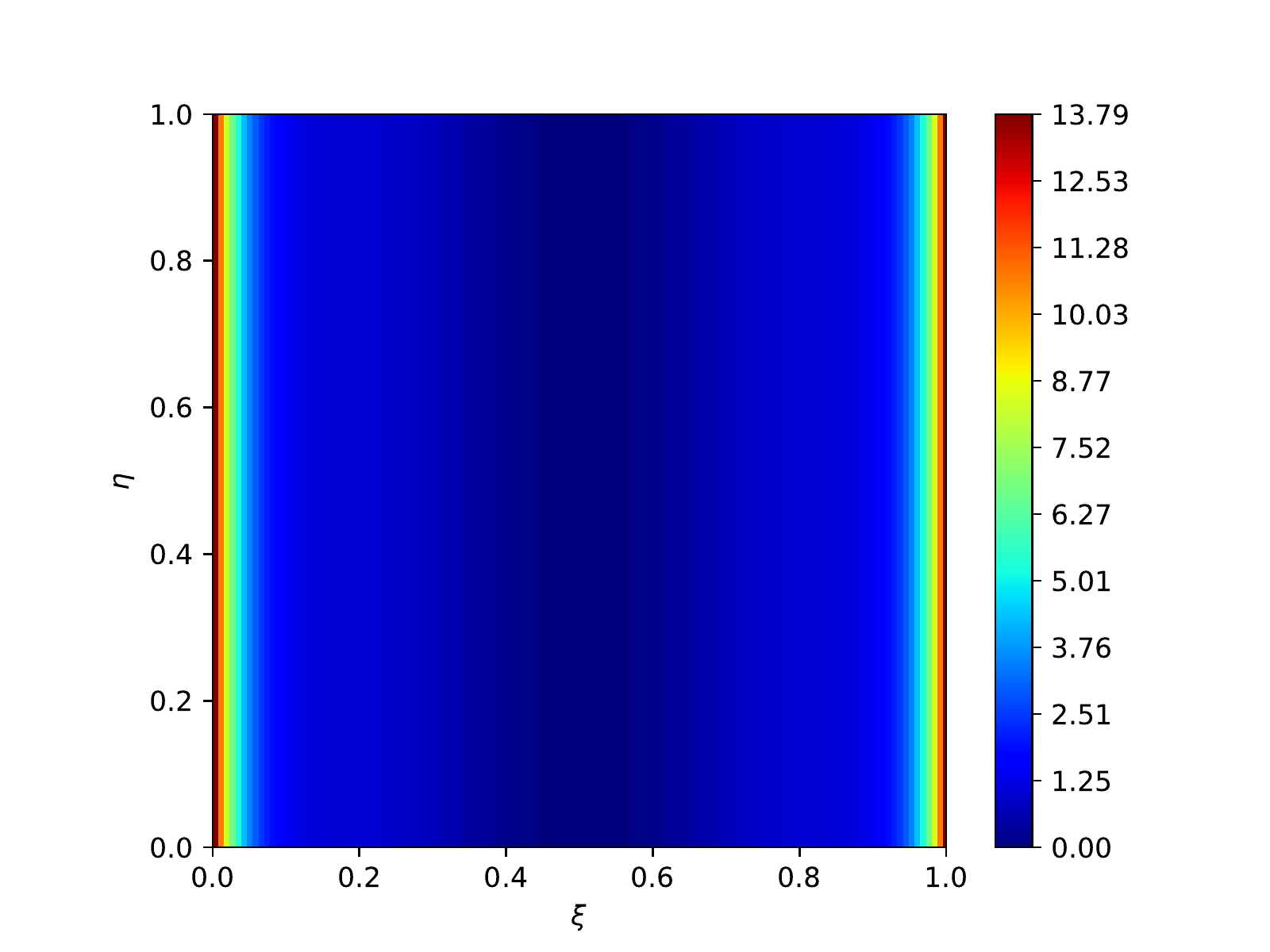}
\caption{$x_{0}$}
\end{subfigure}
\begin{subfigure}{.3\textwidth}
\centering
\includegraphics[width=\linewidth]{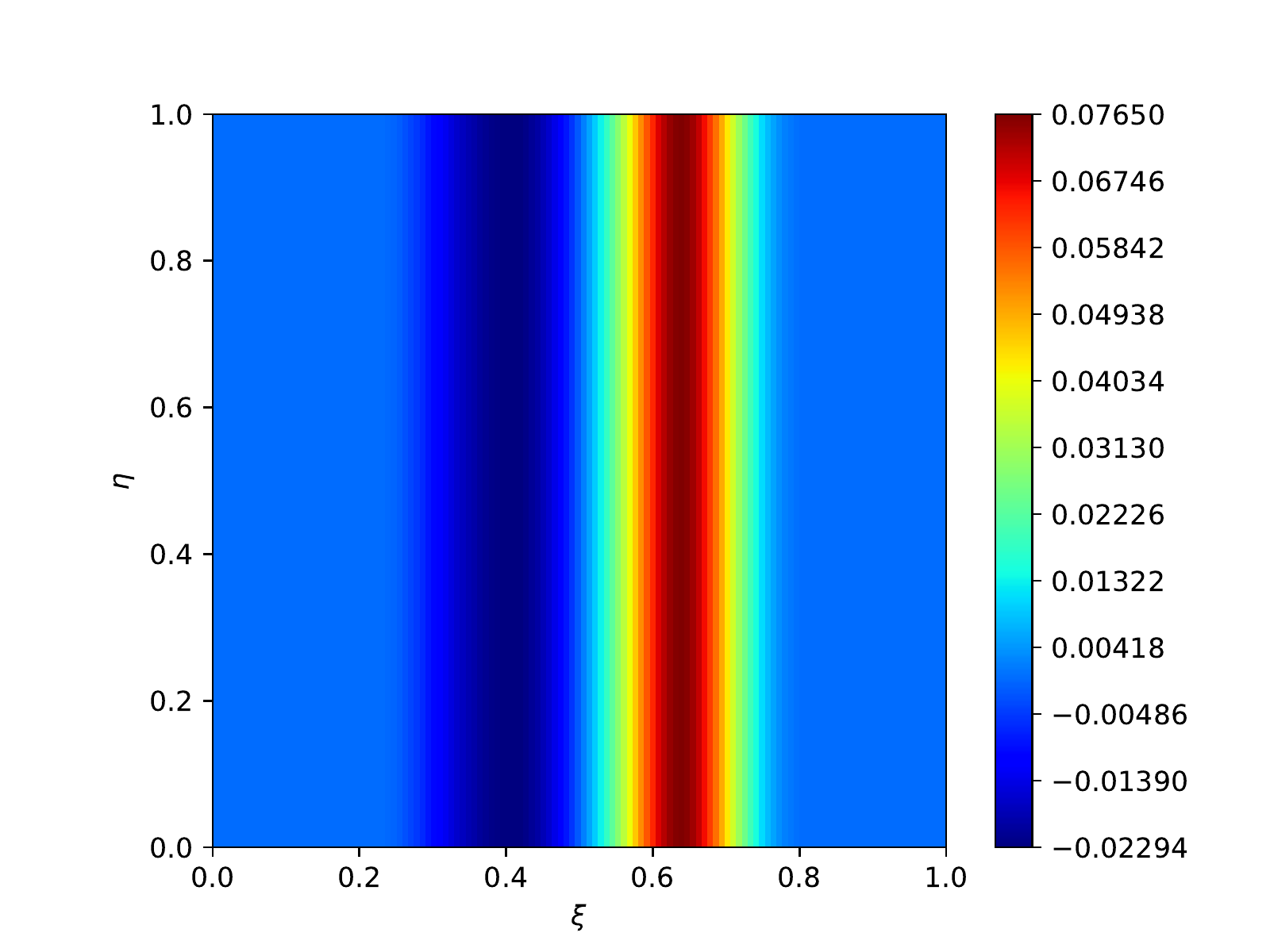}
\caption{$y_0$}
\end{subfigure}
    \caption{(a)-(f) \textcolor{lCol}{$\partial_x\xi$, $\partial_y\xi$, $\partial_x\eta$, $\partial_y\eta$}, $x_0$, and $y_0$ in the curvilinear coordinate system for the mesh of aerofoil ``fx84w097''.}
    \label{fig:geo_info_method_b}
\end{figure}

Considering the fact that the coordinate transformation matrix is univalent for a given mesh, the four components of transformation matrix $\mathbf{T}$ as well as the reference points can be used for geometric encoding, i.e. \textcolor{lCol}{$\partial_x\xi$, $\partial_y\xi$, $\partial_x\eta$, $\partial_y\eta$}, $x_0$, $y_0$. Taking the mesh of aerofoil ``fx84w097'' as an  example, the full encoded geometric information is shown in Fig. \ref{fig:geo_info_method_b}. 
We denote this one as ``Method B'', where

\textcolor{lCol}{
\begin{equation*}
    \mathcal{G}_{_{i_{\rm max}\times j_{\rm max}\times6}}
    =\begin{bmatrix}
    \partial_x\xi & \partial_y\xi & \partial_x\eta & \partial_y\eta & x_0 & y_0
    \end{bmatrix}.
\end{equation*}
}

\noindent Here, the reference coordinates ($x_0$, $y_0$) at $j=1$ or $\eta=0$ contain the full coordinates of the corresponding aerofoil 
(the shape of the $x_0$ and $y_0$ arrays is $i_{\rm max}\times1$). 
To match the network architecture, 
we extend the two variables to a size of $i_{\rm max}\times j_{\rm max}$ to keep the consistency with the shape of other inputs, 
which means the content of the resulting arrays is constant along the $\eta$ direction.

It is beneficial to consider the non-dimensional form of transformation matrix in order to find more meaningful pattern, so we define
\textcolor{lCol}{
\begin{equation}
\left.\begin{aligned}
    \hat{\xi}_x=\partial_x\xi/|\nabla \xi| \\
    \hat{\xi}_y=\partial_y\xi/|\nabla \xi|   \\
    \hat{\eta}_x=\partial_x\eta/|\nabla \eta| \\
    \hat{\eta}_y=\partial_y\eta/|\nabla \eta|
\end{aligned}\right\} 
\end{equation}}

\noindent
Therefore, the geometric information can be determined by nine variables: $J^{-1}$, $\hat{\xi}_x$, $\hat{\xi}_y$, $|\nabla \xi|/J$, $\hat{\eta}_x$, $\hat{\eta}_y$, $|\nabla \eta|/J$, $x_0$, $y_0$,
where $J$ is the transformation Jacobian defined to be

\begin{figure}
\centering
\begin{subfigure}{.3\textwidth}
\centering
\includegraphics[width=\linewidth]{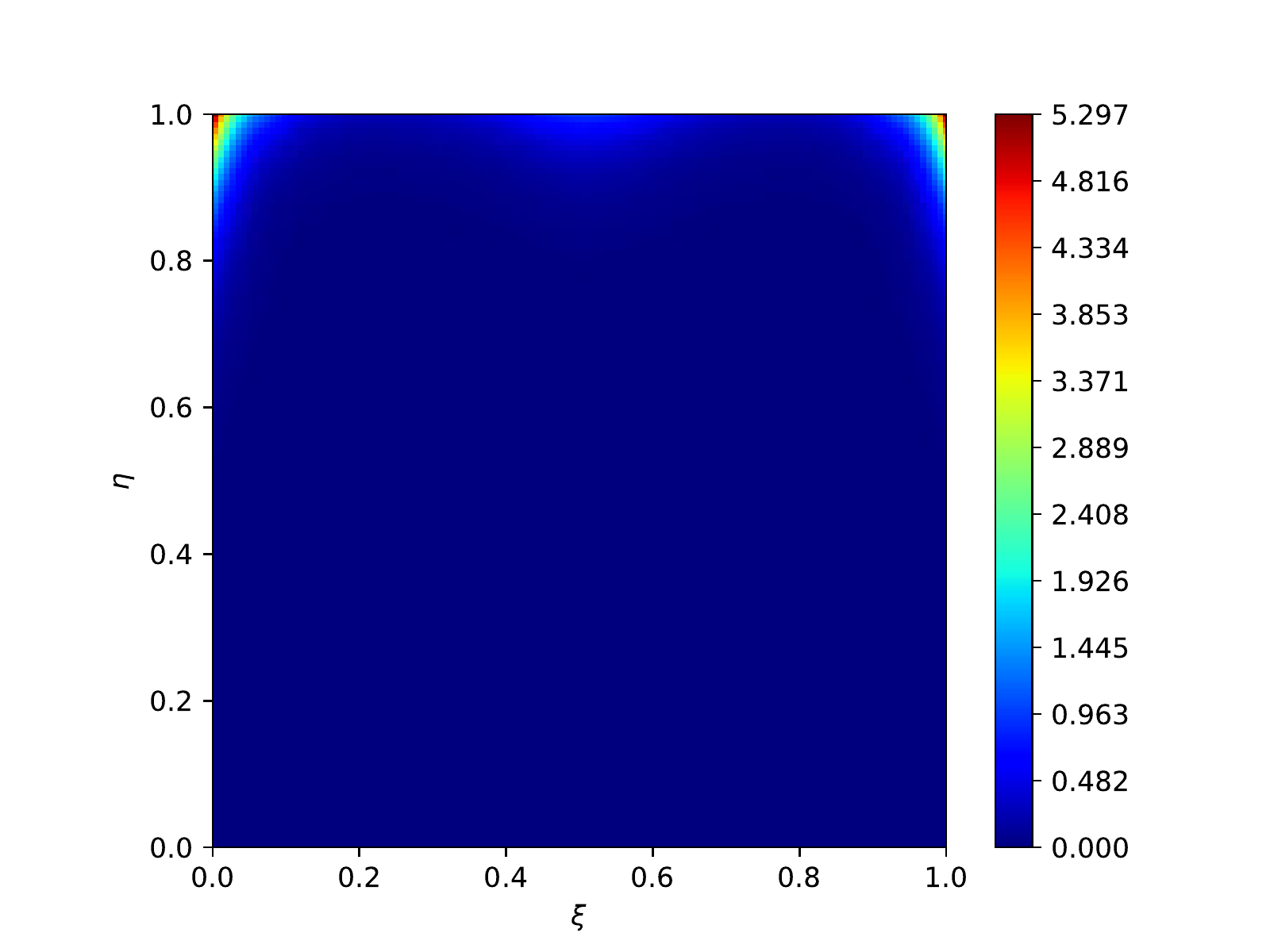}
\caption{$J^{-1}$}
\end{subfigure}
\begin{subfigure}{.3\textwidth}
\centering
\includegraphics[width=\linewidth]{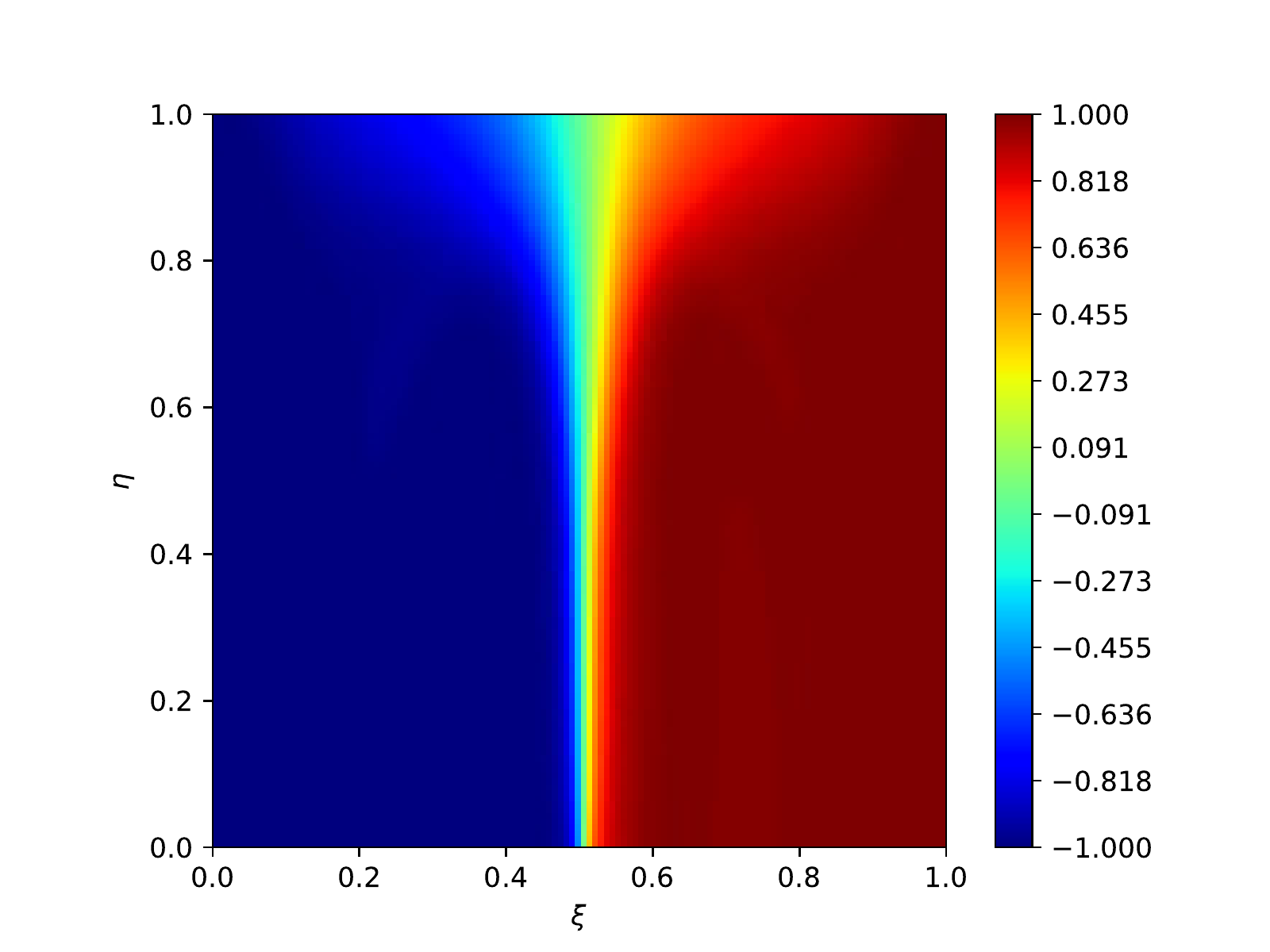}
\caption{$\hat{\xi}_x$}
\end{subfigure}
\begin{subfigure}{.3\textwidth}
\centering
\includegraphics[width=\linewidth]{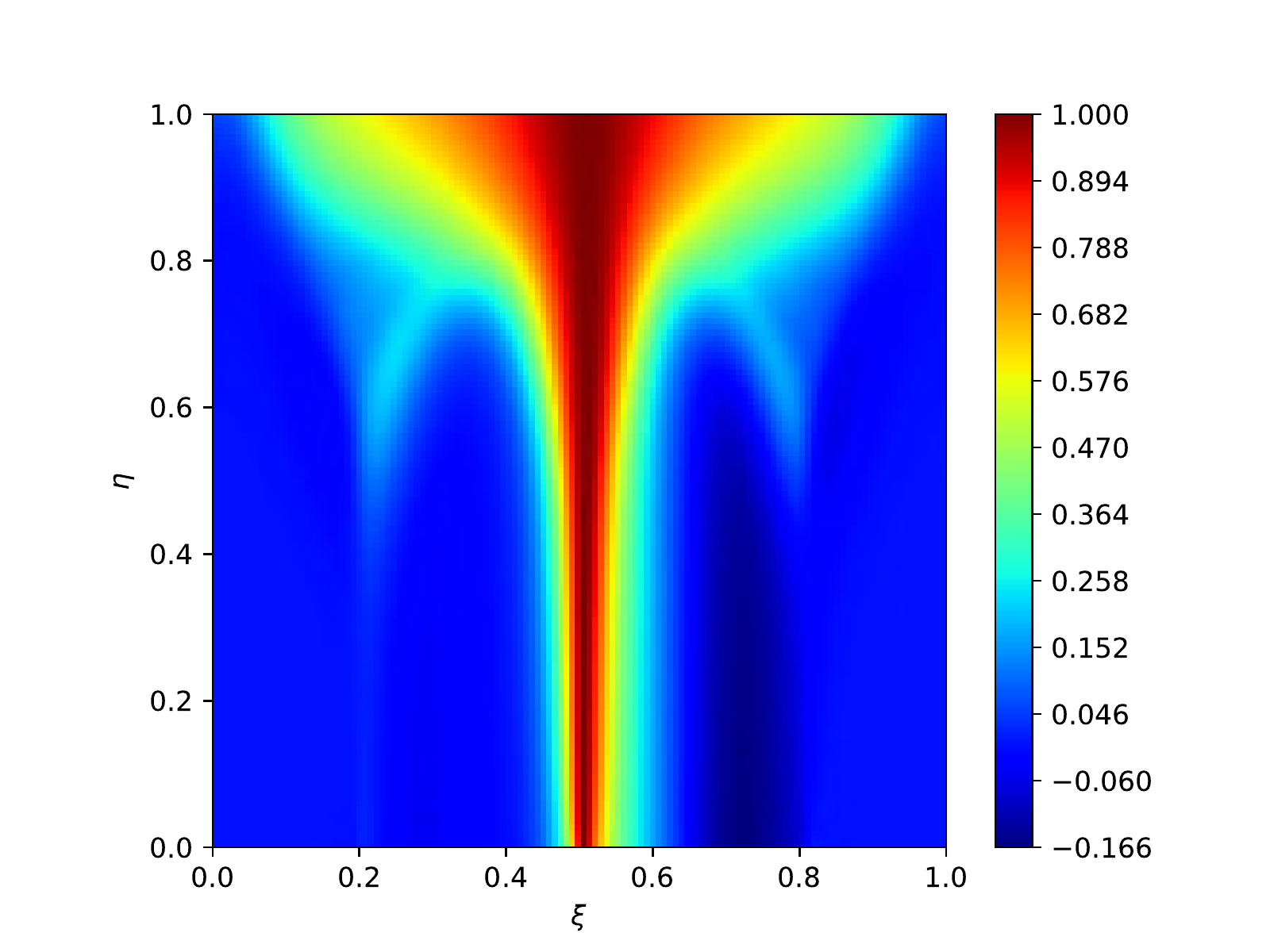}
\caption{$\hat{\xi}_y$}
\end{subfigure}

\begin{subfigure}{.3\textwidth}
\centering
\includegraphics[width=\linewidth]{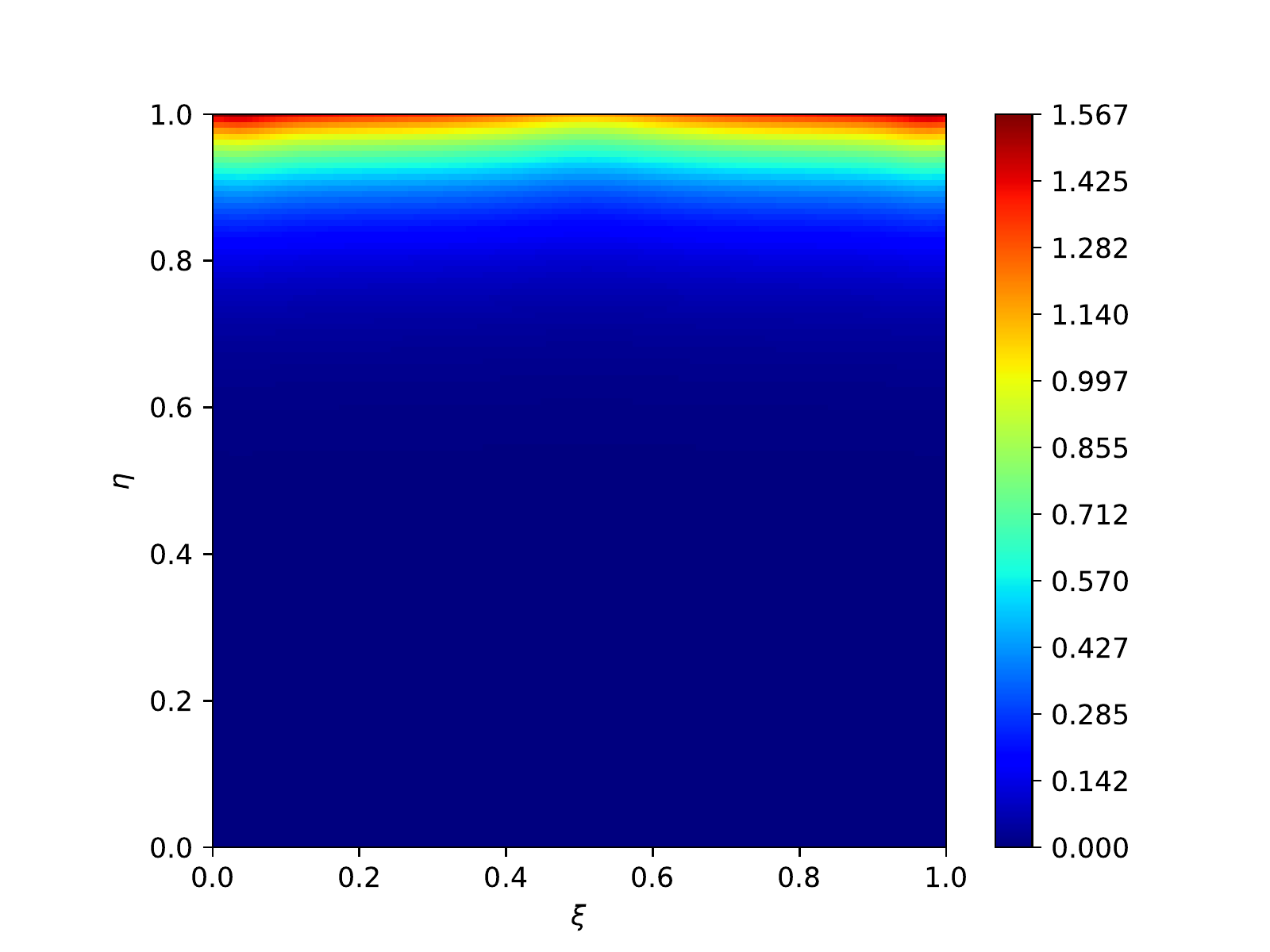}
\caption{$|\nabla \xi|/J$}
\end{subfigure}
\begin{subfigure}{.3\textwidth}
\centering
\includegraphics[width=\linewidth]{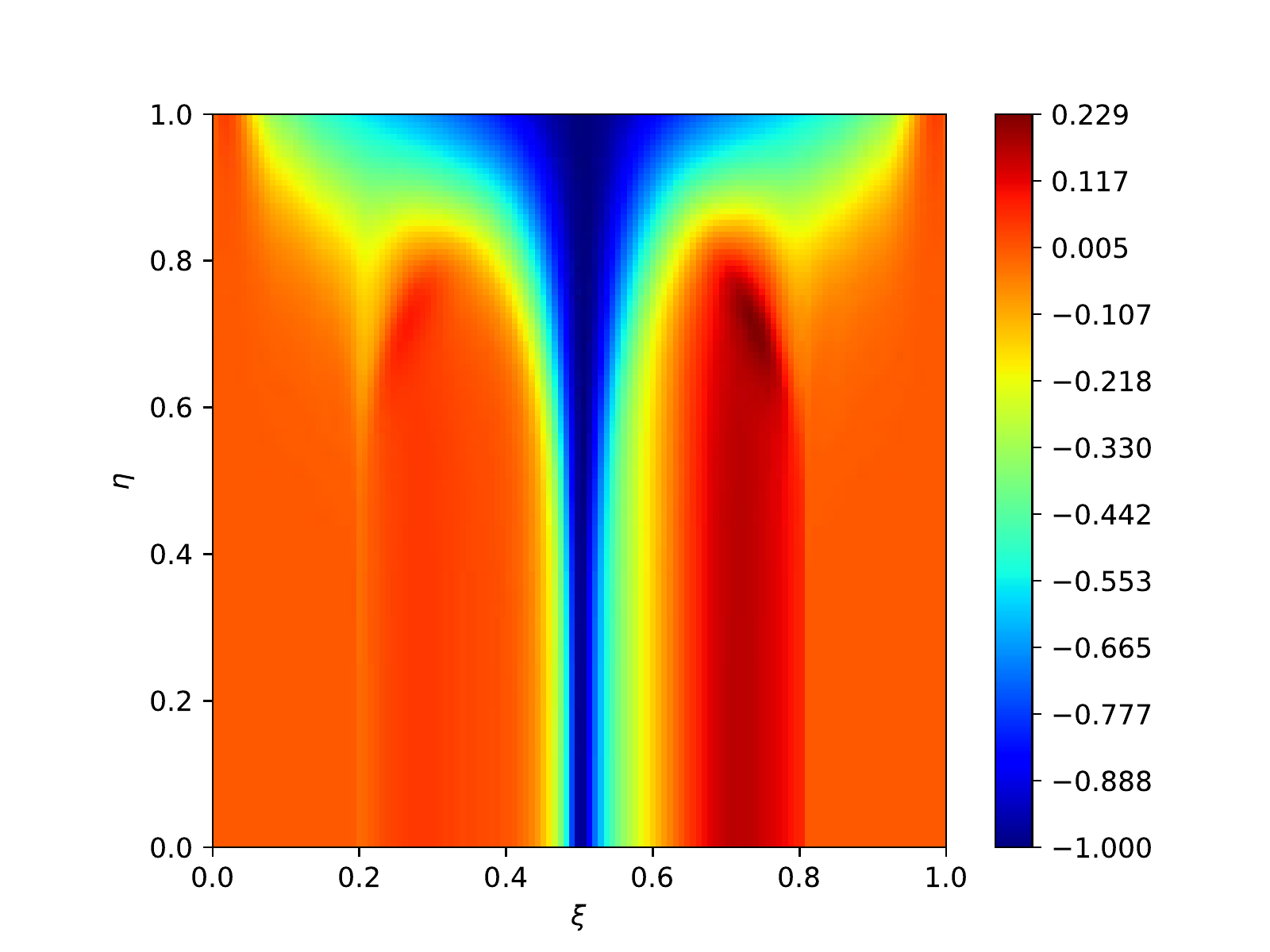}
\caption{$\hat{\eta}_x$}
\end{subfigure}
\begin{subfigure}{.3\textwidth}
\centering
\includegraphics[width=\linewidth]{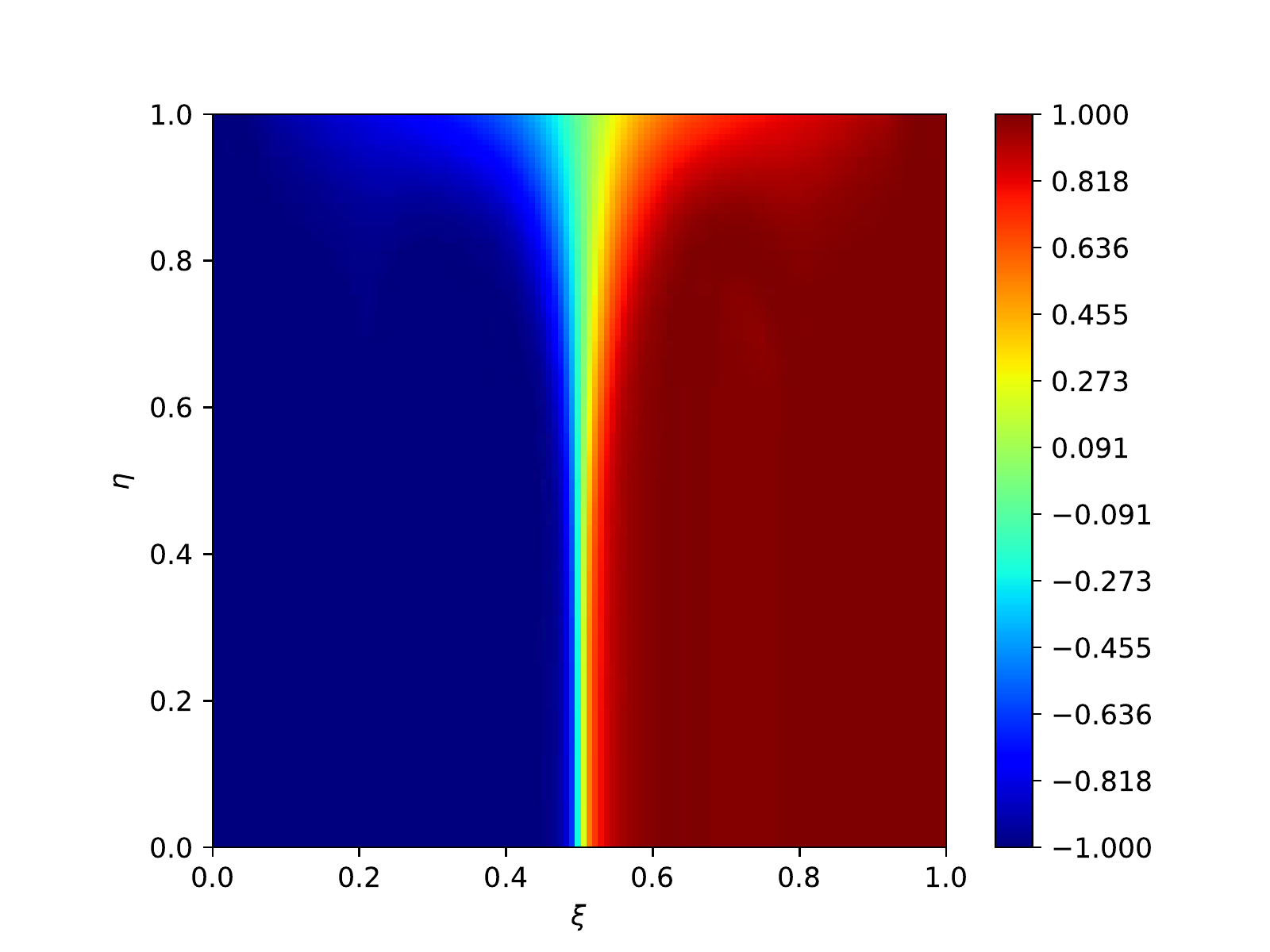}
\caption{$\hat{\eta}_y$}
\end{subfigure}

\begin{subfigure}{.3\textwidth}
\centering
\includegraphics[width=\linewidth]{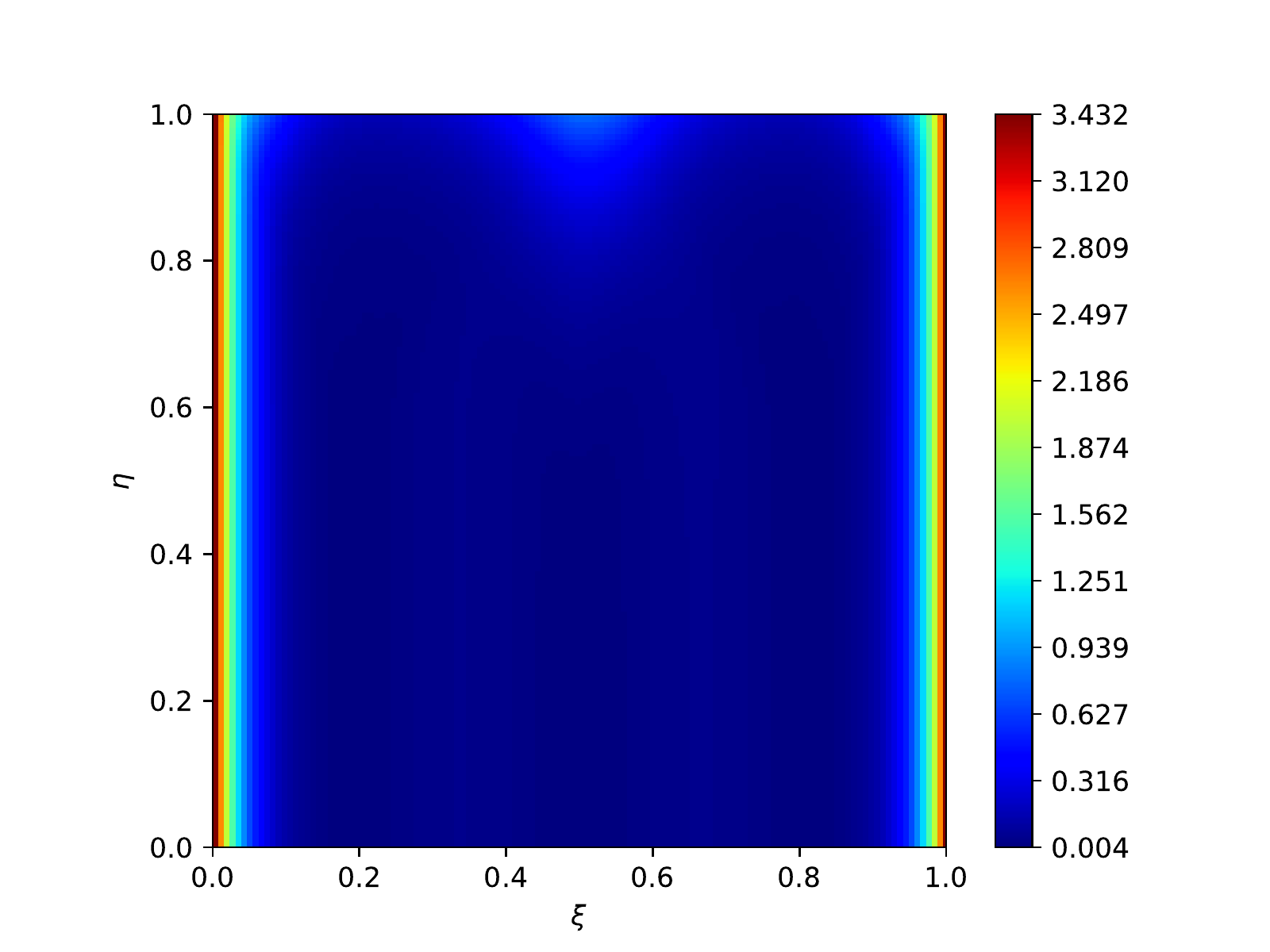}
\caption{$|\nabla \eta|/J$}
\end{subfigure}
\begin{subfigure}{.3\textwidth}
\centering
\includegraphics[width=\linewidth]{coordinates_canonical_xin-eps-converted-to.pdf}
\caption{$x_0$}
\end{subfigure}
\begin{subfigure}{.3\textwidth}
\centering
\includegraphics[width=\linewidth]{coordinates_canonical_yin-eps-converted-to.pdf}
\caption{$y_0$}
\end{subfigure}

\caption{(a)-(i) $J^{-1}$, $\hat{\xi}_x$, $\hat{\xi}_y$, $|\nabla \xi|/J$, $\hat{\eta}_x$, $\hat{\eta}_y$, $|\nabla \eta|/J$, $x_0$, and $y_0$ in the curvilinear coordinate system for the mesh of aerofoil ``fx84w097''.}
    \label{fig:geo_info_method_c}
\end{figure}

\textcolor{lCol}{
\begin{equation*}
    J=
    \begin{vmatrix}
    \partial_x\xi & \partial_y\xi\\
    \partial_x\eta & \partial_y\eta 
    \end{vmatrix}.
    \end{equation*}}
\noindent
As we are dealing with a two-dimensional space, the geometrical interpretation of those variables is as follows: $J^{-1}$ is the area of the grid cell, $\hat{\xi}_x$ the \emph{x} component of unit normal to \emph{i} edge, $\hat{\xi}_y$ the \emph{y} component of unit normal to \emph{i} edge, $|\nabla \xi|/J$ the length of \emph{i} edge, $\hat{\eta}_x$ the \emph{x} component of unit normal to \emph{j} edge, $\hat{\eta}_y$ the \emph{y} component of unit normal to \emph{j} edge, $|\nabla \eta|/J$ the length of \emph{j} edge, and ($x_0$, $y_0$) the reference coordinates on the aerofoil surface, respectively. In fact, the use of non-dimensional metrics is a common practice in the implementation of finite volume approach. For more details on the geometrical interpretation, we refer to \cite{Krist_Biedron_Rumsey1998manual}, \cite{blazek2001computational} and \cite{WangSW2002}.
The nine variables in curvilinear coordinate system are shown in Fig. \ref{fig:geo_info_method_c} taking the mesh of aerofoil ``fx84w097'' for example. We denote this method as ``Method~C'', where

\begin{equation*}
    \mathcal{G}_{_{i_{\rm max}\times j_{\rm max}\times9}} 
    =\begin{bmatrix}
    J^{-1} & \hat{\xi}_x & \hat{\xi}_y & |\nabla \xi|/J & \hat{\eta}_x & \hat{\eta}_y & |\nabla \eta|/J & x_0 & y_0
    \end{bmatrix}.
\end{equation*} 


The three encoding methods are summarized in Table \ref{tab:three_encoding_methods}. For all the three methods, the encoded free stream conditions are Mach number, Reynolds number and the angle of attack (AoA), i.e. 

\begin{equation*}
    \mathbf{x}_{_{i_{\rm max}\times j_{\rm max}\times3}}= 
    \begin{bmatrix}
    \xmach_{\infty} & \alpha_{\infty} & \Rey_{\infty}
    \end{bmatrix}.
\end{equation*}

\noindent
Note that we extend each variable to a size of $i_{\rm max}\times j_{\rm max}$ to match the other inputs in the neural network architecture, which means that each of the three channels in the $\mathbf{x}$ tensor is constant.

\textcolor{lCol}{The RANS equations in generalized coordinates are derived in \ref{appA:CFD_equations}. Hence, for methods B and C, discretized PDEs given by equation \ref{eq:discreteForm} are rewritten into equation \ref{eq:discreteForm2} and equation \ref{eq:discreteForm3}, respectively, and the supervised learning can still be described by equation \ref{eq:training_objective}.}

\begin{table}
  \begin{center}
\def~{\hphantom{0}}
\resizebox{\columnwidth}{!}{  \begin{tabular}{cccc} 
      \toprule
      Method    & Geometric information $\mathcal{G}$ &   free-stream condition $\mathbf{x}$ & No. of channels   \\
      \midrule 
      A    & $x$, $y$ & $\xmach{}_{\infty}$, $\alpha_{\infty}$, $\Rey_{\infty}$ & $N=5$\\
      B    & \textcolor{lCol}{$\partial_x\xi$, $\partial_y\xi$, $\partial_x\eta$, $\partial_y\eta$}, $x_0$, $y_0$ & $\xmach_{\infty}$, $\alpha_{\infty}$, $\Rey{}_{\infty}$ & $N=9$\\
      C    & $J^{-1}$, $\hat{\xi}_x$, $\hat{\xi}_y$, $|\nabla \xi|/J$, $\hat{\eta}_x$, $\hat{\eta}_y$, $|\nabla \eta|/J$, $x_0$, $y_0$  & $\xmach_{\infty}$, $\alpha_{\infty}$, $\Rey_{\infty}$ & $N=12$\\
      \bottomrule
  \end{tabular}
  }
  \caption{Three encoding methods for DNN models.}
  \label{tab:three_encoding_methods}
  \end{center}
\end{table}

\subsection{\textcolor{lCol}{Computational details}}
\noindent
The open source structured-grid code CFL3D version 6.7 is used to obtain the ground truth data \citep{Rumsey1997, Rumsey2010}. The two-dimensional \textcolor{lCol}{RANS} in generalized coordinates \textcolor{lCol}{(see Appendix C)} are numerically solved by the finite-volume method. The convective terms are discretized with third-order upwind scheme and viscous terms with second-order differencing \citep{roe1981fds}.
For low-speed steady cases, low-Mach number preconditioning is employed \citep{weiss_smith1995precond}. Spalart-Allmaras one-equation model is used for turbulence closure \citep{spalart1992}. 

In the present work, we particularly focus on aerofoils with sharp trailing edges, so we choose 990 aerofoil shapes from the UIUC database. 
The open source code Construct2D version 2.1.4 is used to generate a 2D structured curvilinear grid. 
The leading edge of the aerofoil is taken as the origin of the coordinate system and the chord length is $c$, used as the reference length for nondimensionalization. We develop a template with the same grid topology to mesh the domain around aerofoils. 
For a c-grid topology mesh, the resolution is kept the same, i.e. 128 grid cells in the wall-normal direction, and 25 grid points in the wake, and 79 grid points around the aerofoil surface. In the case of an o-grid topology, there are 128 grid cells in the wall-normal direction and 128 cells in the circumferential direction around the aerofoil. 
Typical c-grid and o-grid meshes with $128\times128$ cells are given in Figs. \ref{fig:cmesh_topology}a and \ref{fig:cmesh_topology}b. \textcolor{black}{That means we use $i_{max}=128$ and $j_{max}=128$ in the present study.}
From the viewpoint of practical engineering, the mesh resolution used in the present study is meaningful at the preliminary design stage. Note that our main focus here is to showcase the trained DNN model is capable of providing fast and accurate prediction by comparing with the ground truth results, and the proposed methodology can be easily extended to high-resolution cases. 


\textcolor{lCol}{The number of degrees of freedom to describe the flowfield of interest is 65536 (i.e. $128\times128\times4$), evaluated based on the resolution of the discretized domain and the number of independent variables (i.e. $\rho$, $u$, $v$, and $a$). In the case of the RANS simulation for data generation, an additional transport equation (i.e. the one-equation Sparlart-Allmaras model) is solved to determine the eddy viscosity $\mu_T$ \citep{spalart1992}, so the total number of degrees of freedom for the CFD solver is $128\times128\times5=81920$.}

As shown in Fig. \ref{fig:cmesh_topology}a, when using a c-grid topology mesh, an inflow/outflow boundary based on one-dimensional Riemann invariant is imposed at about 20\emph{c} away from the aerofoil in the (\emph{x}, \emph{y}) plane. The grid stretching is employed to provide higher resolution near the surface and in the wake region and the minimal wall-normal grid spacing is $1\times10^{-5}$ to ensure $y^+_n<1.0$ for the Reynolds numbers within the range from $5\times10^{5}$ to $5\times10^{6}$. In viscous cases, a no-slip adiabatic wall boundary condition is applied on the aerofoil surface. 
The o-grid topology mesh in Fig. \ref{fig:cmesh_topology}b is only used in the inviscid cases, in which the slip wall boundary condition is imposed on the surface. The same inflow/outflow boundary is imposed at about 45\emph{c} away from the surface.

\subsection{\textcolor{lCol}{Ground truth data}}
To assess the reliability and accuracy of the learned model thoroughly, we consider four different cases to generate ground-truth flow fields for training, validation and test sets:

The first scenario is compressible flow over aerofoils, i.e. 990 different aerofoils, at a fixed free stream condition $\xmach_{\infty}=0.4$, $\Rey_{\infty}=1\times10^{6}$, and $\alpha_{\infty}=2.5^{\circ}$.
\textcolor{lCol}{The ground truth simulations are performed for all the aerofoils (i.e. ``one simulation for one mesh'') at the given free stream condition. Thus, we obtain 990 flow-field samples in total.} 
The motivation of this scenario is to isolate and evaluate the effect of changing geometric information only, so the input channels for free-stream conditions are turned off. 

The second case is low Mach number (i.e. $\xmach_{\infty}=0.1$) viscous flow over aerofoils with varying angles of attack and Reynolds numbers. 
\textcolor{lCol}{The free-stream condition for a ground truth simulation is sampled from a uniform probability distribution over the range of Reynolds numbers $\Rey_{\infty}\in[0.5, 5]$ million and angles of attack [-22.5, 22.5].}
This case is originally proposed by \cite{thuerey2018deep} and is used as a benchmark case in our study. All the input channels are active, although the one for Mach number is constant.

The third one is compressible inviscid flow over the aerofoil ``naca0012'', which is an interpolation case using small dataset proposed by \cite{deAvila_icml2020_6414}.

The fourth is transonic viscous flow over aerofoils with various angles of attack, Mach numbers and Reynolds numbers, which is our ultimate goal, i.e. to train a model that accurately infers the RANS solutions of compressible viscous flows including shock wave and boundary layer interactions \cite{Holder1955reynolds}. 
\textcolor{lCol}{The free-stream condition for a ground truth simulation is randomly selected from a range of Mach numbers $\xmach_{\infty}\in[0.55, 0.8]$, Reynolds numbers $\Rey_{\infty}\in[0.5, 5]$ million, and angles of attack [$-0.5^{\circ}$, $8.0^{\circ}$], following a uniform distribution.} 

The parameters used for the four cases are summarized in Table \ref{tab:testcases_summary}.

The typical data generation routine is as follows:

(1) Free-stream conditions and the aerofoil shapes are randomly selected from the range of interest as shown in Table \ref{tab:testcases_summary}.

(2) $\mathcal{G}=g(\mathbf{s})$ is technically realized by mesh generation. In Cases 1, 2 and 4, once an aerofoil $\mathbf{s}$ is chosen, a c-grid topology mesh is automatically generated based on the previously mentioned template. Note that in the current procedure, one aerofoil only has one unique mesh discretization, that means, we eventually have 990 c-grid topology meshes for 990 aerofoils. 
In Case 3, as we only consider the aerofoil ``naca0012'', an o-grid topology mesh is generated.

(3) CFL3D code is run with the generated mesh. Once converged, the flow field (i.e. density, velocity components in \emph{x} and \emph{y} directions and the speed of sound), the corresponding geometric information and free stream conditions are saved as a \textcolor{black}{labeled data sample} for training, validation and test. All of the saved variables are in the canonical space or computational space ($\xi$, $\eta$). Note also that the aerofoil shape information has been included or fully represented by the grid line $j=1$ or $\eta=0$.

Unless noted otherwise, we use a 80\% to 20\% random split for training and validation datasets. We found validation sets of several hundred samples to yield stable estimates, and hence
use an upper limit of 400 as the maximal size of the validation dataset. The validation set allows for an unbiased evaluation of the quality of the trained model during training, for example, to detect overfitting. In addition, as learning rate decay is used, the variance of the learning iterations gradually decreases, which lets the training process fine-tune the final state of the model.
To later on evaluate the capabilities of the trained models with respect to generalization, we use an additional set of 20 aerofoil shapes that were not used for training, to generate a test dataset with 20 samples (using the same range of Mach, Reynolds numbers and angles of attack as described in Table \ref{tab:testcases_summary}).

\begin{table}
  \begin{center}
\def~{\hphantom{0}}
   \resizebox{\columnwidth}{!}{\begin{tabular}{cccc}
      \toprule
      Case    & No. of aerofoils & Training/vali./test & Free-stream ($_{\infty}$)\\
      \midrule 
      1    & 970/20 & 780/190/20 & $\xmach=0.4$, $\Rey=1\times10^{6}$, $\alpha=2.5^{\circ}$\\
\cmidrule(lr{.5em}){1-4} 
      2    & 970/20 & 9300/400/20 & $\xmach=0.1$, $Re\in[5\times10^5, 5\times10^{6}]$, \\ 
           &     &                 & $\alpha\in[-22.5^{\circ}, 22.5^{\circ}]$\\
\cmidrule(lr{.5em}){1-4} 
      3    & 1   &   168/0/63 & $\xmach_{train}\in\{0.2,0.3,0.35,0.4,0.5,0.55,0.6,0.7\}$, \\
           &     &            & $\xmach_{test}\in\{0.25,0.45,0.65\}$,\\
           &     &            & $\alpha\in\{-10^{\circ},-9^{\circ},...,9^{\circ},10^{\circ}\}$\\
\cmidrule(lr{.5em}){1-4} 
      4    & 970/20 &  9300/400/20 & $\xmach\in[0.55, 0.8]$, $\Rey{}\in[5\times10^5, 5\times10^{6}]$, \\ 
           &     &                 & $\alpha\in[-0.5^{\circ}, 8^{\circ}]$\\
      \bottomrule
  \end{tabular}
  }
  \caption{Dataset sizes and parameterisations for the four cases of the present study. Here, only the largest dataset considered for each is listed. Varying the number of samples is evaluated in \ref{appA:dataset_size}.}
  \label{tab:testcases_summary}
  \end{center}
\end{table}



\subsection{Preprocessing}
As neural networks inherently rely on specific ranges of input values, pre-processing of the data is crucial for obtaining a high inference accuracy. Thus, all input channels and target flow field data in the training dataset are processed in the $128\times128$ curvilinear coordinate system with uniform grid spacing as mentioned above, and are normalised to the range from 0 to 1 in order to minimise the errors from limited precision in the training phase. To do so, we first find the maximum and minimum values for each variable in the entire training dataset, e.g. $\xmach_{\infty, \rm max}$ and $\xmach_{\infty, \rm min}$, $\xi_{x, \rm max}$ and $\xi_{x, min}$, $u_{\rm max}$ and $u_{\rm min}$. Then we get the final normalized form: 
\begin{equation*}
\tilde{\xmach}_{\infty}=(\xmach_{\infty}-\xmach_{\infty, \rm min})/(\xmach_{\infty, \rm max}-\xmach_{\infty, \rm min}+\epsilon)
\end{equation*}
\begin{equation*}
\tilde{\xi}_{x}=(\xi_{x}-\xi_{x, \rm min})/(\xi_{x, \rm max}-\xi_{x, \rm min}+\epsilon)
\end{equation*}
\begin{equation*}
\tilde{u}=(u-u_{\rm min})/(u_{\rm max}-u_{\rm min}+\epsilon),
\end{equation*}

\noindent
where, $\epsilon$ is a small positive value, e.g. $1\times10^{-20}$. As most of the variations of computational grid occur in vicinity of aerofoils, we manually set $x_{\rm min}=0$ and $x_{\rm max}=1.25$.


\subsection{Neural network architecture}\label{sec:NNarchitecture}
\noindent
The deep neural network model is based on a U-net architecture \citep{Ronneberger2015unet}, a convolutional network originally used for the fast and precise segmentation of images, and later used for the inference of flow fields and the development of DNN models \citep{thuerey2018deep, chen_cakal_hu_thuerey_2021}. 

\begin{figure}
    \centering
    \includegraphics[width=.999\textwidth]
    {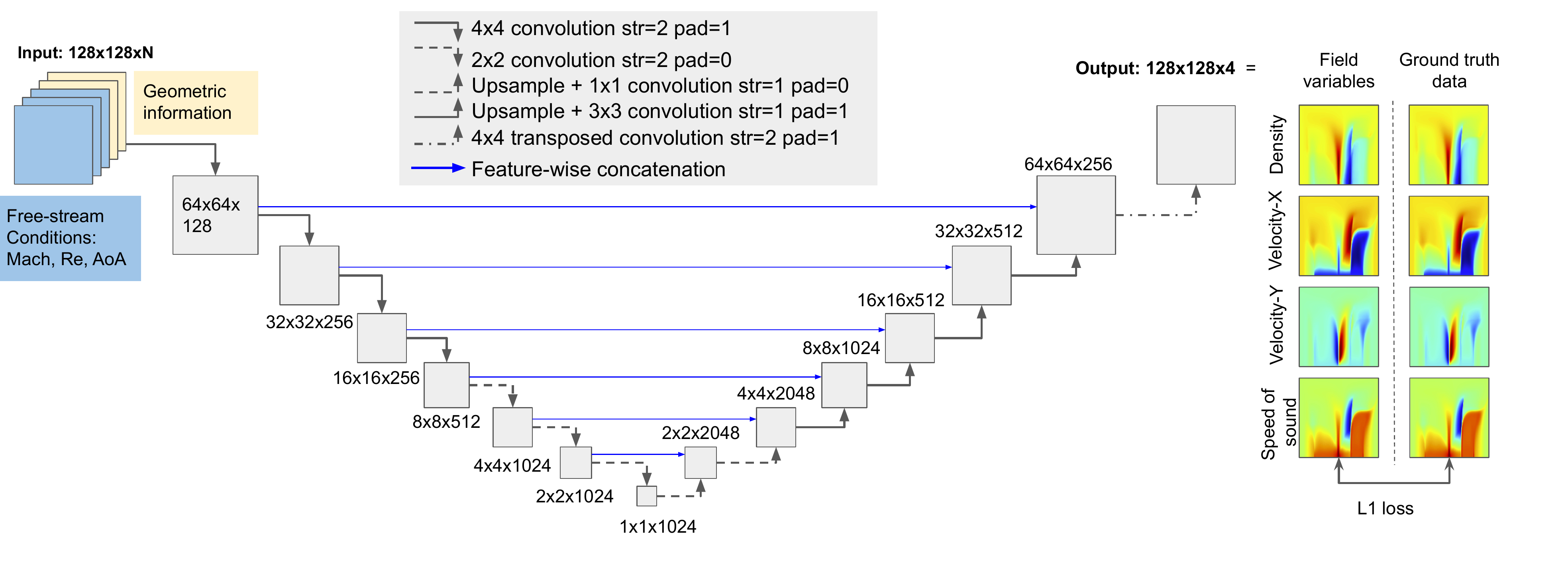}
    \caption{An overview of the deep neural network architecture with input and output specifications. It contains $3.09\times10^7$ weights.}
    \label{fig:architecture}
\end{figure}

As inputs for the learning task \textcolor{lCol}{$\Breve{\mathbf{y}}=\Breve{f}(\mathbf{x},\mathcal{G};\mathbf{w})$},
we consider the inflow boundary conditions, i.e. 
$\mathbf{x} = [\xmach_{\infty}$, $\alpha_{\infty}$, $\Rey_{\infty}]$, and the geometric information in the curvilinear coordinate system for $\mathcal{G}$. 
This is illustrated in Fig. \ref{fig:architecture}.
Hence, 
the input becomes $128\times128\times N$, where \emph{N} is the number of the input channels (see Table \ref{tab:three_encoding_methods}). 
In the encoding part, 7 convolutional blocks are used to transform the input (i.e. $128^2\times N$) into a single data point with 1024 features. The decoder part of the network is designed symmetrically with another 7 layers in order to reconstruct the outputs with the desired dimension, i.e. $128^2\times4$, corresponding to the flow field variables $\mathbf{y}=[\rho, u, v, a]$ (i.e. density, velocity components in \emph{x} and \emph{y} directions and the speed of sound) on the $128\times128$ grid. 
Leaky ReLU activation functions with a slope of 0.2 is used in the encoding layers, and regular ReLU activations in the decoding layers. 

In order to assess the performance of the DNN models, we have tested three different models with varying weight counts in \ref{appA:network_size}, and found the network with a weight count of $3.09\times10^7$ to yield a good performance across all our tests. 

\textcolor{lCol}{The present method could potentially be extended to three-dimensional cases, provided that the three-dimensional coordinate transformation matrix components are calculated \citep[]{Krist_Biedron_Rumsey1998manual}, and correspondingly three-dimensional convolutions are used in the U-net. As the input features represent volumetric data, an additional data compression step could be beneficial, e.g., via a learned autoencoder network \cite[]{Wiewel2018}}.

\subsection{Training and test accuracy}
\noindent The neural network is trained with the Adam optimizer in PyTorch \cite[]{kingma2014adam, paszke2019pytorch}. An $L_1$ difference 
\textcolor{lCol}{$L_1=\mathopen|\mathbf{y}-\Breve{\mathbf{y}}\mathopen|$}  
is used for the loss calculation, where $\mathbf{y}$ denotes the ground truth solutions from the training data set and \textcolor{lCol}{$\Breve{\mathbf{y}}$} the DNN output. 
For most of the cases, the training runs converge after 1200 epochs with a learning rate $6\times10^{-4}$ and a batch size of 5 (unless otherwise mentioned).
The learning rate was set to decrease to 10\% of its initial value over the course of the second half of the training iterations, which helps to fine-tune the final state of the model and decrease the variance in the performance.

Though we focus on the geometrical encoding method for accurate predictions, we are aware that the robustness of the deep neural network model is an important topic in both machine learning and aerodynamics communities. 
For aerodynamic situations that include shock-induced separations and stalls, it is important to assess the uncertainty of predictions. 
In the present study, as the neural network and data generation processes are deterministic, uncertainties are caused by the random initialization and batch selection for the non-linear learning process.
Thus, we perform five or more training runs with different random seeds for each setting \cite[]{thuerey2018deep}. In the following sections, we provide the mean prediction and the standard deviation to quantify the variance of the predictions.

\section{Results and discussion}\label{sec:results_and_discussion}
\noindent
\subsection{A comparison of the three methods}
\noindent Table \ref{tab:allcases_methods_accuracy} summarizes the training, validation and test losses for the models trained with methods A, B, and C in the four cases. 
The smallest validation and test losses are highlighted in bold.
In Case 1, which is the dataset with a fixed free-stream condition, methods B and C show very close accuracy in terms of the validation and test losses. In Case 2 which at the quasi-incompressible regime, method C outperforms others with the smallest validation and test losses. Case 4 is at transonic viscous regime, the prediction by the model using method C is the most accurate.
On a separate note, the dataset for Case 3 is generated following the settings of the ``interpolation'' case proposed by \cite{deAvila_icml2020_6414}. 
Looking at the $L_1$ loss of the predictions, the three methods are very close to each other. This can be attributed to the single geometry (i.e. naca0012), which is used in Case 3.

It should be mentioned that we also assessed the effects on performance with different scales of networks and amounts of flow samples in the dataset 
(see \ref{appA:network_size} and \ref{appA:dataset_size}), and results show the trends are consistent with Table \ref{tab:allcases_methods_accuracy}. 
Therefore, we will focus on the large-scale neural network models (i.e. $3.09\times10^7$ weights) trained with method C for the detailed discussion in \S\ref{sec:results_and_discussion}.

\textcolor{lCol}{Noticing the fact that methods B and C share common features, i.e. the replication and expansion of the airfoil surface coordinates in the $\eta$ direction, 
it might seem that [$x_0$, $y_0$] could suffice as 
input geometric features to achieve high accuracy. However, by doing so, the inference problem would not be well-posed from a mathematical viewpoint, i.e. the solution would not be unique. For example, one can generate infinitely many CFD meshes based on a single aerofoil profile. From this single input of [$x_0$, $y_0$] the different solutions could not be distinguished. While it is true that the flowfield solution in the continuous space is unique for the given aerofoil profile and free-stream conditions, the discretized solution of PDEs must rely on the corresponding CFD meshes. Therefore, the full information on the coordinate transformation matrix is deemed essential.}

\begin{table}
  \begin{center}
\def~{\hphantom{0}}
  \resizebox{\columnwidth}{!}{\begin{tabular}{ccccr} 
      \toprule
     Case & Method    & Training ($L_1$) &   Vali. ($L_1 \pm SD$) & Test ($L_1 \pm SD$) \\
      \midrule 
       &  A  & $2.80\times10^{-4}$ &  $(9.45 \pm 1.00)\times10^{-4}$ &   $(6.75 \pm 0.09)\times10^{-4}$ \\
    1   &  B &   $2.34\times10^{-4}$ &  $(8.59 \pm 0.95)\times10^{-4}$ &  $(5.11 \pm0.10)\times10^{-4}$\\
       & C &  $2.57\times10^{-4}$ &  $\mathbf{(7.06 \pm 1.06)\times10^{-4}}$ &  $\mathbf{(5.04 \pm 0.10)\times10^{-4}}$  \\
\cmidrule(lr{.5em}){1-5} 
       &  A    &  $4.26\times10^{-4}$ &  $(9.58\pm0.33)\times10^{-4}$ &  $(7.17\pm 0.20 )\times10^{-4}$  \\
    2   &  B    &  $3.39\times10^{-4}$ &   $(8.12\pm0.45)\times10^{-4}$ &  $\mathbf{(4.90\pm0.07)\times10^{-4}}$\\
       &  C    &  $3.57\times10^{-4}$ &  $\mathbf{(8.04\pm0.44)\times10^{-4}}$  &  $(5.26\pm0.13)\times10^{-4}$  \\       
\cmidrule(lr{.5em}){1-5} 
       &  A    &  $1.89\times10^{-4}$  &     &  $(7.63\pm0.24)\times10^{-4}$   \\  
    3   &  B    &  $1.86\times10^{-4}$ &     &  $(7.18\pm0.14)\times10^{-4}$\\    
       &  C    &  $1.82\times10^{-4}$  &     &  $(7.62\pm0.29)\times10^{-4}$  \\           
      \cmidrule(lr{.5em}){1-5} 
       &  A    &  $3.51\times10^{-4}$ &  $(7.46 \pm 0.36)\times10^{-4}$ &  $(6.96\pm0.16)\times10^{-4}$  \\
   4   &  B    &   $2.99\times10^{-4}$ &        ($6.54 \pm 0.62) \times10^{-4}$ &    ${(5.77 \pm 0.06)\times10^{-4}}$  \\
       &  C    &  $2.99\times10^{-4}$ &  $\mathbf{(5.91 \pm 0.21) \times10^{-4}}$ &   $\mathbf{(5.65 \pm 0.09)\times10^{-4}}$  \\          
      \bottomrule
  \end{tabular}
  }
  \caption{Training, validation and test losses in the four cases using three encoding methods.}
  \label{tab:allcases_methods_accuracy}
  \end{center}
\end{table}

\subsection{Case 1: a fixed free-stream condition}
\noindent

\begin{figure}
    \centering
    \includegraphics[width=1\textwidth]{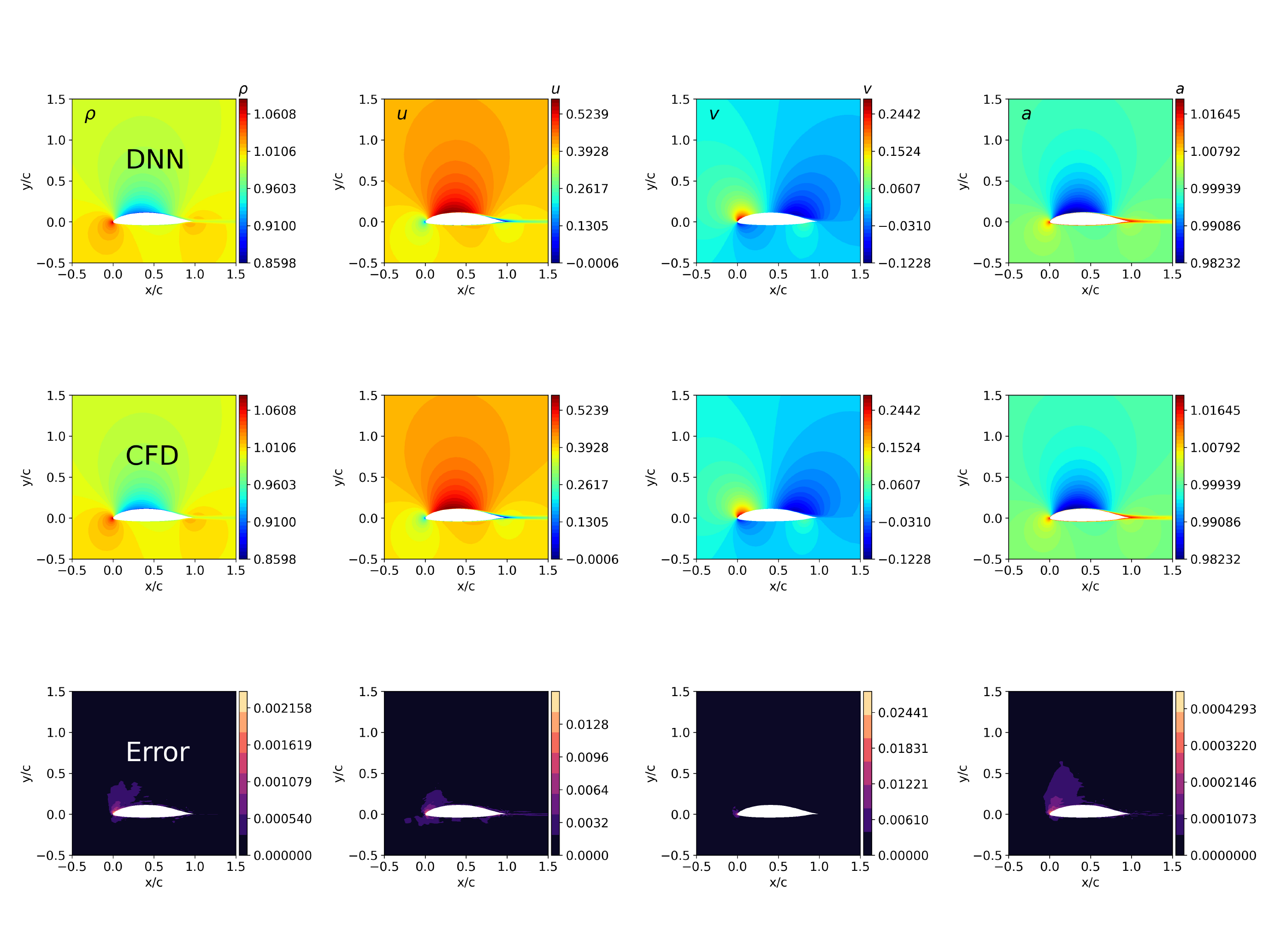}
    \caption{The comparison of flow fields for the aerofoil ``ah94156'' in Case 1 predicted by DNN (top row) and CFD (middle row) and the relative errors (bottom row).}
    \label{fig:Exp1_flowfield}
\end{figure}

\begin{figure}

\begin{subfigure}{.45\textwidth}
\centering
\includegraphics[width=\linewidth]{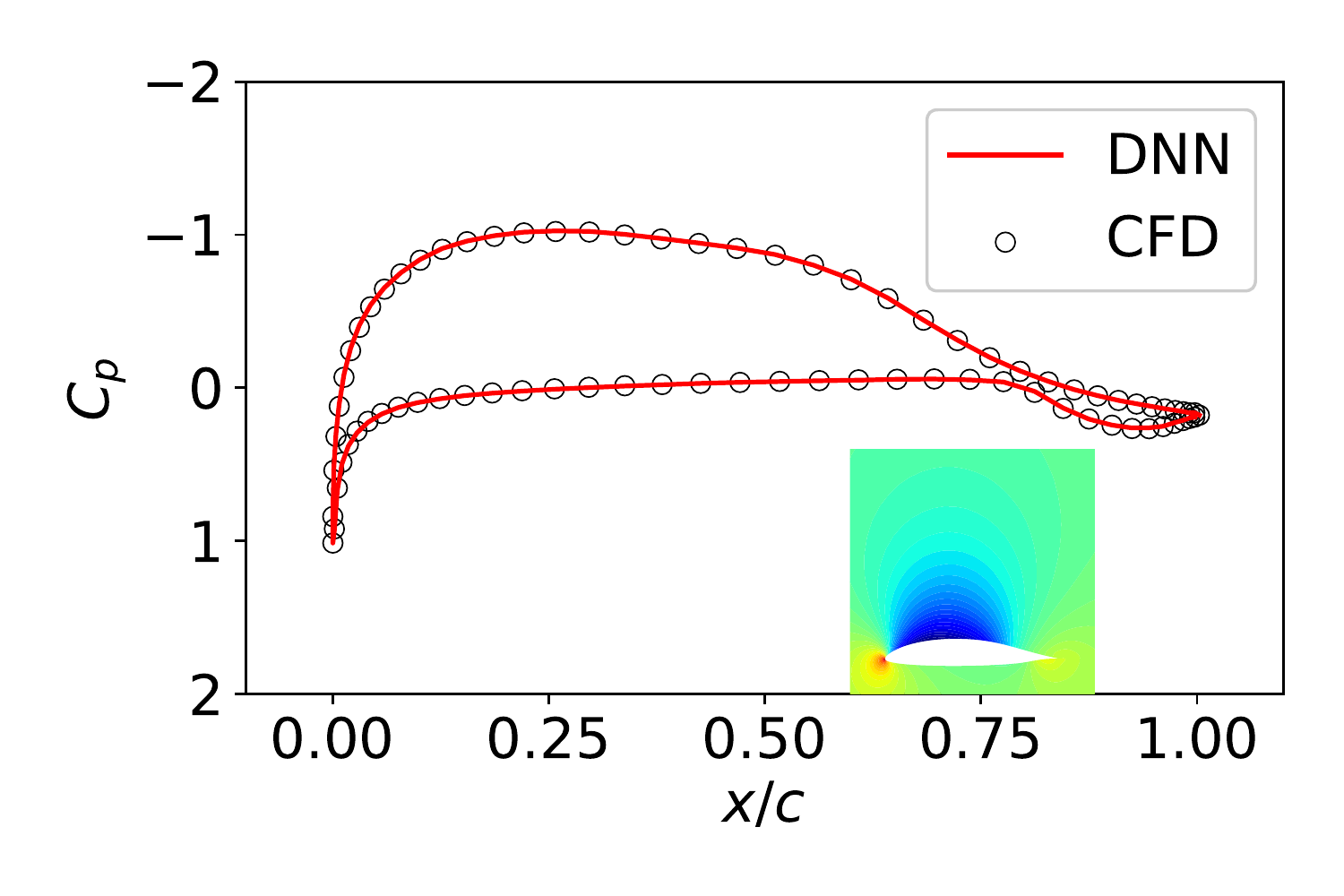} 
\caption{ah94156}
\end{subfigure}
\begin{subfigure}{.45\textwidth}
\centering
\includegraphics[width=\linewidth]{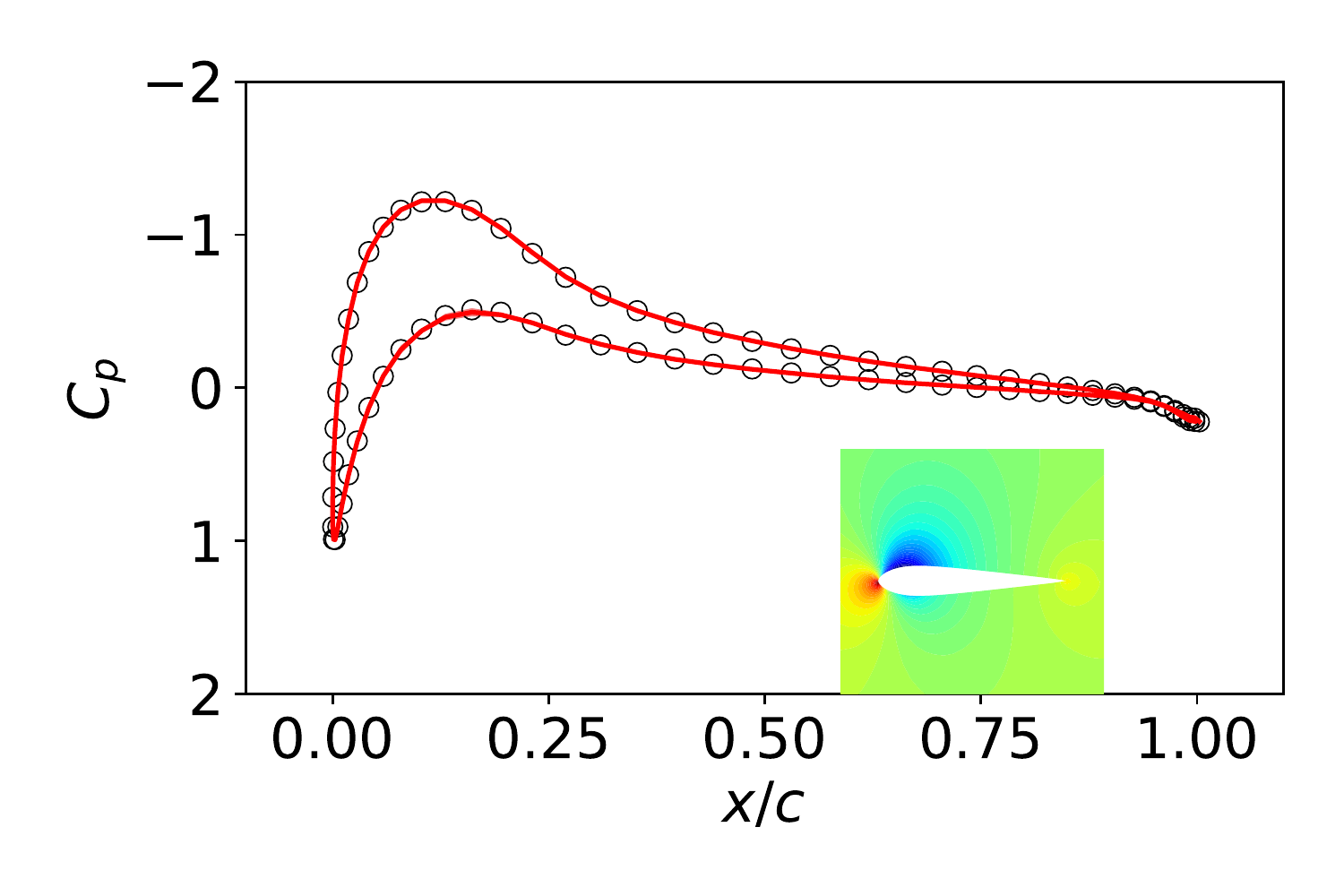} 
\caption{e473}
\end{subfigure}

\begin{subfigure}{.45\textwidth}
\centering
\includegraphics[width=\linewidth]{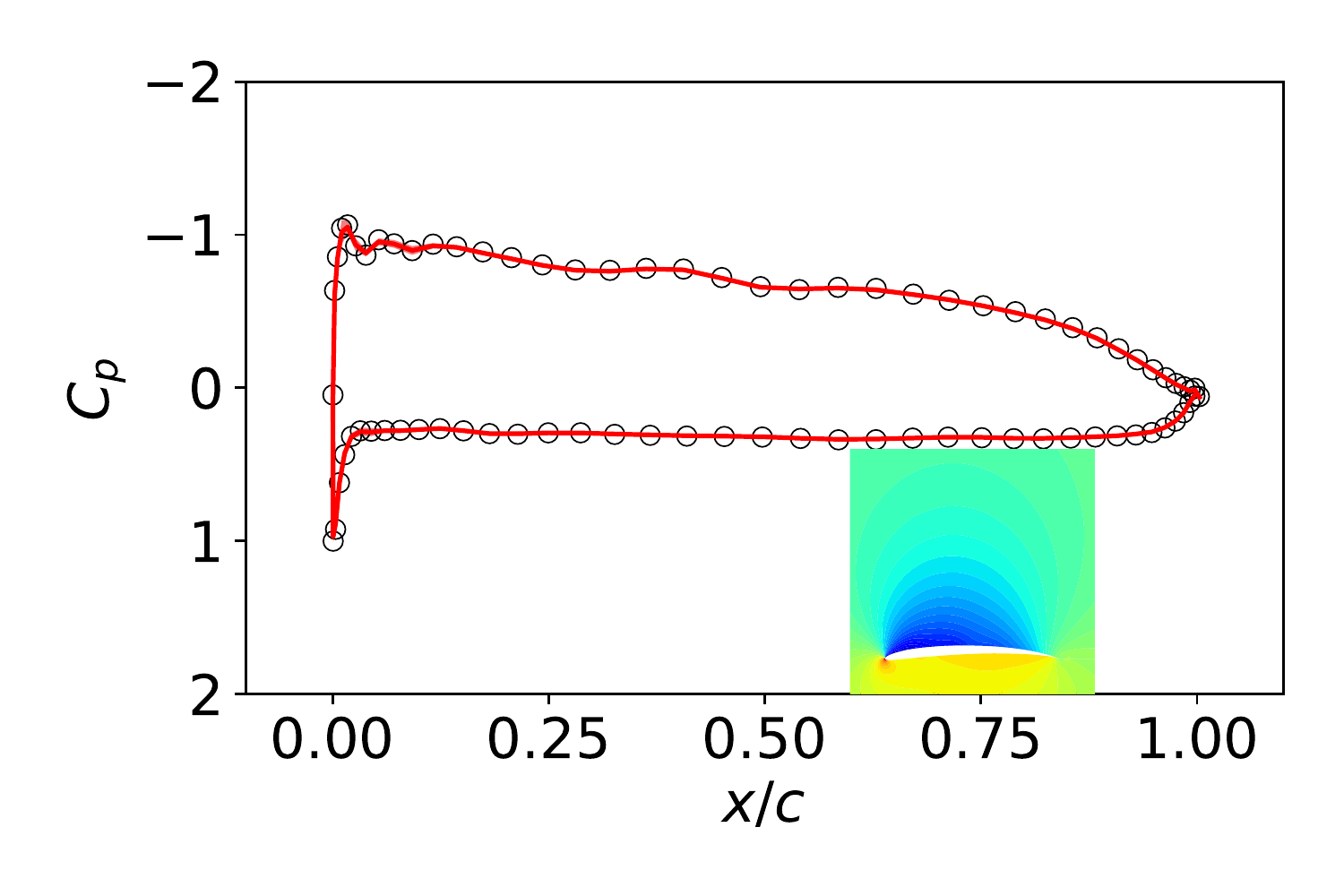} 
\caption{e59}
\end{subfigure}
\begin{subfigure}{.45\textwidth}
\centering
\includegraphics[width=\linewidth]{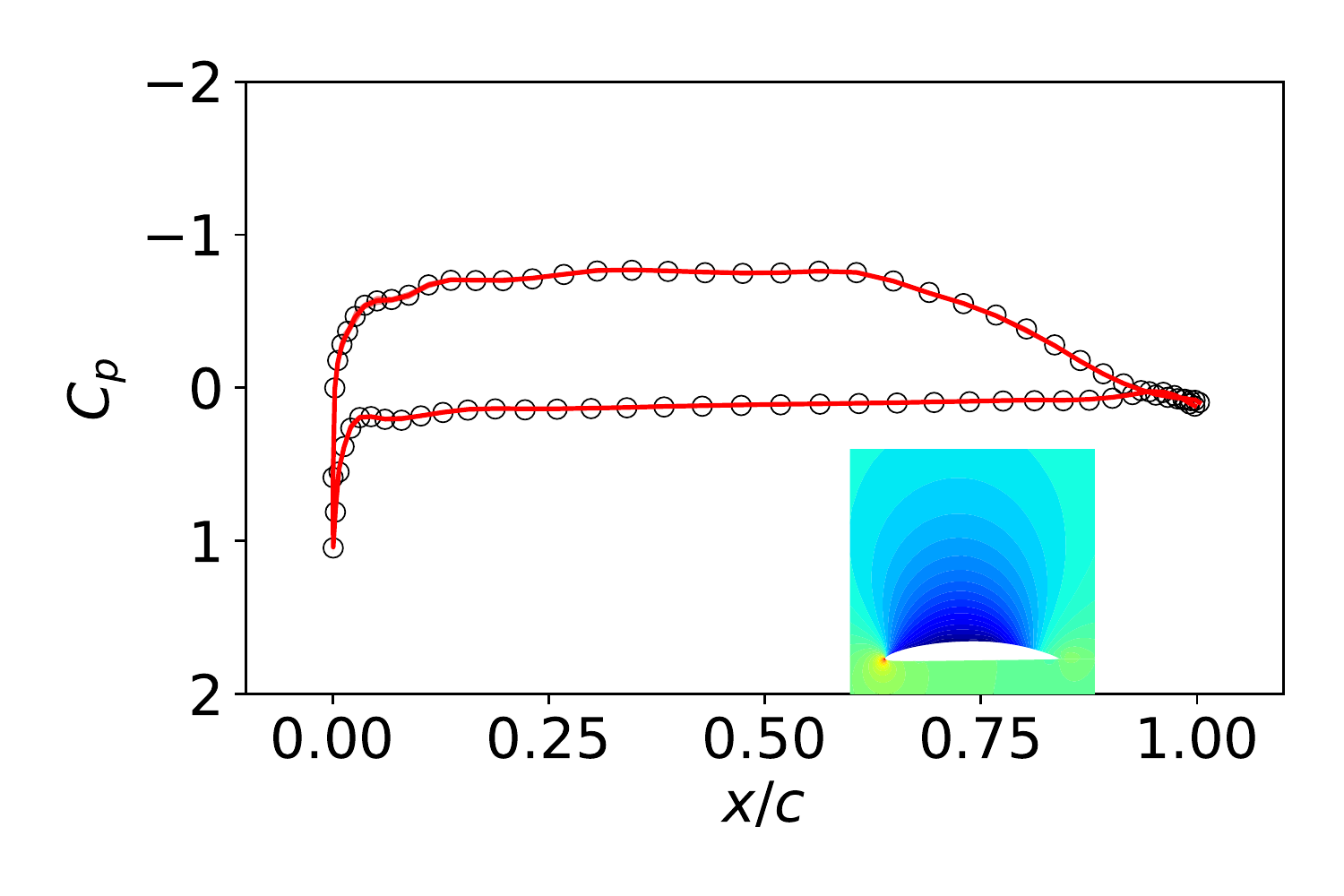} 
\caption{goe07k}
\end{subfigure}

\caption{Distributions of pressure coefficient in Case 1: a) ah94156, b) e473, c) e59, and d) goe07k. 
The red lines represent the DNN results with the shaded region visualizing $\pm3 SD$, and the black symbols represent the reference data. 
The embedded plots show the corresponding pressure fields.}
\label{fig:Exp1_Cp}
\end{figure}

\begin{figure}

\begin{subfigure}{.45\textwidth}
\centering
\includegraphics[width=\linewidth]{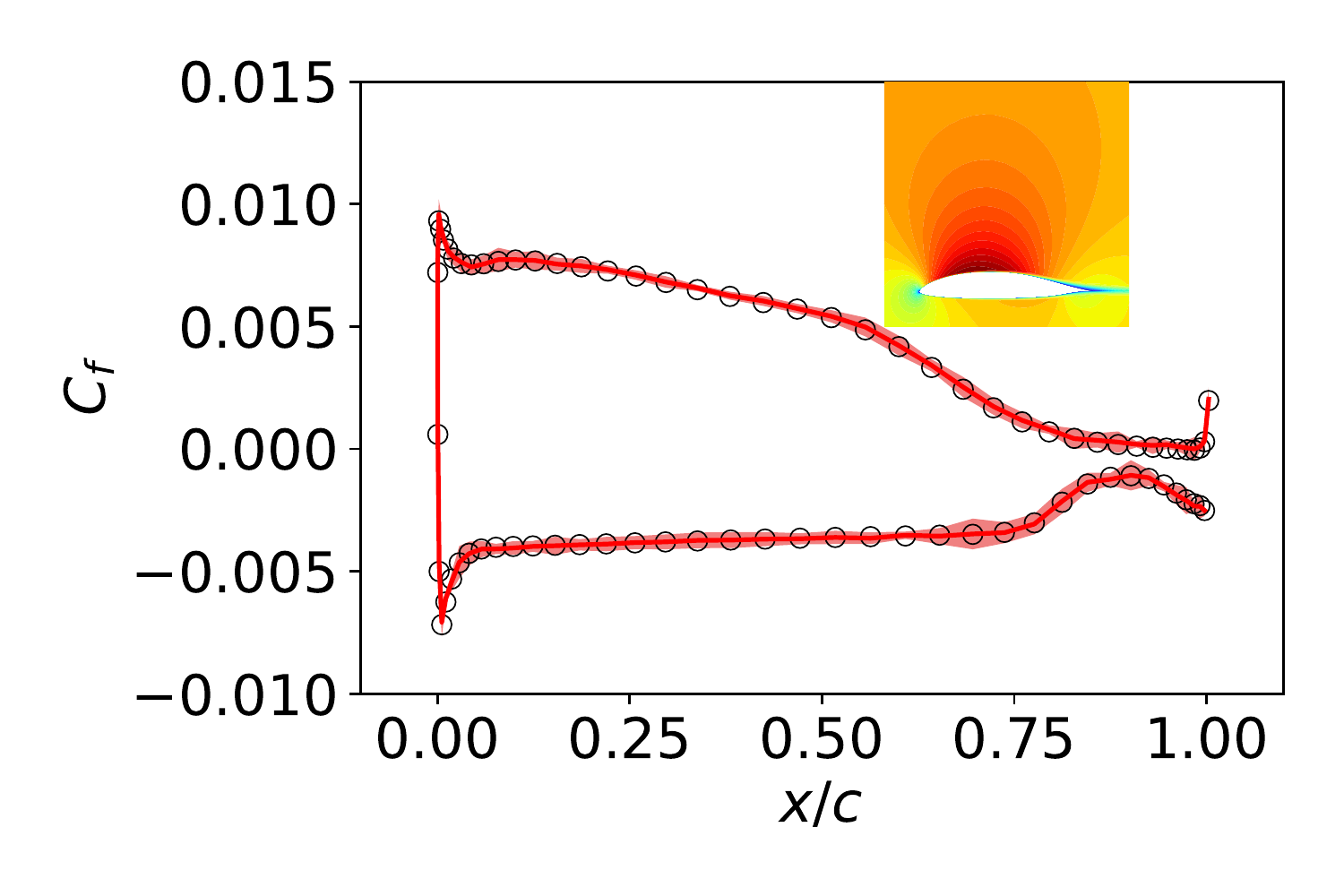}
\caption{ah94156}
\end{subfigure}
\begin{subfigure}{.45\textwidth}
\centering
\includegraphics[width=\linewidth]{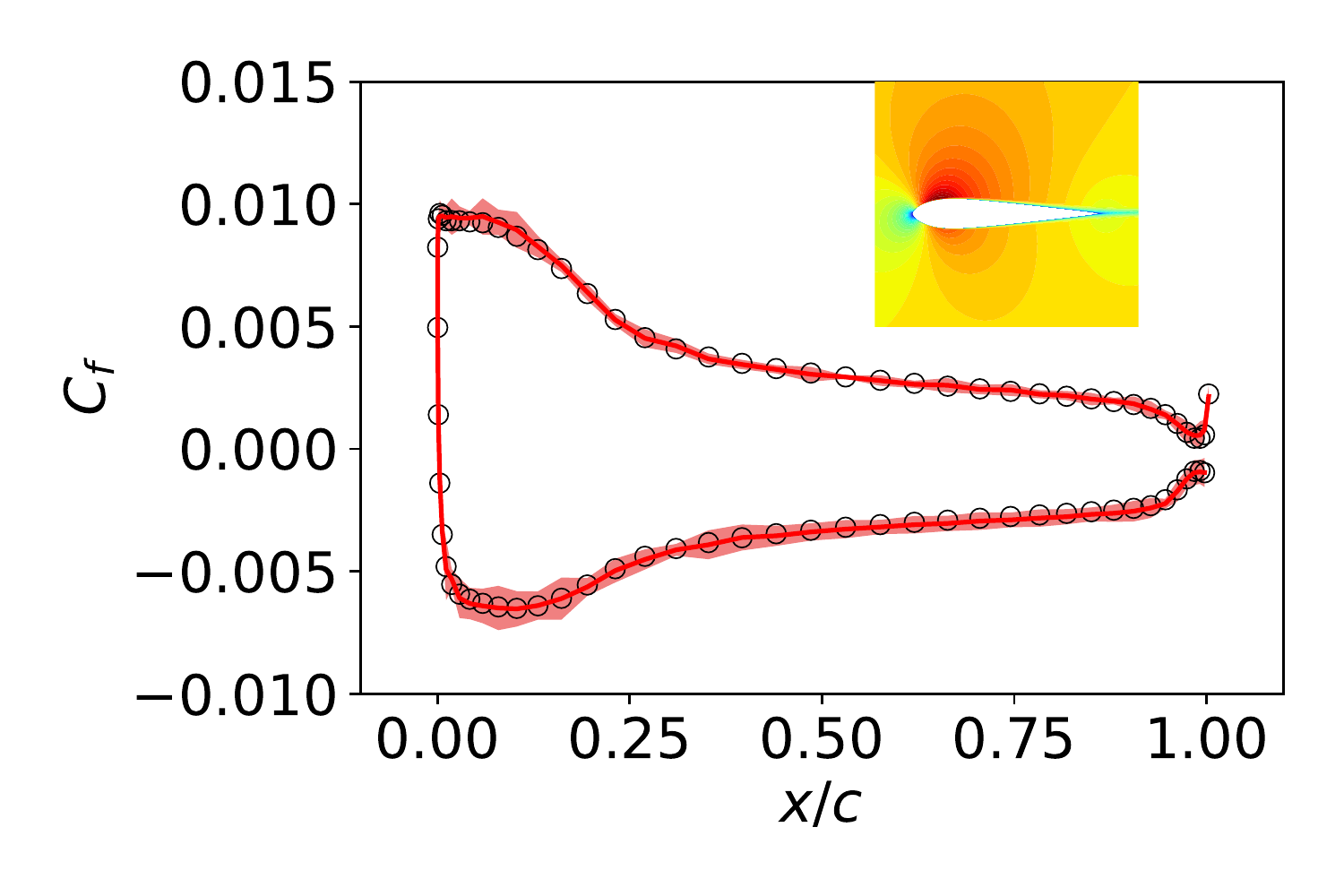}
\caption{e473}
\end{subfigure}

\begin{subfigure}{.45\textwidth}
\centering
\includegraphics[width=\linewidth]{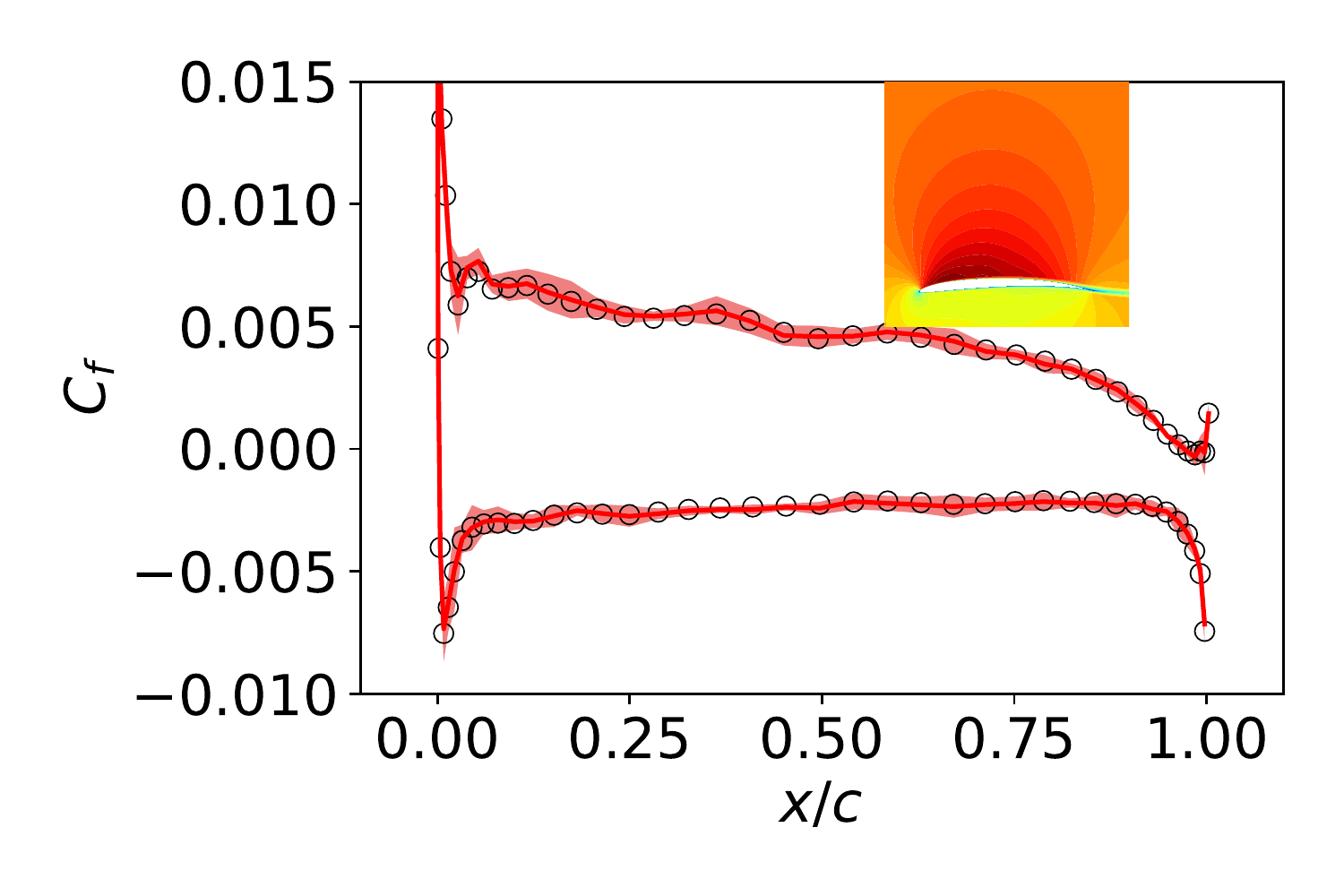}
\caption{e59}
\end{subfigure}
\begin{subfigure}{.45\textwidth}
\centering
\includegraphics[width=\linewidth]{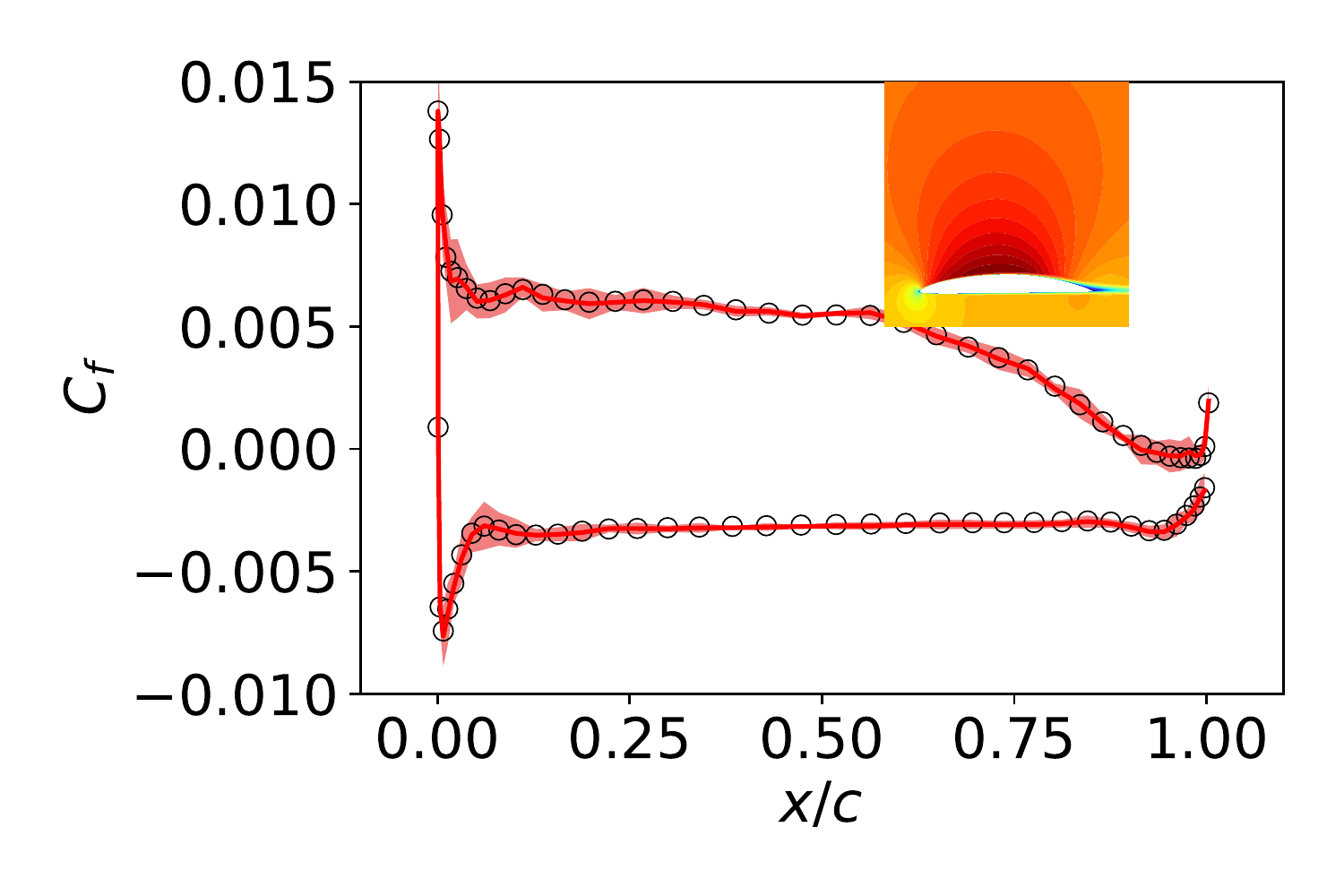}
\caption{goe07k}
\end{subfigure}
\caption{Distributions of skin friction coefficient in Case 1: a) ah94156, b) e473, c) e59, and d) goe07k. 
The red lines represent the DNN results with the shaded region visualizing $\pm3 SD$, and the black symbols represent the reference data.
The embedded plots show the corresponding \emph{x}-component velocity fields.}
\label{fig:Exp1_Cf}
\end{figure}

\begin{figure}

\begin{subfigure}{.3\textwidth}
\centering
\includegraphics[width=\linewidth]{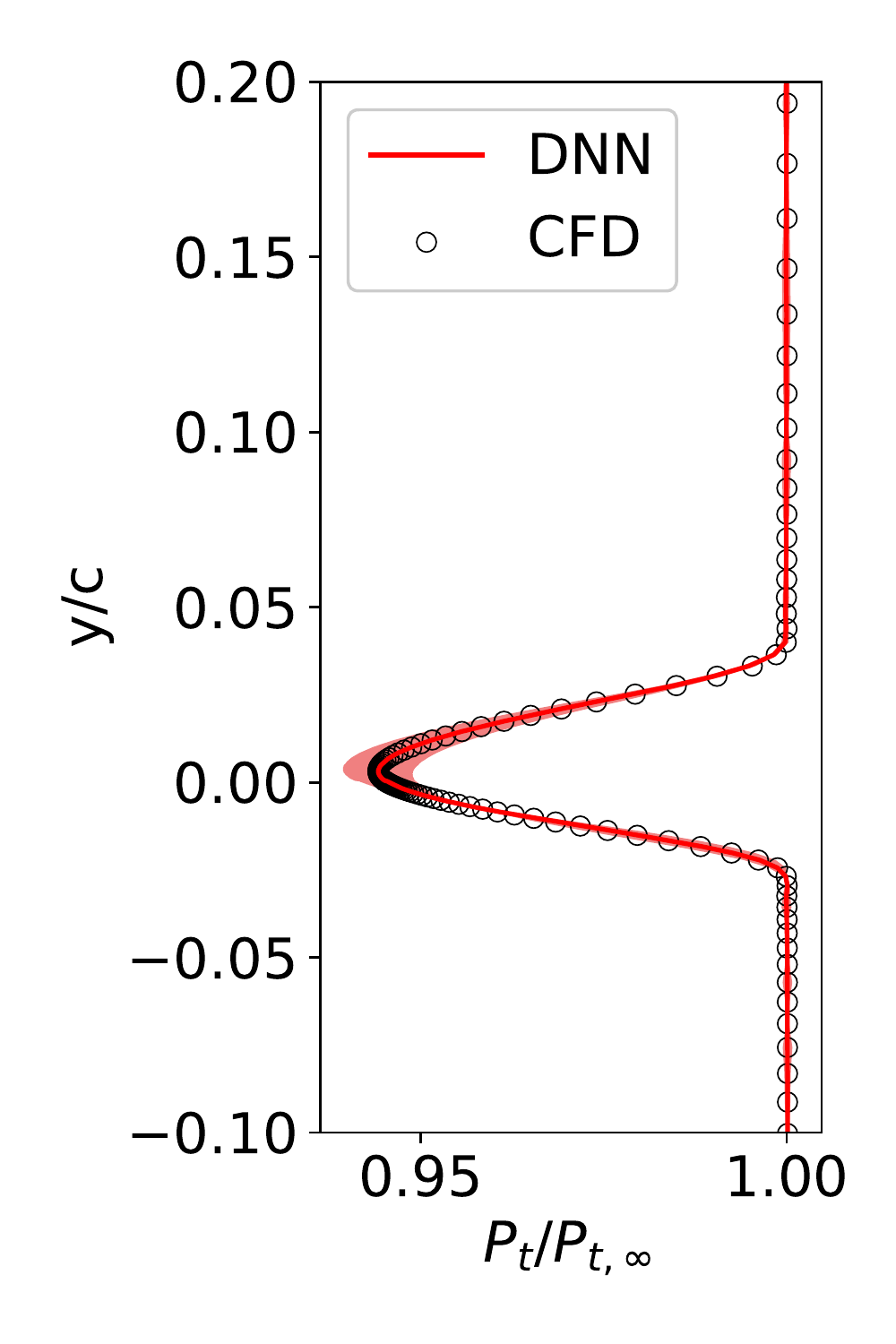}
\caption{ah94156}
\end{subfigure}
\begin{subfigure}{.3\textwidth}
\centering
\includegraphics[width=\linewidth]{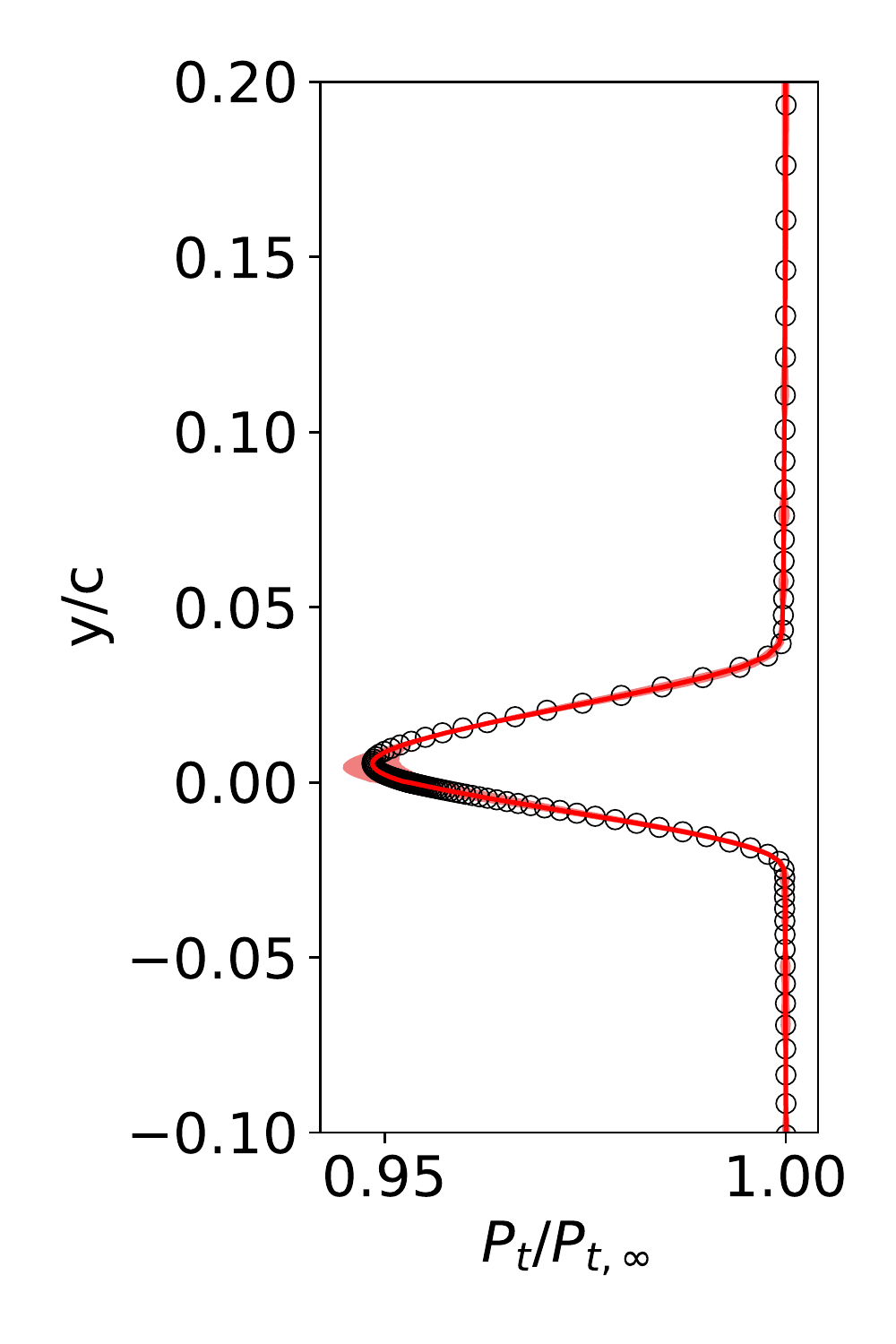}
\caption{e473}
\end{subfigure}

\begin{subfigure}{.3\textwidth}
\centering
\includegraphics[width=\linewidth]{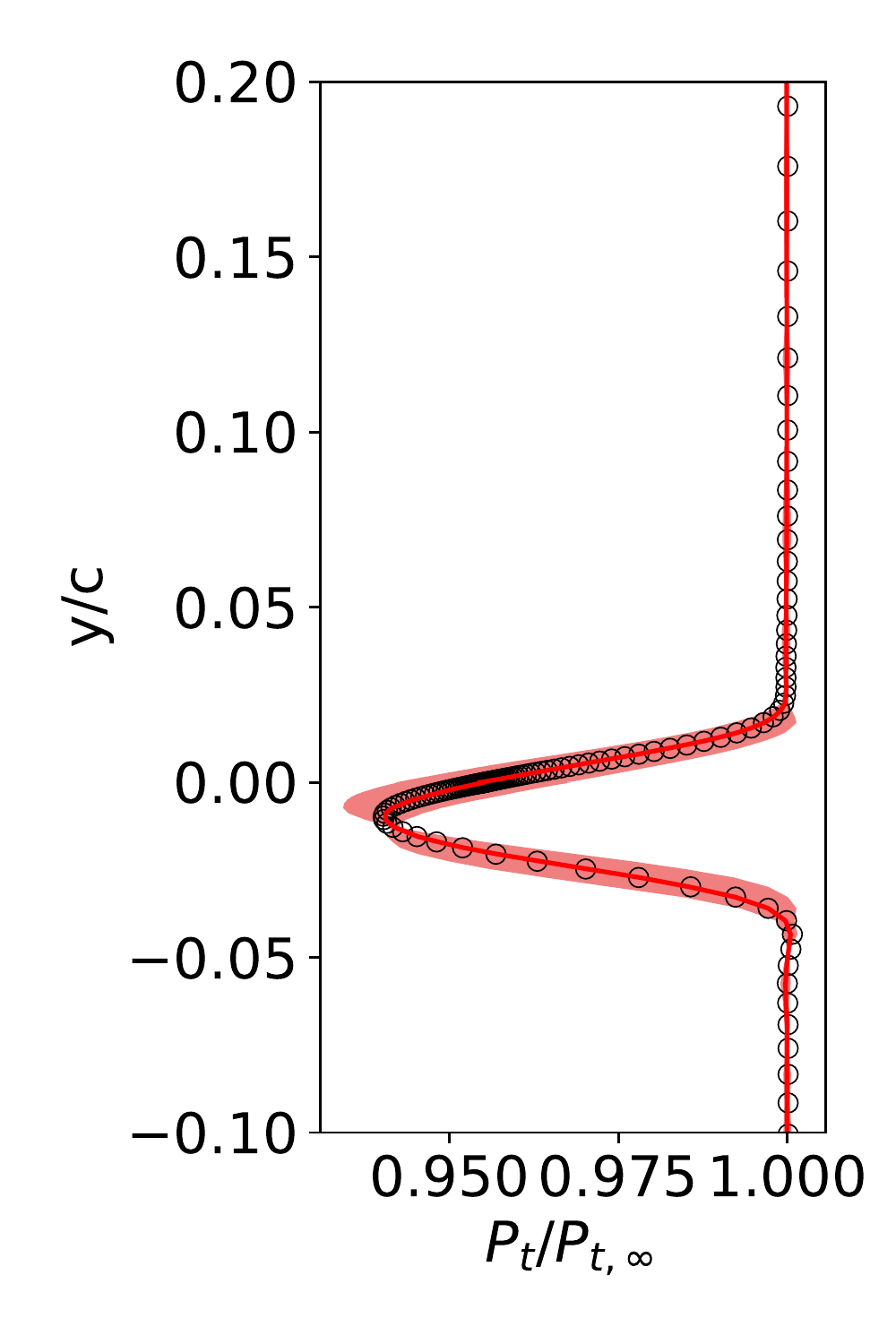}
\caption{e59}
\end{subfigure}
\begin{subfigure}{.3\textwidth}
\centering
\includegraphics[width=\linewidth]{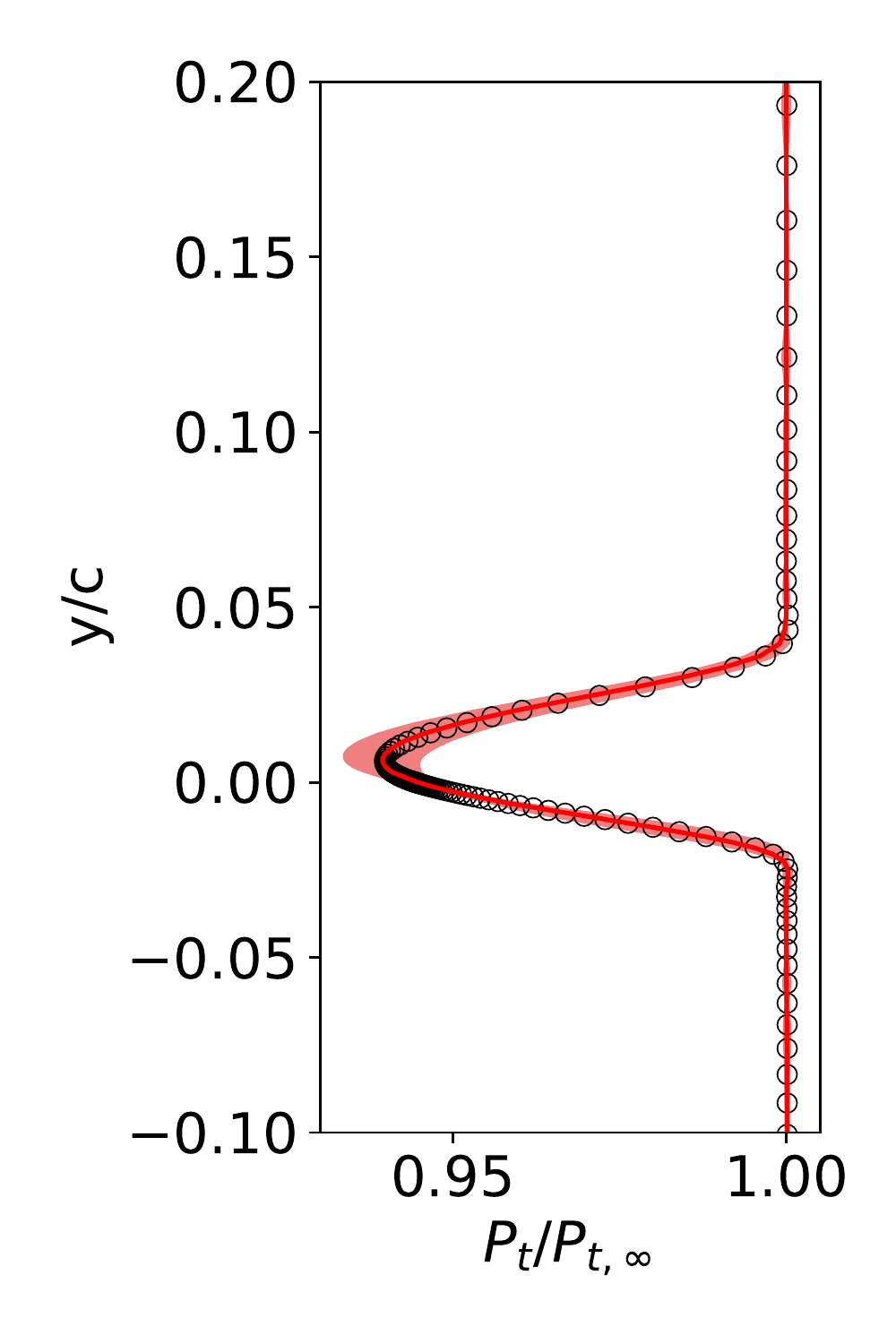}
\caption{goe07k}
\end{subfigure}
\caption{Total pressure profiles at $x/c=1.2$ in Case 1. 
The red lines represent the DNN results with the shaded region visualizing $\pm3 SD$, and the black symbols represent the reference data.
}\label{fig:Exp1_wake}
\end{figure}

\noindent
\subsubsection{Overview}
As shown in Table \ref{tab:testcases_summary}, for Case 1 there are 990 flow fields that are obtained with c-grid topology grids at a fixed free-stream condition, 
i.e. $\xmach_{\infty}=0.4$, $\Rey_{\infty}=1\times10^6$, and $\alpha_{\infty}=2.5^{\circ}$. 
Each solution uses a different discretization of the physical space.
In the training phase, 780 flow fields are used as training set and 190 for validation set. 
After this, additional 20 aerofoil shapes that did not appear for training are used to
generate a test dataset with 20 samples at the same free-stream condition. 
Case 1 is designed to isolate and evaluate the effect of changing geometric information.
In most instances, the boundary layers on both aerofoil surfaces 
remain attached at the current free-stream condition. The average $L_1$ loss for the test set using method C is $4.90\times10^{-4}$, indicating a high accuracy of the trained DNN model.

\subsubsection{Flow field prediction}
One of the advantages of the proposed method is that the DNN model predicts a full flow field. 
Taking the aerofoil ``ah94156'' as an example, Fig. \ref{fig:Exp1_flowfield} shows the density, \emph{x} and \emph{y} components of velocity as well as the speed of sound predicted by DNN (top row) and the CFD solver (middle row). The relative errors at each individual pixel are shown in the bottom row, which are defined as 
\textcolor{lCol}{$e_f=|\Breve{f}-f|/(\frac{1}{S}\sum\limits_{m=1}^S{|f_m|})$}, where \emph{S} is the number of pixels of images (\emph{S}=$128\times128$ here), \emph{f} the function under consideration, \textcolor{lCol}{$\Breve{f}$} the approximation of the DNN.
There is no visible difference in predictions by DNN and CFL3D. 

We choose four representative shapes from 20 test results to investigate details: ``ah94156'', ``e473'', ``e59'' and ``goe97k'',
corresponding to thick, symmetrical, thin, flat-bottom aerofoils, respectively. 

As the pressure coefficient, skin friction coefficient and wake profile are critical for engineering designs, we evaluate our DNNs with respect to these quantities.
Figure \ref{fig:Exp1_Cp} gives the distributions of pressure coefficient defined as $C_p=(p_w - p_{\infty})/(p_{t,\infty}-p_{\infty})$, reflecting a typical behavior of attached boundary layers. 
In particular, in Figs. \ref{fig:Exp1_Cp}c and \ref{fig:Exp1_Cp}d,
the mild-mannered pressure distribution on the lower surface (or pressure side) is due to the flat design of aerofoils. 
As ``e59'' is a thin aerofoil with a sharper leading edge, the suction peak occurs in very close proximity to the leading edge as shown in Fig. \ref{fig:Exp1_Cp}c. 

The skin friction coefficient is defined as $C_f=\tau_w/(p_{t,\infty}-p_{\infty})$.
The comparisons of skin friction coefficient distributions in Fig. \ref{fig:Exp1_Cf} are also encouraging. 
We define the positive shear in the clock-wise direction.
The distribution follows the behavior of a fully turbulent flow over an aerofoil, starting from a high shear stress at the leading edge then decaying gradually. 
Some slight volatility in the vicinity of the leading edge on the suction side observed 
in Figs. \ref{fig:Exp1_Cf}c and \ref{fig:Exp1_Cf}d are caused by the changes in pressure gradient.

Figure \ref{fig:Exp1_wake} shows the total pressure profiles $p_{t}/p_{t,\infty}$ measured at 20\% of chord length downstream of trailing edges. As the flows are mostly attached, 
the wake profiles are relatively homogeneous in the outer region but exhibit sharp deficit peaks. 

The sensitivity analysis is conducted to identify the effect on $C_p$, $C_f$ distributions and wake total pressure profiles due to different network training runs. From Figs. \ref{fig:Exp1_Cp} to \ref{fig:Exp1_wake}, the shade regions show the $\pm3 SD$ uncertainties around the mean, which indicates the present predictive model performs robustly. The shaded regions of uncertainty are small, which is reflected in that the standard deviations of the pressure and viscous drags account for less than $5\%$ and $1\%$ of the corresponding mean total drags in all the four test aerofoils.

These results highlight that the DNN model captures fine details of the flow fields. It also confirms that the geometric information has been encoded such that the DNN can employ it to infer the desired solutions. Hence, this approach provides a good basis for further studies with more complex free-stream conditions.

\subsection{Case 2: low Mach number viscous flow}
\noindent
\subsubsection{Overview}

Next we target a benchmark case for the inference of incompressible RANS results \citep{thuerey2018deep}:
The free-stream conditions of RANS are selected randomly 
in a range of Reynolds numbers $\Rey_{\infty}\in[0.5, 5]$ million, 
and angles of attack in the range of [-22.5, 22.5], and follows a uniform distribution.
In the present setup, we run a simulation by randomly choosing 
one of the 970 aerofoils with domain discretizations in the training data set,
and randomly choosing the free-stream condition from the above described range. 
We use an additional set of 20 aerofoil shapes that were not used for training, to
generate a test dataset with 20 samples using the same range of Reynolds numbers and angles of attack.
To make a reasonable comparison, we set the free-stream Mach number to 0.1 in the current compressible solver, and employ a low-Mach number preconditioning to improve the convergence \citep{weiss_smith1995precond}. 
The mean test loss ($L_1$) for the whole test set is $4.82\times10^{-4}$. The test losses of four representative conditions/shapes from the test dataset are shown in Table \ref{tab:exp_2_summary}. 

The mean relative error for the test set is defined to be:
\begin{equation*}
e_f=\frac{1}{T}\sum\limits_{n=1}^{T}({\sum\limits_{m=1}^S{|\Breve{f}_{m,n}-f_{m,n}|}/\sum\limits_{m=1}^S{|f_{m}|}}), 
\end{equation*}
where \emph{T} is the number of samples in the test set (\emph{T}=20 here), \emph{S} is the number of pixels of images (\emph{S}=$128\times128$ here), \emph{f} the function under consideration, $\Breve{f}$ the approximation of the DNN.
The best prediction by the trained model in \cite{thuerey2018deep} achieves relative errors of 2.15\% for the \emph{x} velocity, 2.6\% for the \emph{y} velocity and 14.76\% for pressure values. 
In the present work, the corresponding relative errors are 0.0438\% for density, 0.137\% for the \emph{x} velocity and 0.0712\% for the \emph{y} velocity, 0.0933\% for the speed of sound and 0.178\% for pressure. 
Hence, improvements of more than an order of magnitude compared to the results by \cite{thuerey2018deep} have been achieved, with almost two orders of magnitude for the pressure fields.

\subsubsection{Flow field prediction}
We select the four representative conditions/shapes from the test dataset for detailed discussions as shown in Table \ref{tab:exp_2_summary}. Aerofoil ``e342'' experiences a high positive angle of attack. In the case of aerofoil ``goe07k'', the angle of attack is slightly negative, and aerofoil ``mue139'' are at a relatively high Reynolds number of the described range. In all of the three cases, the predictions exhibit small $L_1$ losses, i.e. the order of magnitude of $10^{-4}$. Note that even in the worst case in terms of $L_1$ loss, i.e. aerofoil ``goe398'' at a high negative angle of attack of $-22.00^{\circ}$, the DNN predicts an accurate flow field with $L_1\approx1.33\times10^{-3}$. 

\begin{figure}
    \centering
    \includegraphics[width=1\textwidth]{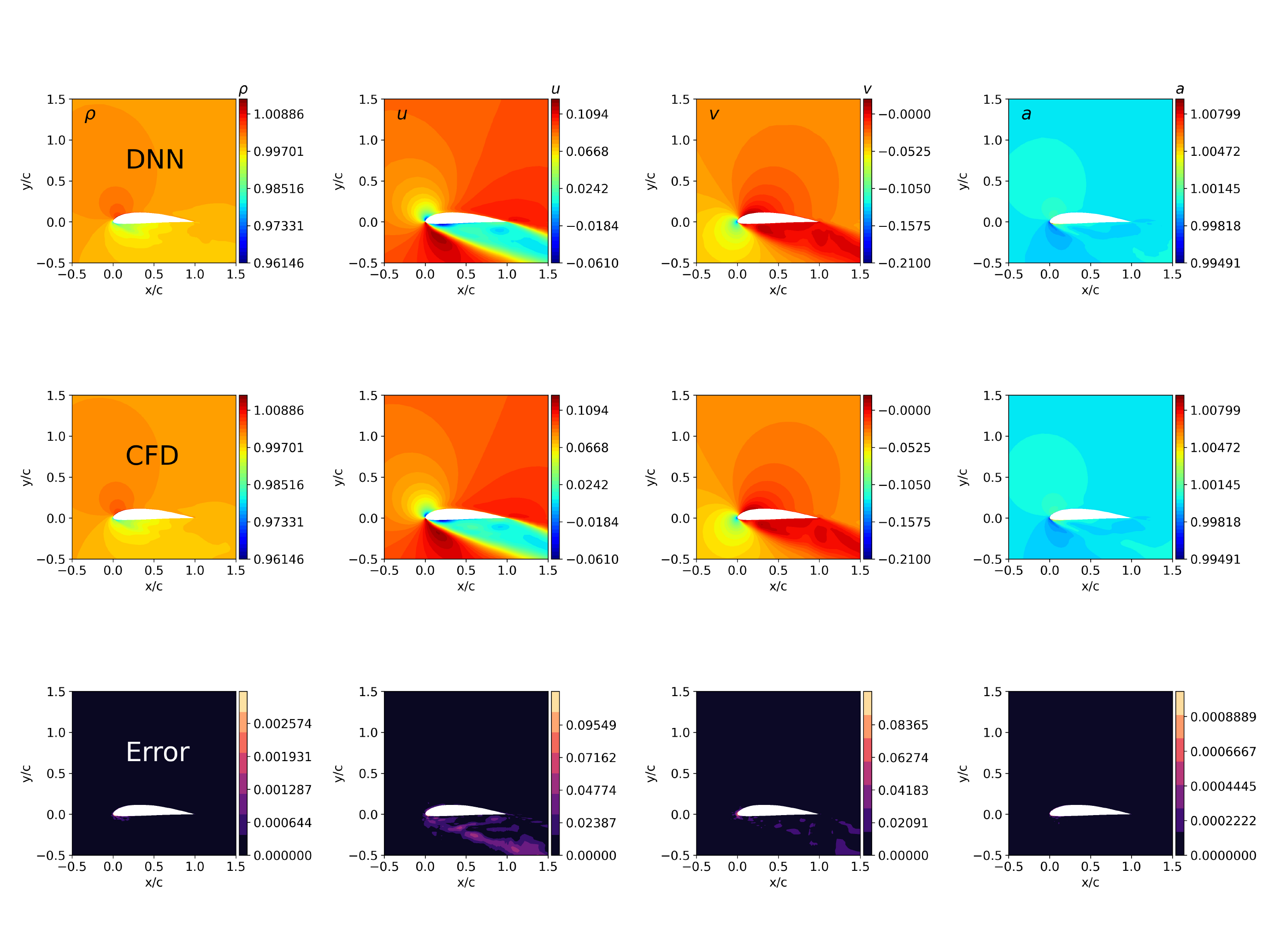}
    \caption{The comparison of flow fields for the aerofoil ``goe398'' in Case 2 predicted by DNN (top row) and CFD (middle row) and the relative errors (bottom row).}
    \label{fig:Exp2_flowfield}
\end{figure}

Figure \ref{fig:Exp2_flowfield} shows the fields of density, \emph{x} and \emph{y} components of velocity as well as the speed of sound predicted by DNN (top row) and the CFD solver (middle row). The relative errors at each individual pixel are shown in the bottom row
The predictions by DNN and CFD solver are not apparent, and the only visible difference resides in \emph{u} in the separated region, which is shown in the relative error contour.

\begin{figure}

\begin{subfigure}{.45\textwidth}
\centering
\includegraphics[width=\linewidth]{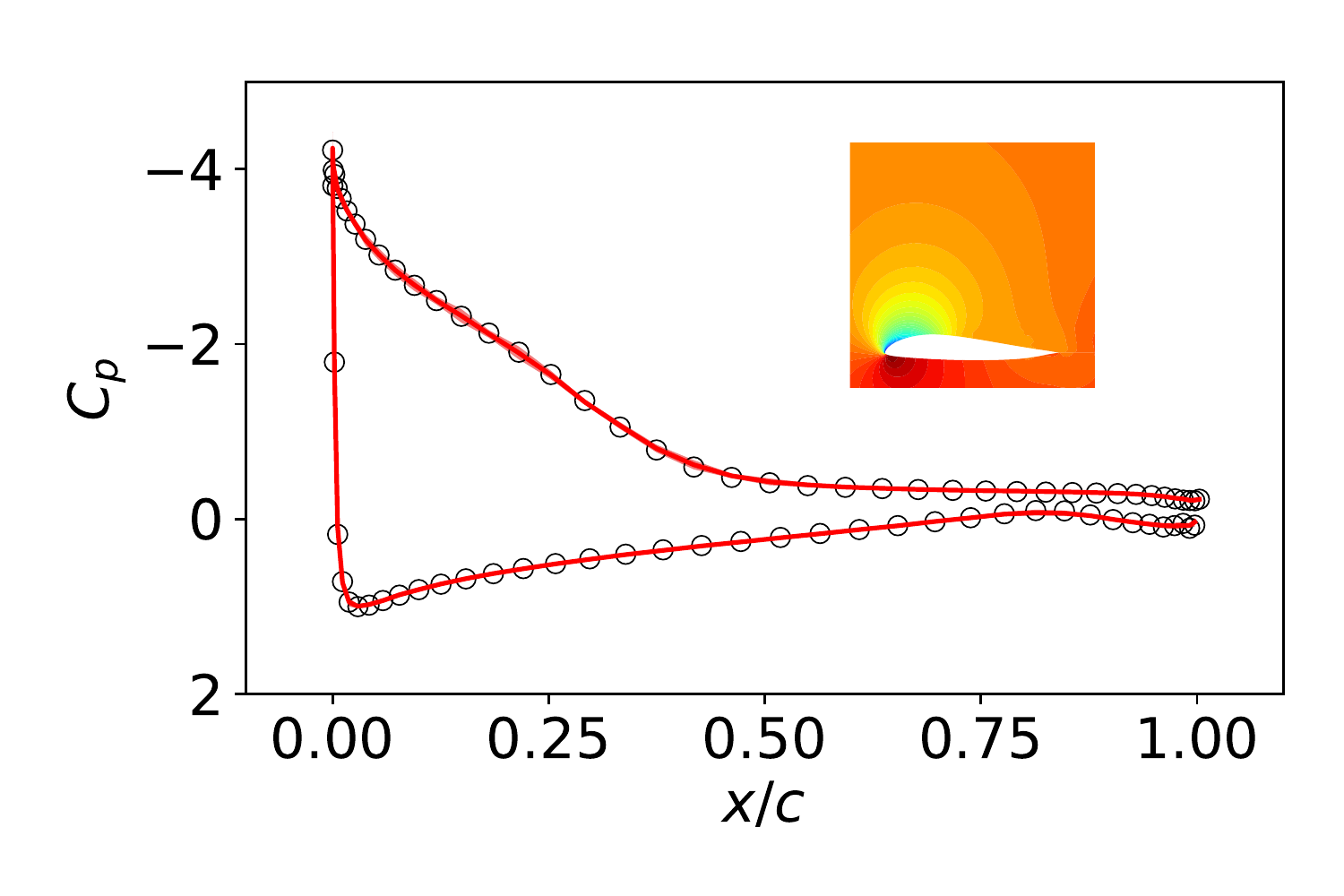}
\caption{e342}
\end{subfigure}
\begin{subfigure}{.45\textwidth}
\centering
\includegraphics[width=\linewidth]{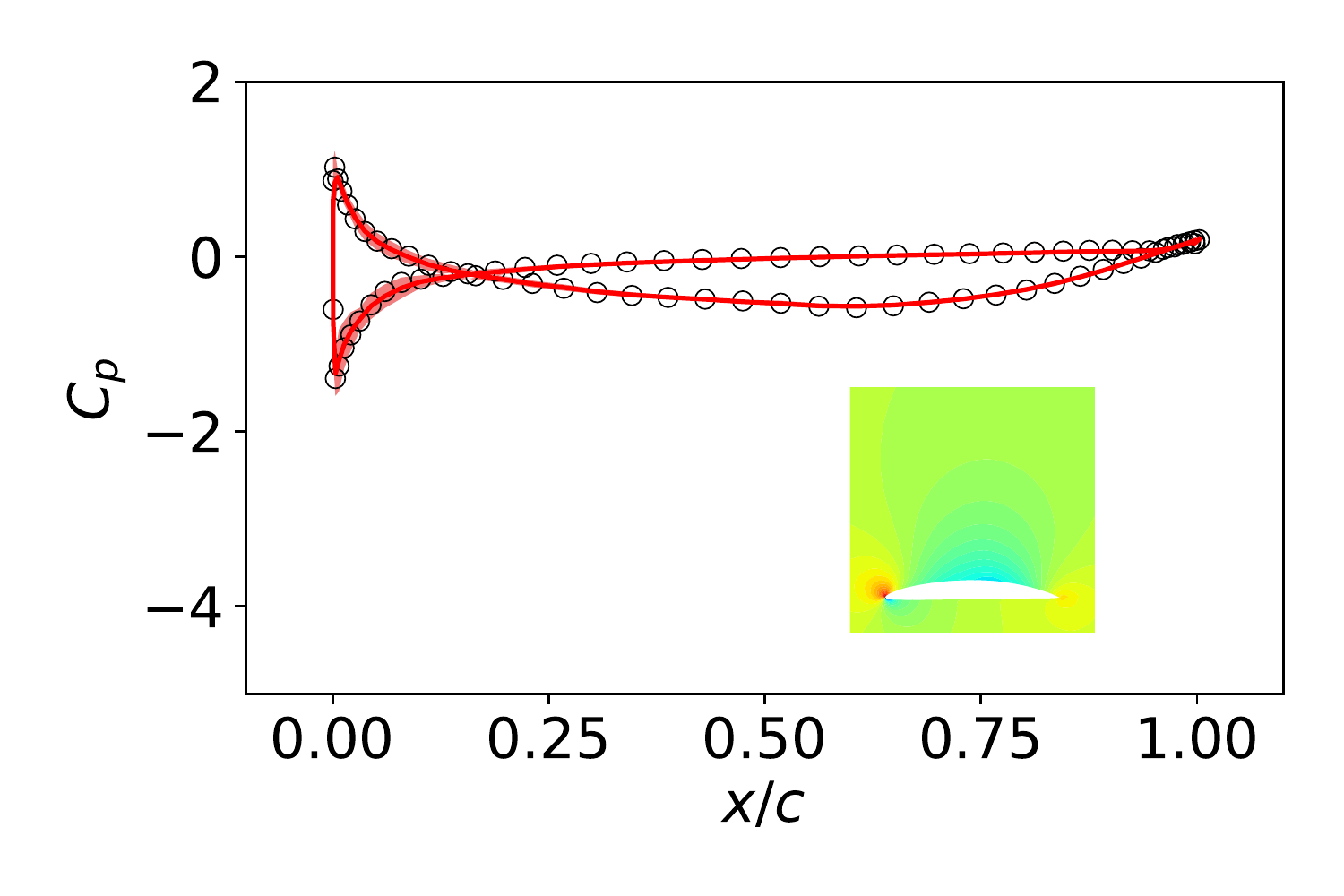}
\caption{goe07k}
\end{subfigure}

\begin{subfigure}{.45\textwidth}
\centering
\includegraphics[width=\linewidth]{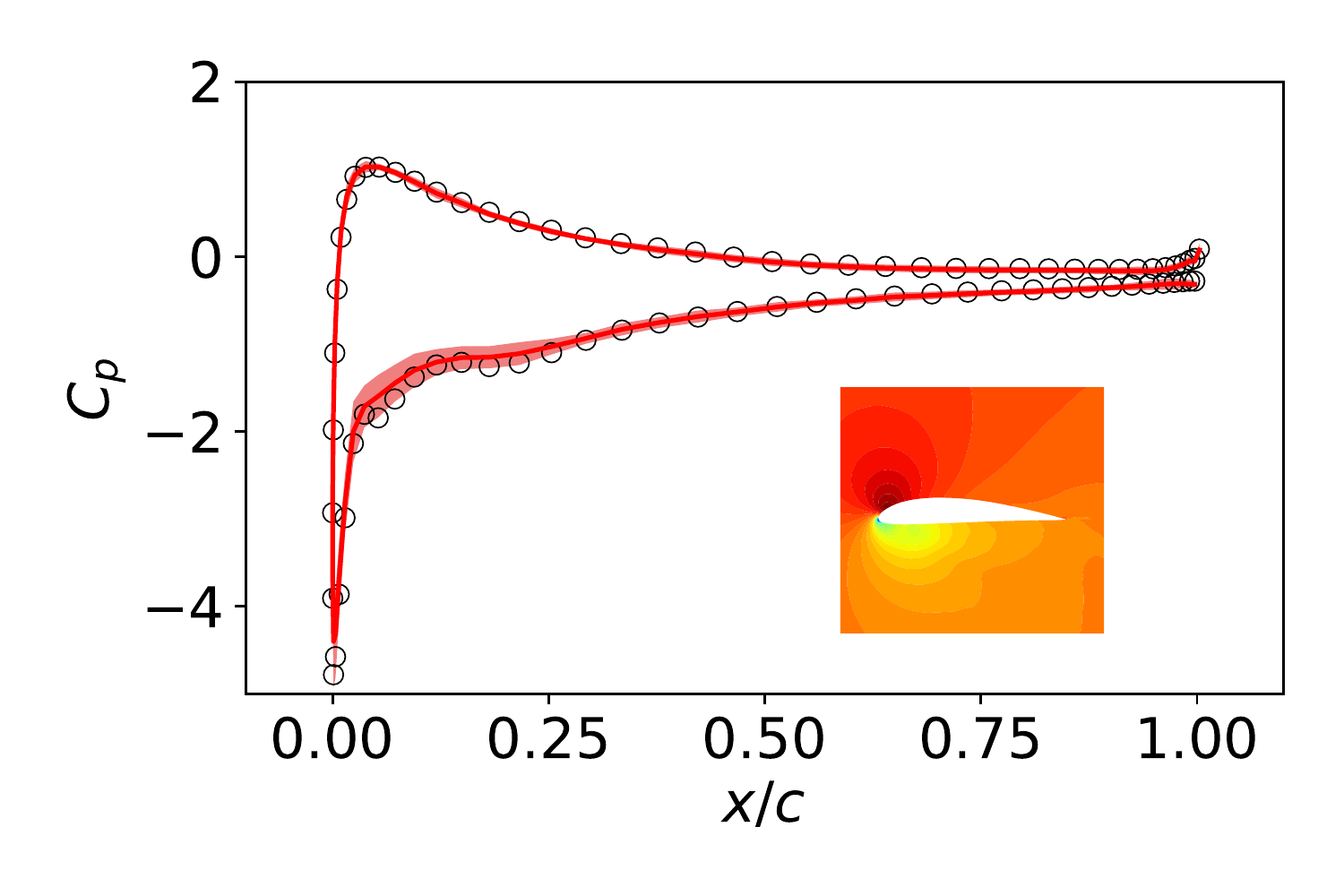}
\caption{goe398}
\end{subfigure}
\begin{subfigure}{.45\textwidth}
\centering
\includegraphics[width=\linewidth]{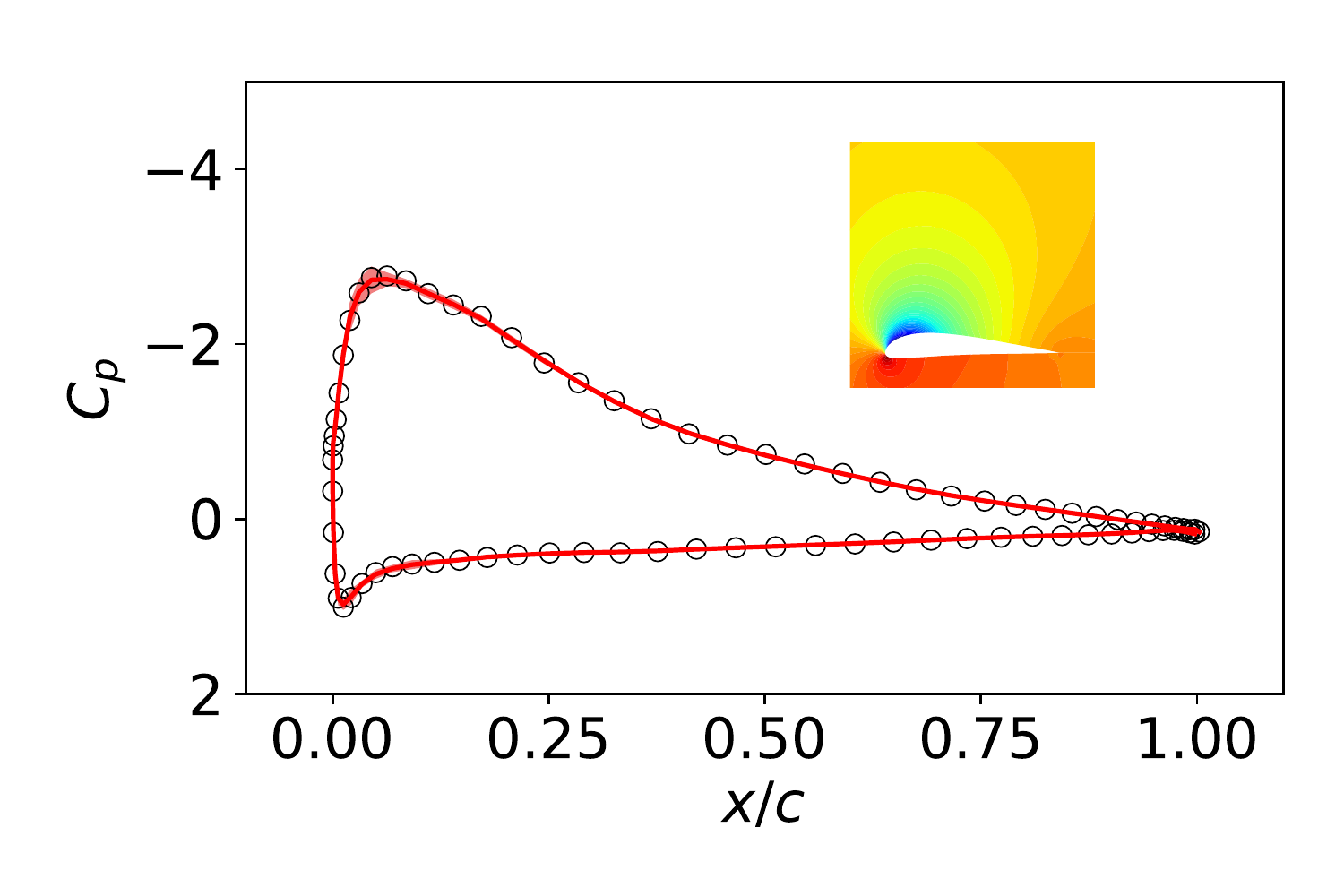}
\caption{mue139}
\end{subfigure}

\caption{Distributions of pressure coefficient in Case 2: a) e342, b) goe07k, c) goe398, and d) mue139. 
The red lines represent the DNN results with the shaded region visualizing $\pm3 SD$, and the black symbols represent the reference data. 
The embedded plots show the corresponding pressure fields. The free-stream conditions are listed in Table \ref{tab:exp_2_summary}.
}\label{fig:Exp2_Cp}
\end{figure}

Figure \ref{fig:Exp2_Cp} shows pressure coefficients $C_p=(p_w - p_{\infty})/(p_{t,\infty}-p_{\infty})$ 
as well as the pressure field around the aerofoil predicted by the DNN. To keep consistency with the upper and lower surfaces of the aerofoils, the \emph{y}-axis is inverted for positive angles of attack. The distributions predicted by DNN agree well with the reference results. Especially, the strong suction peaks are well captured in the DNN's predictions as shown in Figs. \ref{fig:Exp2_Cp}a and \ref{fig:Exp2_Cp}c.

\begin{figure}

\begin{subfigure}{.45\textwidth}
\centering
\includegraphics[width=\linewidth]{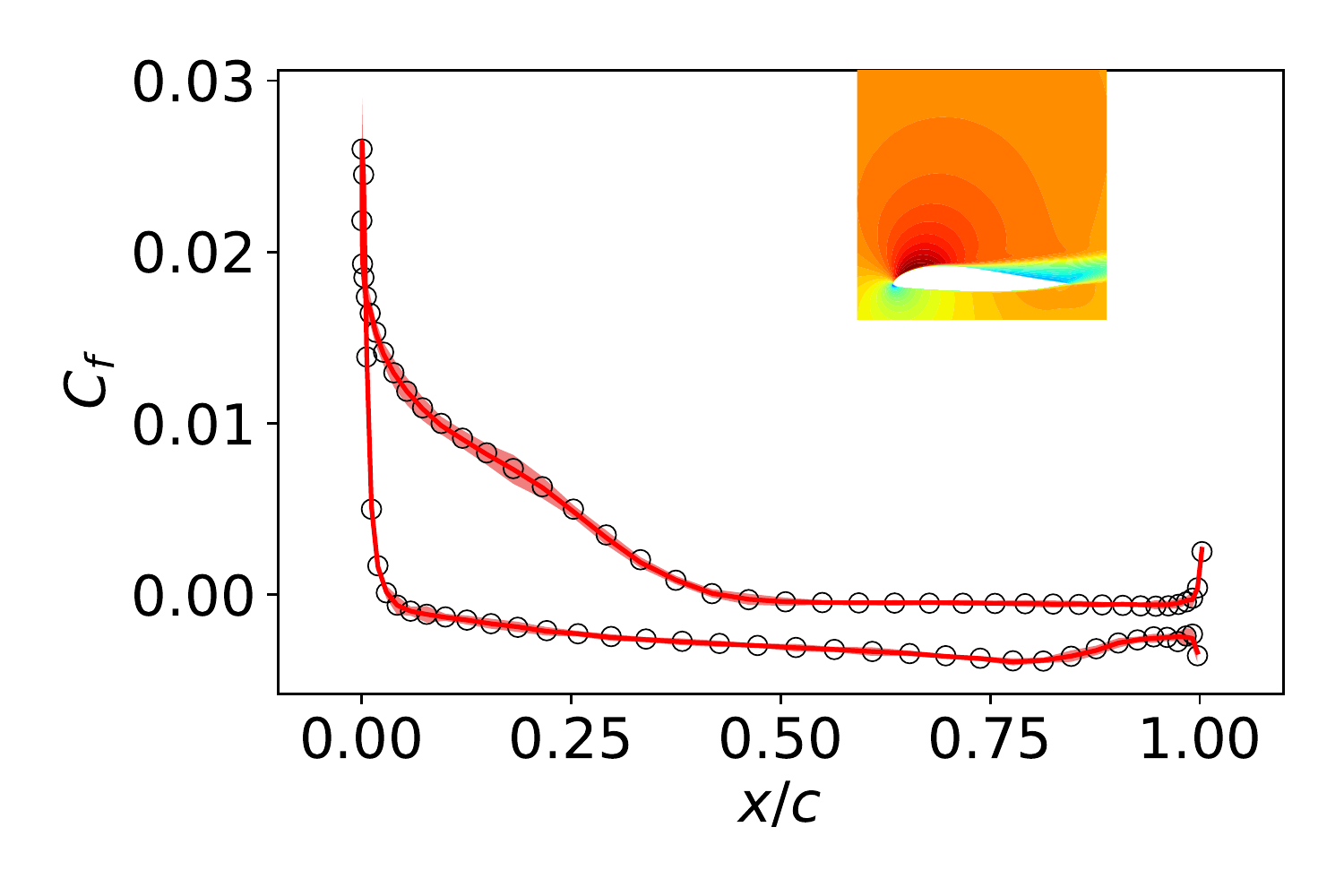}
\caption{e342}
\end{subfigure}
\begin{subfigure}{.45\textwidth}
\centering
\includegraphics[width=\linewidth]{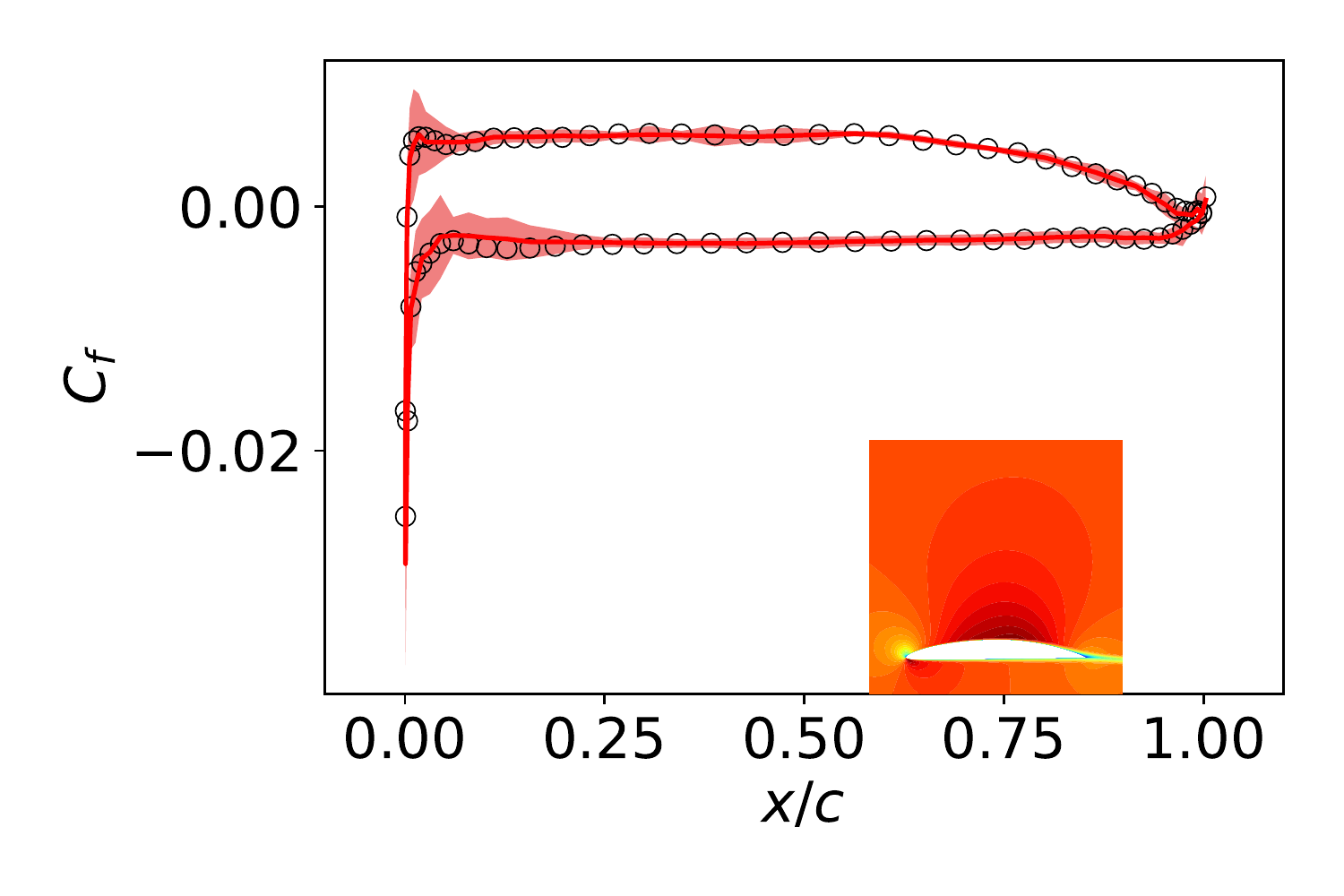}
\caption{goe07k}
\end{subfigure}

\begin{subfigure}{.45\textwidth}
\centering
\includegraphics[width=\linewidth]{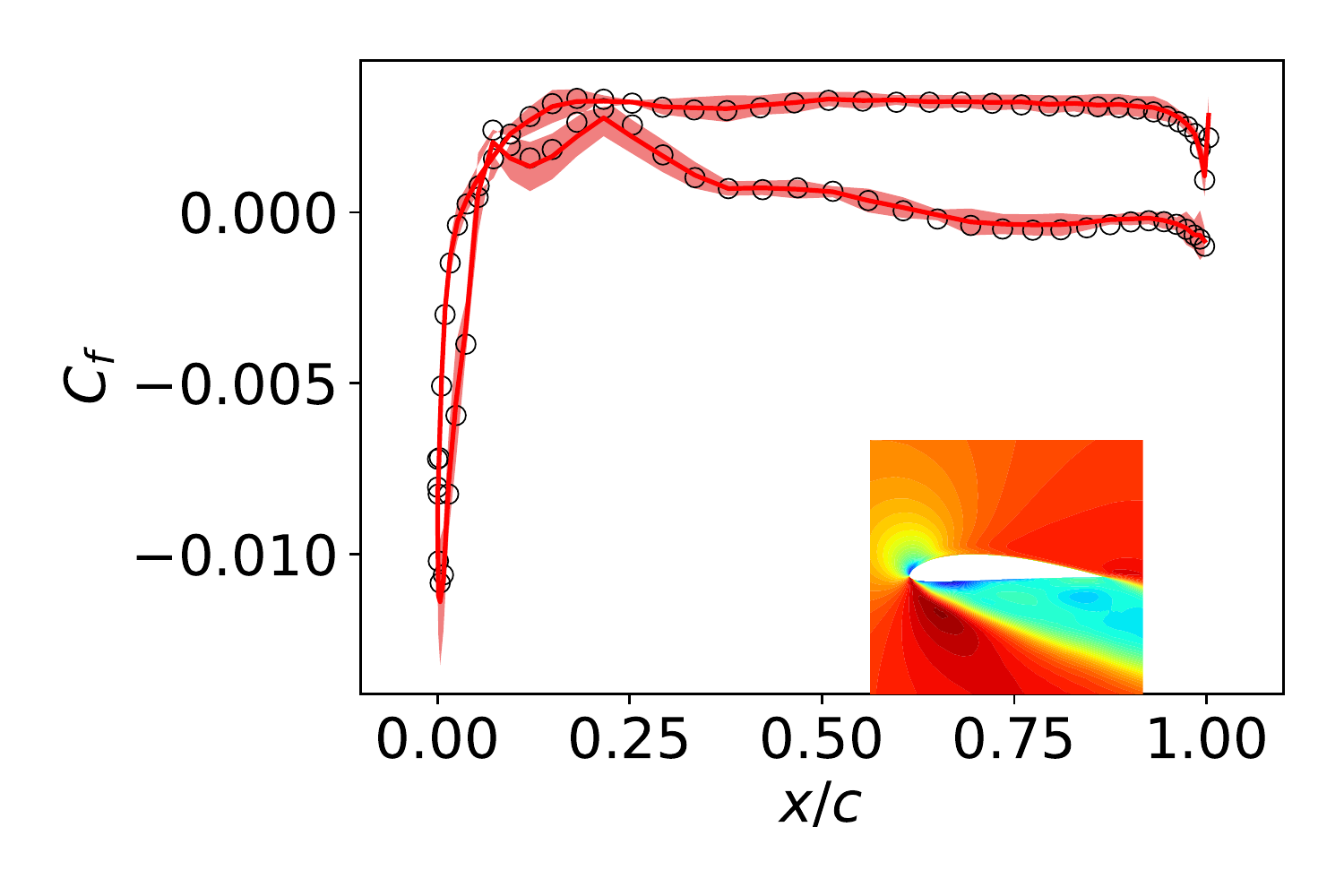}
\caption{goe398}
\end{subfigure}
\begin{subfigure}{.45\textwidth}
\centering
\includegraphics[width=\linewidth]{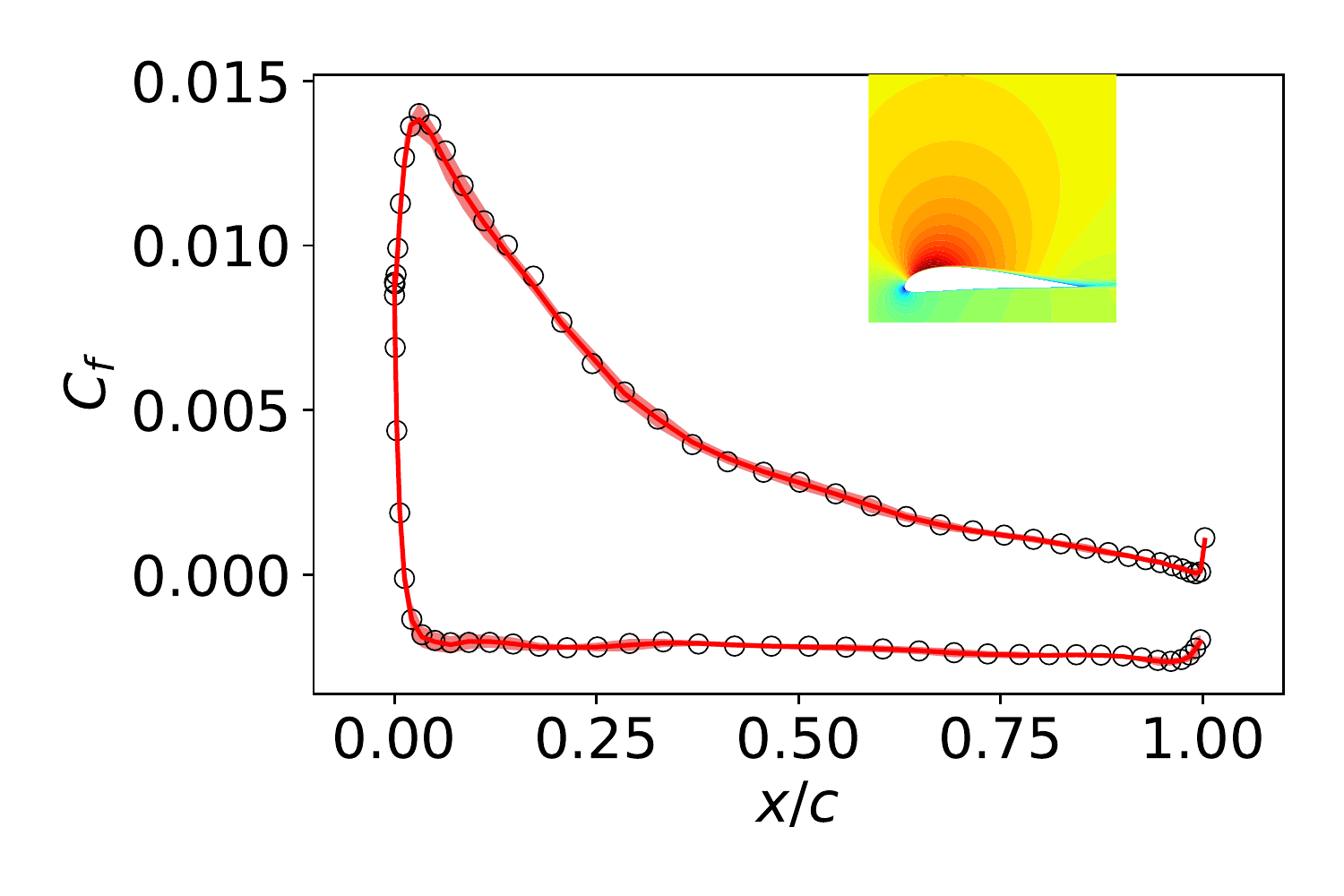}
\caption{mue139}
\end{subfigure}

\caption{Distributions of skin friction coefficient in Case 2: a) e342, b) goe07k, c) goe398, and d) mue139. 
The red lines represent the DNN results with the shaded region visualizing $\pm3 SD$, and the black symbols represent the reference data.
The embedded plots show the corresponding \emph{x}-component velocity fields. The free-stream conditions are listed in Table \ref{tab:exp_2_summary}.
}\label{fig:Exp2_Cf}
\end{figure}

Skin friction coefficients along the four test aerofoils are shown in Fig. \ref{fig:Exp2_Cf} and a reasonable agreement has been achieved with the results predicted by the solver CFL3D. We define the positive shear in the clock-wise direction.  
In Fig. \ref{fig:Exp2_Cf}a, the skin friction coefficient on the upper surface of aerofoil ``e342'' at $\alpha_{\infty}=14.7^{\circ}$ shows a peak in at the leading edge and slowly declined to a level of zero indicate at $x/c=0.4$ due to the occurrence of separation as evidenced in the accompanying Fig. \emph{x}-component velocity field.
In the cases of aerofoils ``goe07k'' and ``mue139'' in Figs. \ref{fig:Exp2_Cf}b and \ref{fig:Exp2_Cf}d, the skin friction peaks occur near the leading edges and the distributions don't change signs, indicating boundary layers on both upper and lower surfaces remain attached, which is consistent with the corresponding \emph{x}-component velocity fields.
Figure \ref{fig:Exp2_Cf}c shows a typical skin friction distribution with a large-scale separation at leading edge, which is reflected in the abrupt change at $x/c=0.05$ from a negative peak to a positive value. 

\begin{figure}

\begin{subfigure}{.3\textwidth}
\centering
\includegraphics[width=\linewidth]{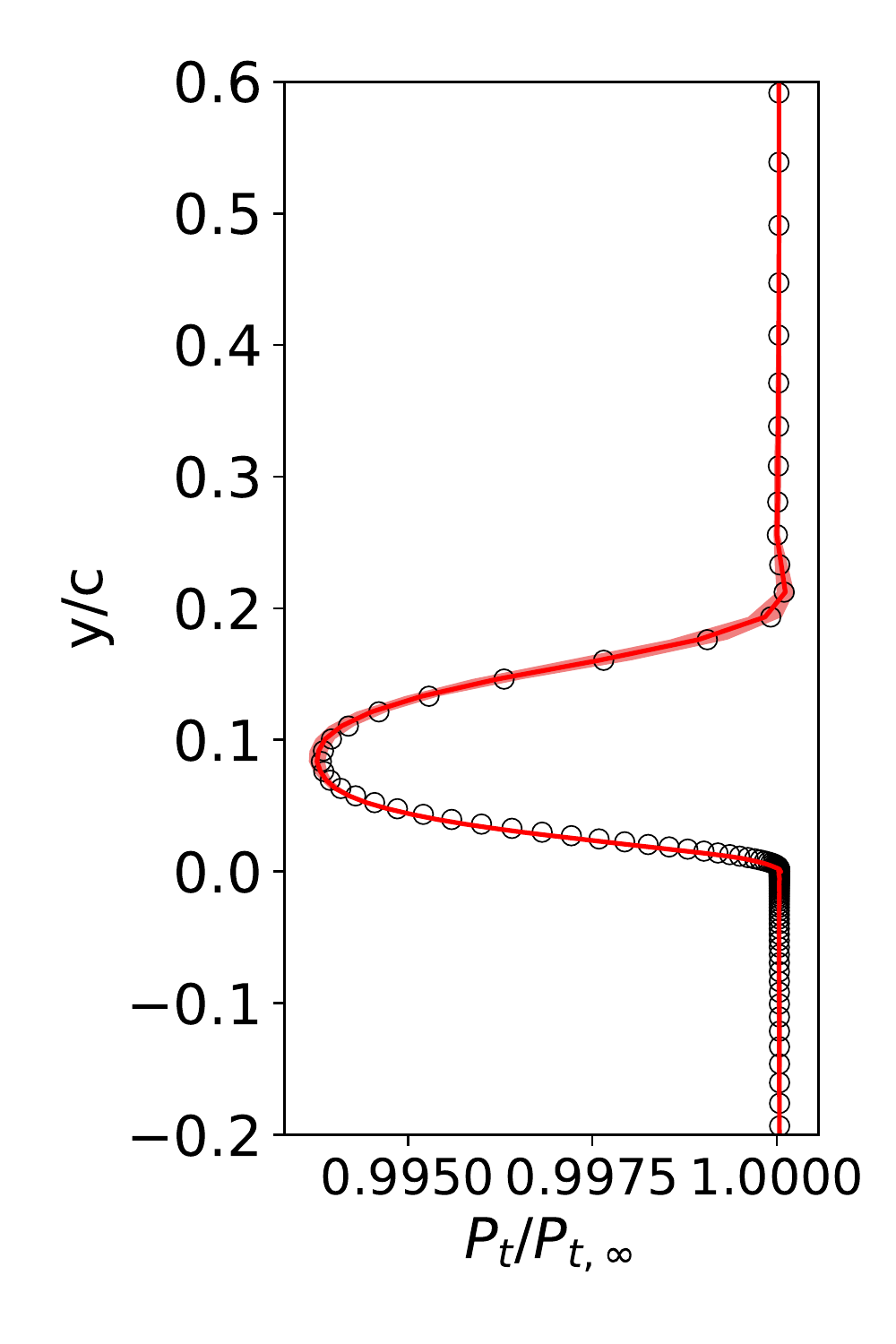}
\caption{e342}
\end{subfigure}
\begin{subfigure}{.3\textwidth}
\centering
\includegraphics[width=\linewidth]{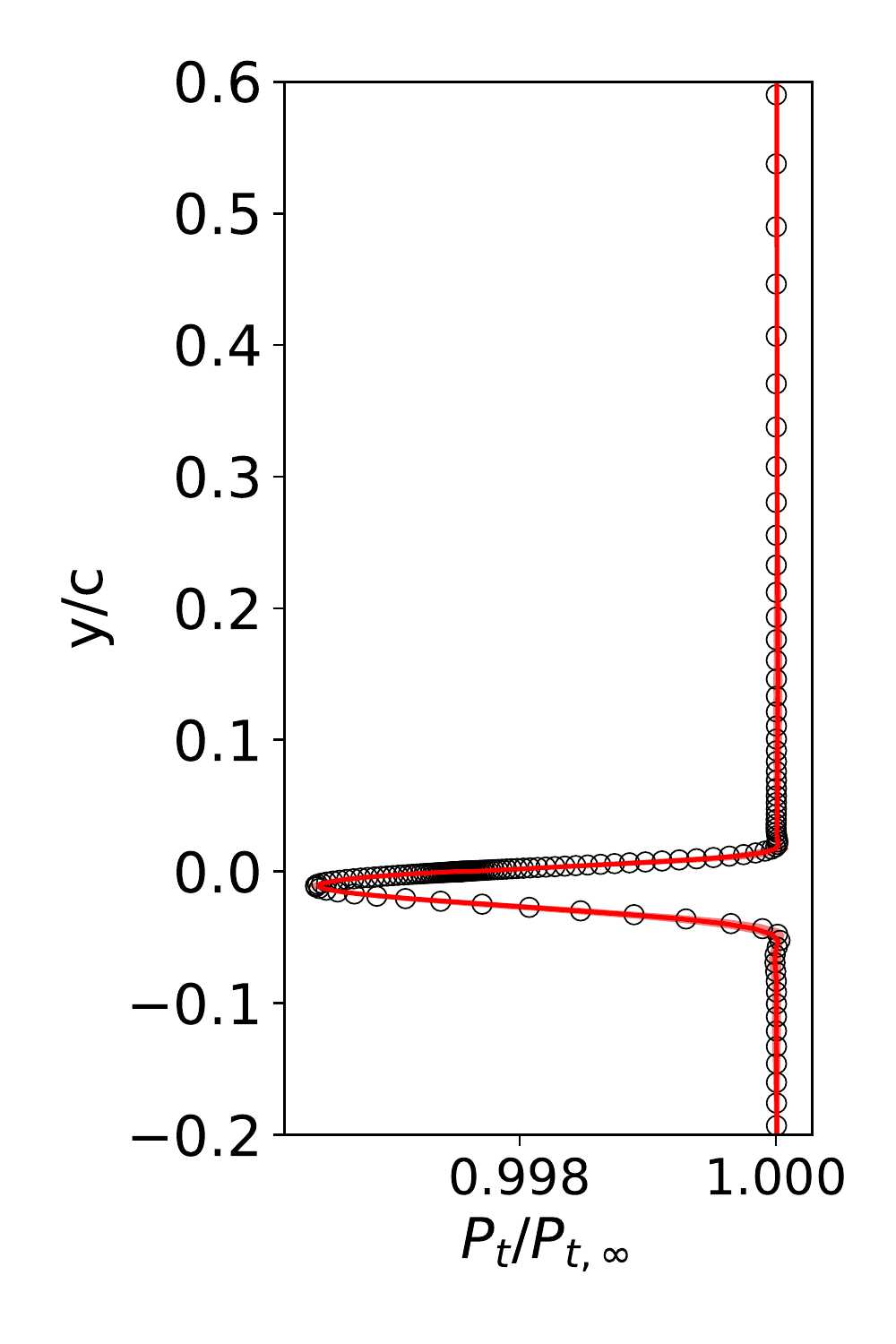}
\caption{goe07k}
\end{subfigure}

\begin{subfigure}{.3\textwidth}
\centering
\includegraphics[width=\linewidth]{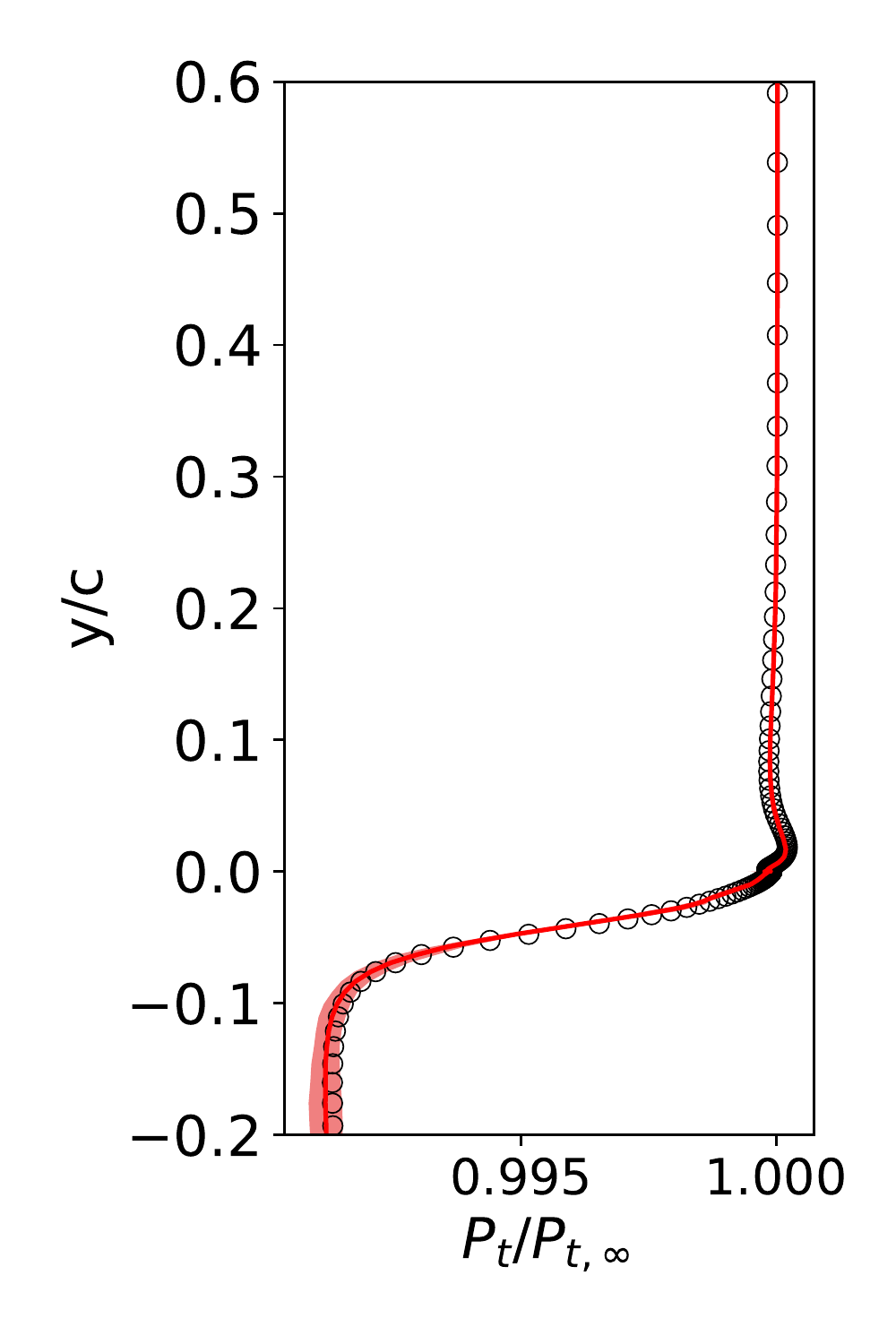}
\caption{goe398}
\end{subfigure}
\begin{subfigure}{.3\textwidth}
\centering
\includegraphics[width=\linewidth]{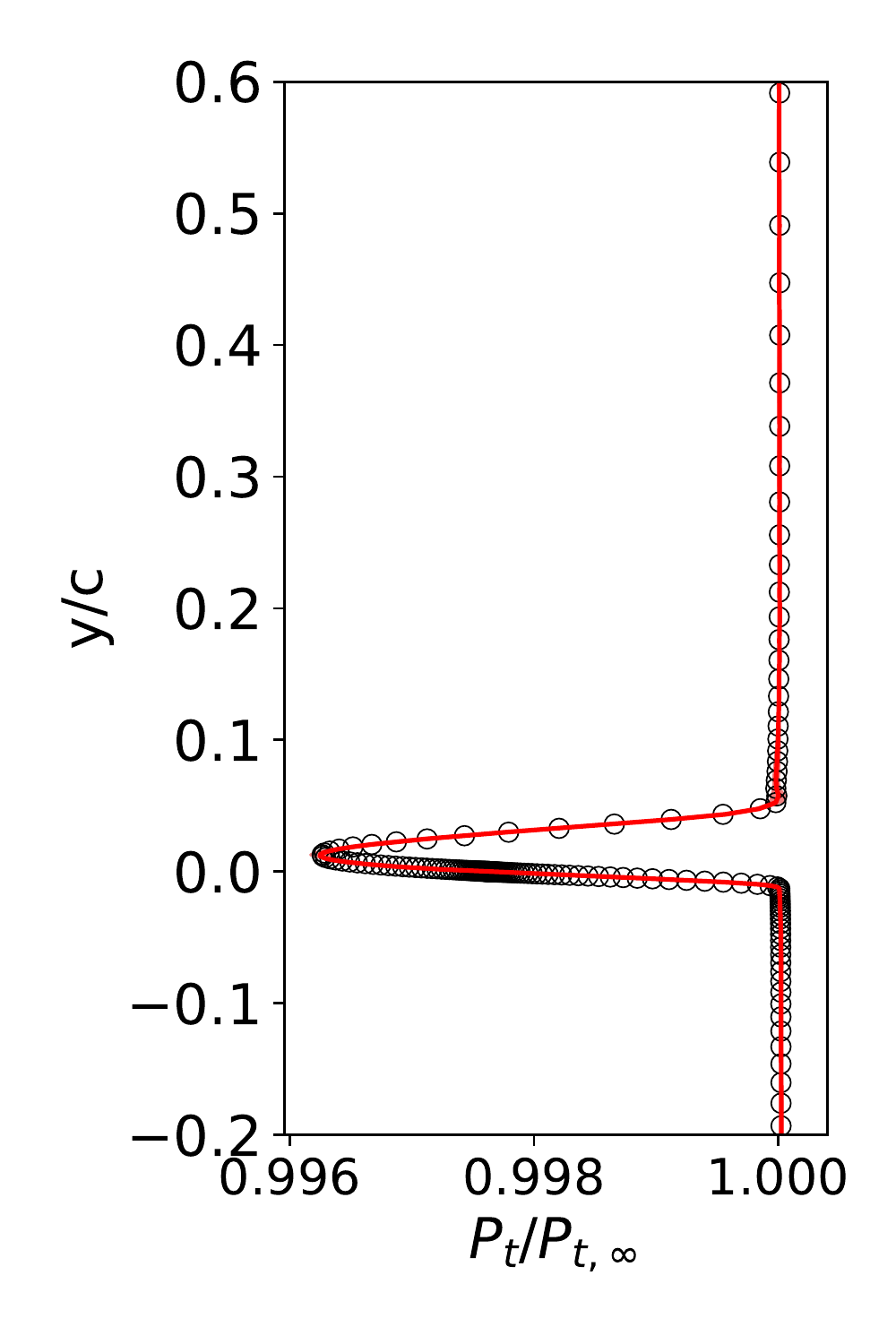}
\caption{mue139}
\end{subfigure}

\caption{Total pressure profiles at $x/c=1.2$ in Case 2. 
The red lines represent the DNN results with the shaded region visualizing $\pm3 SD$, and the black symbols represent the reference data.
}\label{fig:Exp2_wake}
\end{figure}

Figure \ref{fig:Exp2_wake} shows the total pressure profiles $p_{t}/p_{t,\infty}$ measured at 20\% of the chord length downstream of the trailing edges. The profiles are affected by the wake mixing as well as the separated flows. As shown in Fig. \ref{fig:Exp2_wake}a, due to the separation at a high angle of attack, the wake deficit is bigger than those with attached flows shown in Figs. \ref{fig:Exp2_wake}b and \ref{fig:Exp2_wake}d. In the case of a very high negative angle of attack case for ``goe398'' with $\alpha_{\infty}=-22.0^{\circ}$, the wake profile is clearly divided into a high and a low speed regions. These selected cases illustrate that the essential behavior of low Mach number pressure fields are successfully captured by the DNN model.

The shaded regions show $\pm3 SD$ to visualize the uncertainty in Figs. \ref{fig:Exp2_Cp}, \ref{fig:Exp2_Cf} and \ref{fig:Exp2_wake}. From Figs. \ref{fig:Exp2_Cp}c and \ref{fig:Exp2_Cf}c, it can be seen that the due to the complex nature in this condition, i.e. massive flow separation starting from the leading edge, $C_p$ and $C_f$ on the lower surface are slightly more sensitive. It is reflected in that the standard deviations in the pressure and viscous drags take up $1.05\%$ and $0.02\%$ of the mean total drag, respectively. 
The massive separation also leads to a wider sensitivity shade region around the lower part of the wake profile in Fig. \ref{fig:Exp2_wake}. 
In Fig. \ref{fig:Exp2_Cf}b, $\pm3 SD$ shades are wider around the leading edge of the aerofoil ``goe07k'' although the flow is well attached, which is likely because the aerofoil has a sharper leading edge which rarely appears in the training dataset. As a result, the standard deviations in the pressure and viscous drags take up $10.33\%$ and $1.31\%$ of the mean total drag, respectively.
\label{sec:sample1}
\begin{table}
  \begin{center}
\def~{\hphantom{0}}
  \begin{tabular}{lcccc} 
      \toprule
      Aerofoil    & $\xmach_{\infty}$ & $\alpha_{\infty}$ & $\Rey_{\infty}$ & Test loss ($L_1 \pm SD$) \\
      \midrule 
      e342     & 0.1 & $14.70^{\circ}$ & $2.703\times10^6$ &   $(3.68\pm0.11)\times10^{-4}$  \\ %
      goe07k     & 0.1 & $-2.33^{\circ}$ & $0.993\times10^6$ &  $(4.69\pm0.18)\times10^{-4}$  \\ %
      goe398 & 0.1 & $-22.00^{\circ}$ & $4.638\times10^6$ &    $(13.30\pm0.26)\times10^{-4}$  \\%
      mue139 & 0.1 & $8.78^{\circ}$ & $4.704\times10^6$ &    
      $(2.95\pm0.06)\times10^{-4}$  \\%
      \bottomrule

  \end{tabular}
  \caption{L1 losses of four representative conditions/shapes in the test set of Case 2.}
  \label{tab:exp_2_summary}
  \end{center}
\end{table}

\subsection{Case 3: compressible inviscid flow}
\noindent
\subsubsection{Overview}
In this section, we consider compressible inviscid flows over a given aerofoil, i.e. ``naca0012'', at a set of free-stream conditions. This case is originally proposed by \cite{deAvila_icml2020_6414} as a showcase of their deep learning method based on the coupling of an adjoint CFD solver and the GNN. 

To make reasonable comparisons, we generate the flow field data with similar mesh resolution. In particular, the o-grid topology mesh for the aerofoil naca0012 is considered as shown in Fig. \ref{fig:cmesh_topology}b. 
The free-stream conditions are given in Table \ref{tab:testcases_summary}. As can be seen from the table, there are much less samples in the training and test sets, which poses additional challenges on the learning and inference task. 
Thus, it represents a good case to show that the proposed method is capable of providing high fidelity results even when the amount of training data is relatively small.

We train the models for more epochs with a batch size of 7. A learning rate $5\times10^{-5}$ is used for 15000 epochs, and then it gradually decays to $5\times10^{-6}$ for another 15000 epochs. Although we train the DNN model with an $L_1$ loss of the flow variables, i.e. density, \emph{x}-velocity, \emph{y}-velocity and speed of sound, we now calculate a root mean square error (RMSE) to assess predictions following \cite{deAvila_icml2020_6414}: 

\textcolor{lCol}{
\begin{equation*}
    RMSE = \sqrt{\frac{1}{T\times S} \sum\limits_{n=1}^T\sum\limits_{m=1}^S{\frac{1}{3}\Big( (\Breve{u}_{m,n}-u_{m,n})^2+(\Breve{v}_{m,n}-v_{m,n})^2+(\Breve{p}_{m,n}-p_{m,n})^2} \Big)   }.
\end{equation*}}

In previous work \citep{deAvila_icml2020_6414}, the GNN achieves a RMSE of $1.4\times10^{-2}$ and the hybrid CFD-GNN method gives $1.8\times10^{-2}$. By contrast, the present study which is purely based on the DNN predicts flow field accurately with a very small RMSE, i.e. $5.04\times10^{-3}$ with method C and $4.63\times10^{-3}$ with method B.

\subsubsection{Flow field prediction}
\begin{figure}
\begin{subfigure}{.45\textwidth}
\centering
\includegraphics[width=\linewidth]{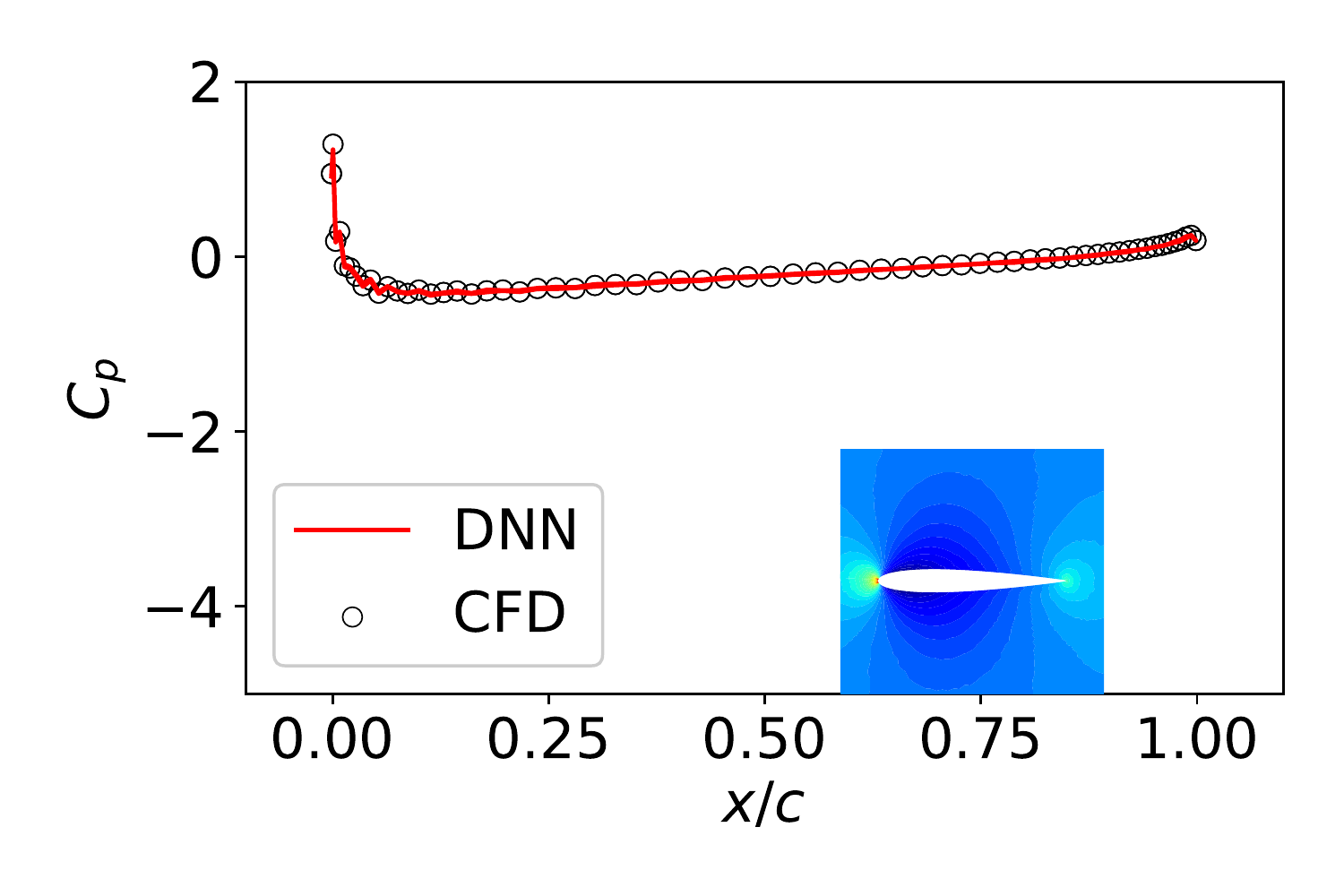}
\caption{}
\end{subfigure}
\begin{subfigure}{.45\textwidth}
\centering
\includegraphics[width=\linewidth]{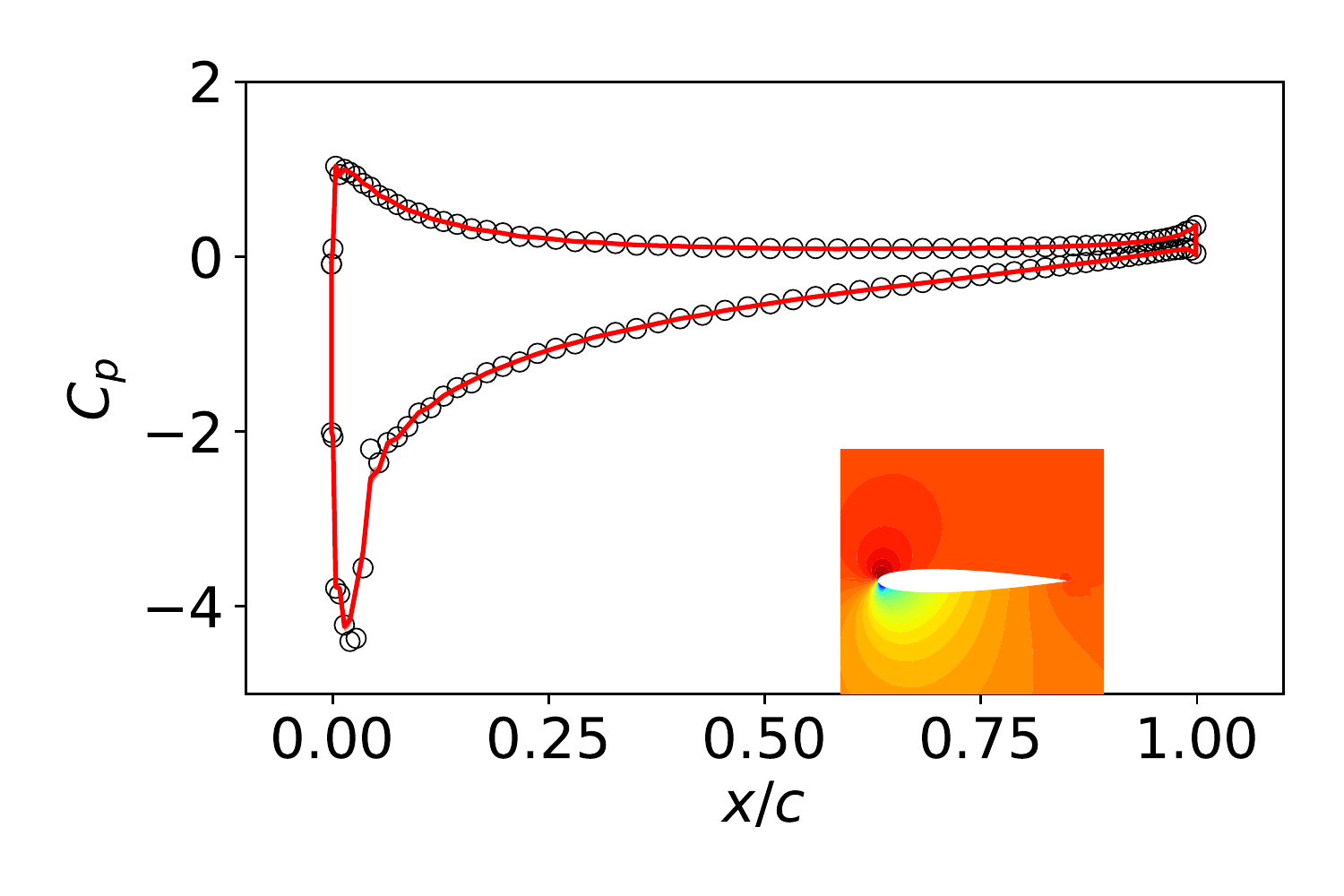}
\caption{}
\end{subfigure}

\begin{subfigure}{.45\textwidth}
\centering
\includegraphics[width=\linewidth]{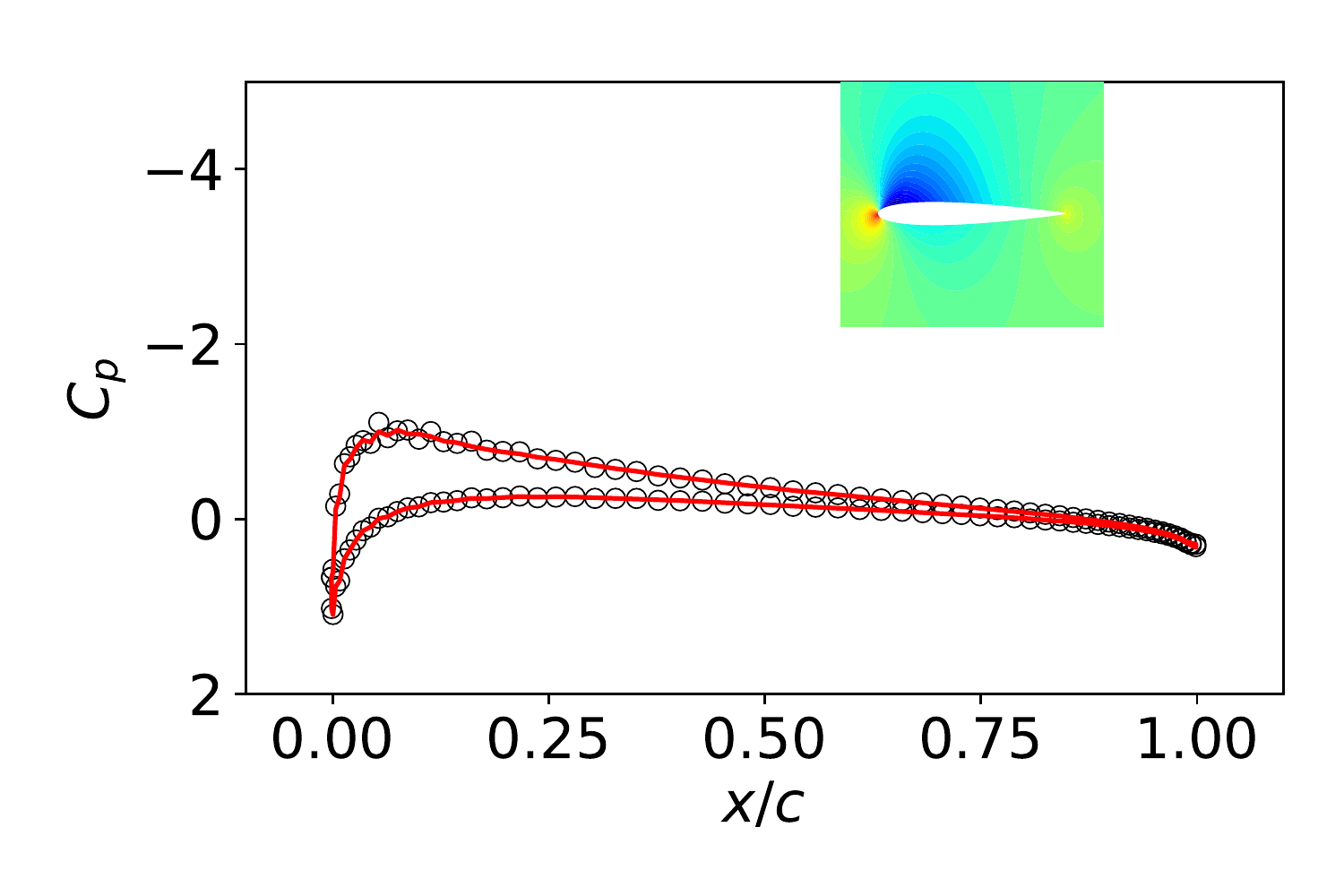}
\caption{}
\end{subfigure}
\begin{subfigure}{.45\textwidth}
\centering
\includegraphics[width=\linewidth]{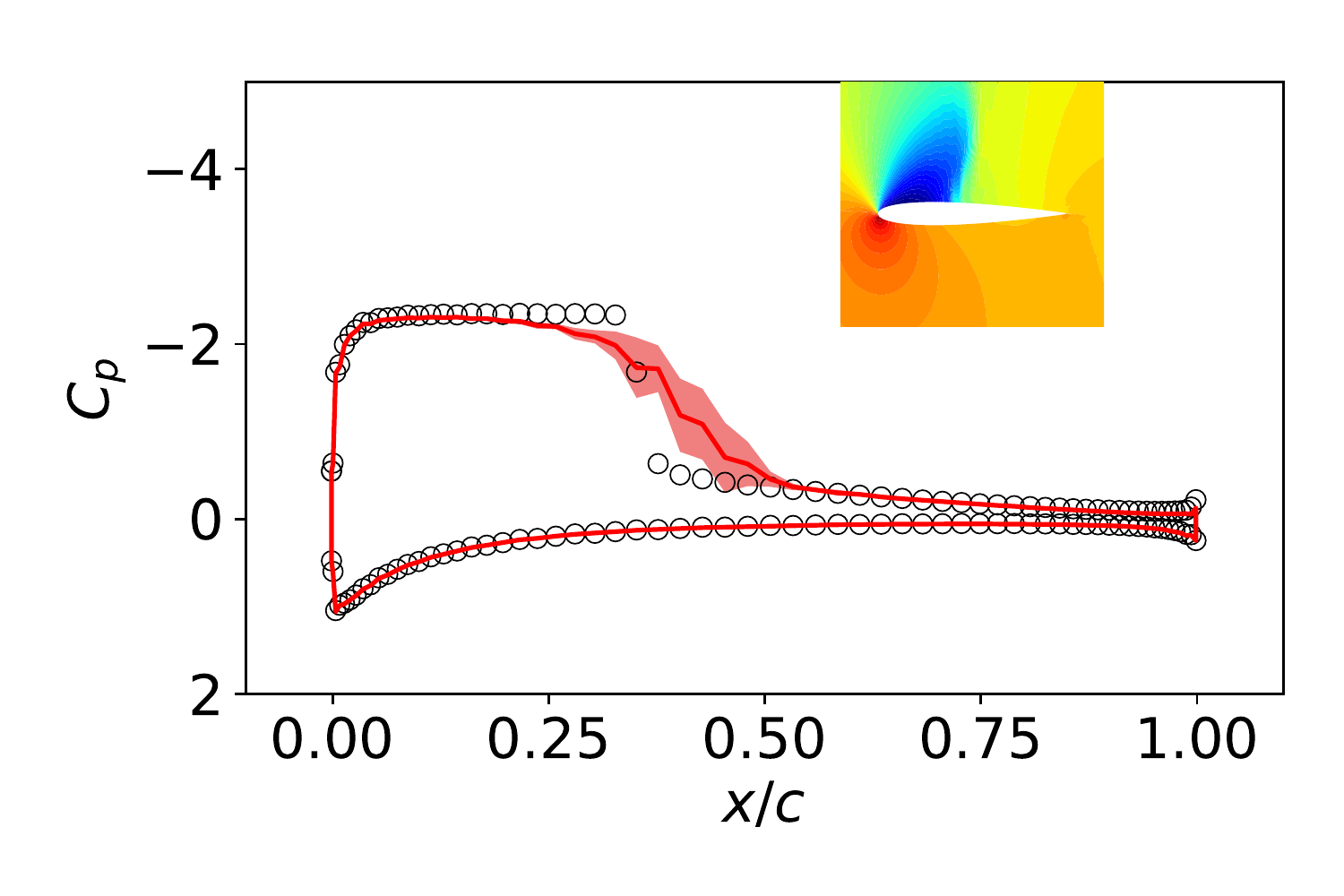}
\caption{}
\end{subfigure}
\caption{Distributions of pressure coefficient $C_p=(p_w - p_{\infty})/(p_{t,\infty}-p_{\infty})$ in Case 3: a) $\xmach_{\infty}=0.25$, $\alpha_{\infty}=0^{\circ}$, b) $\xmach_{\infty}=0.45$, $\alpha_{\infty}=-8^{\circ}$, c) $\xmach_{\infty}=0.65$, $\alpha_{\infty}=2^{\circ}$, and d) $\xmach_{\infty}=0.65$, $\alpha_{\infty}=10^{\circ}$. 
The red lines represent the DNN results with the shaded region visualizing $\pm1 SD$, and the black symbols represent the reference data. The embedded plots show the corresponding pressure fields.}
\label{fig:Exp4_Cp}
\end{figure}

The high accuracy of the predictions of the DNN is reaffirmed
by the distribution pressure coefficients as shown in Figs. \ref{fig:Exp4_Cp}(a-\emph{c}). The worst prediction in terms of $L_1$ loss among the present test cases happens at $10^{\circ}$ and $\xmach_{\infty}=0.65$, as shown in Fig. \ref{fig:Exp4_Cp}d. 
Despite the strong gradient region caused by the shock wave being sub-optimally resolved and wider shaded region of uncertainty around the shock, the overall distribution of pressure coefficient is still predicted reasonably well, especially with respect to the shock location on the suction side. 
In addition, the stagnation point agrees well with the one computed by the CFD solver CFL3D.

\subsection{Case 4: transonic viscous flow}
\begin{table}
  \begin{center}
\def~{\hphantom{0}}
  \begin{tabular}{lcccc} 
      \toprule
      Aerofoil    & $\xmach_{\infty}$ & $\alpha_{\infty}$ & $\Rey_{\infty}$ & Test loss ($L_1 \pm SD$) \\
      \midrule 
      e221     & 0.71 & $3.43^{\circ}$ & $0.661\times10^6$ &  $(11.52 \pm 0.15)\times10^{-4}$ \\ 
      e473     & 0.55 & $0.29^{\circ}$ & $1.294\times10^6$ &  $(2.67 \pm 0.12)\times10^{-4}$\\ 
      fx84w097 & 0.78 & $4.2^{\circ}$ & $1.501\times10^6$ &  $(3.13 \pm 0.02)\times10^{-4}$  \\
      mue139 & $0.68$ & $-0.02^{\circ}$ & $3.178\times10^6$ & $(12.15 \pm 0.80)\times10^{-4}$ \\ 
      \bottomrule
  \end{tabular}
  \caption{L1 losses of four representative conditions/shapes in the test set of Case 4.}
  \label{tab:exp_3_summary}
  \end{center}
\end{table}

\noindent
\subsubsection{Overview}
In this section, we consider transonic viscous flows over aerofoils. The inference of RANS solutions randomly distribute in a range of Mach numbers $\xmach_{\infty}\in[0.55, 0.8]$, Reynolds numbers $\Rey_{\infty}\in[0.5, 5]$ million, and angles of attack in the range of [$-0.5^{\circ}$, $8.0^{\circ}$], following a uniform distribution. The range of free-stream conditions described above is chosen according to the early study of shock wave/boundary layer interaction on aerofoils by \cite{Holder1955reynolds}.

As for Case 2, we randomly choose aerofoil shape and free-stream condition from the training range, and use 20 additional aerofoils as test set.
Note that the flow field is intrinsically unsteady when reaching transonic ``buffet'' boundary \citep{Deck2005oat15a, chen_xu_lu_2010}. In order to obtain meaningful ensemble averaged flow field, we run 24000 iterations with Courant number 1.0 for each case and we average the intermediate results over the last 8000 iterations. 
For a more fundamental discussion on the non-uniqueness at transonic regimes, we refer to \cite{Jameson1991euler} and \cite{Ou2014airfoils}.

In the present work, the relative errors are 0.255\% for density, 0.289\% for the \emph{x} velocity and 0.0996\% for the \emph{y} velocity, 0.0621\% for the speed of sound and 0.269\% for pressure. The mean test loss ($L_1$) for the whole test set is $5.54\times10^{-4}$.

\subsubsection{Flow field prediction}

\begin{figure}
    \centering
    \includegraphics[width=1\textwidth]{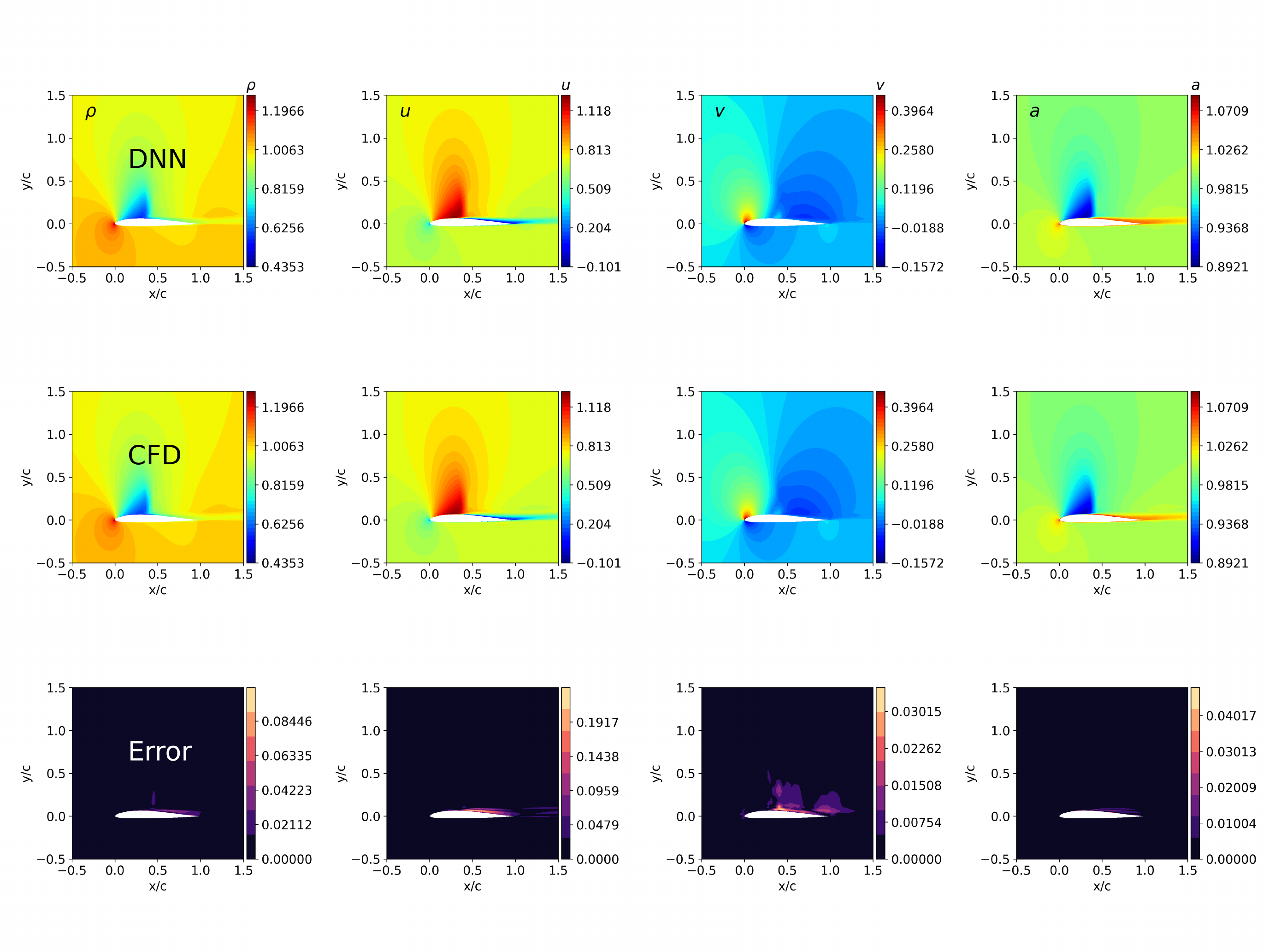}
    \caption{The comparison of flow fields for the aerofoil ``e221'' in Case 4 predicted by DNN (top row), CFD (middle row) and relative errors (the bottom row).}
    \label{fig:Exp3_flowfield}
\end{figure}

\begin{figure}

\begin{subfigure}{.45\textwidth}
\centering
\includegraphics[width=\linewidth]{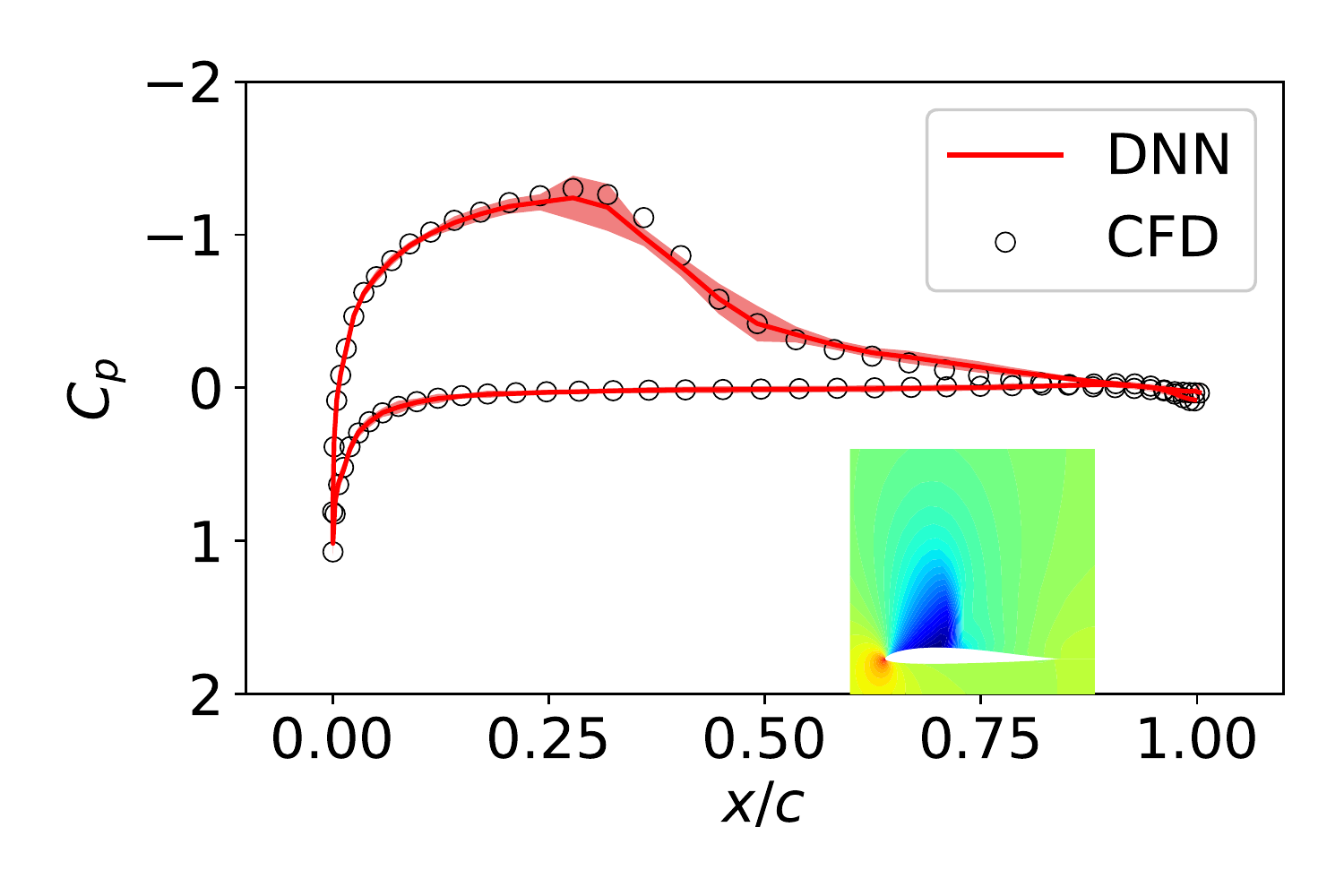}
\caption{e221}
\end{subfigure}
\begin{subfigure}{.45\textwidth}
\centering
\includegraphics[width=\linewidth]{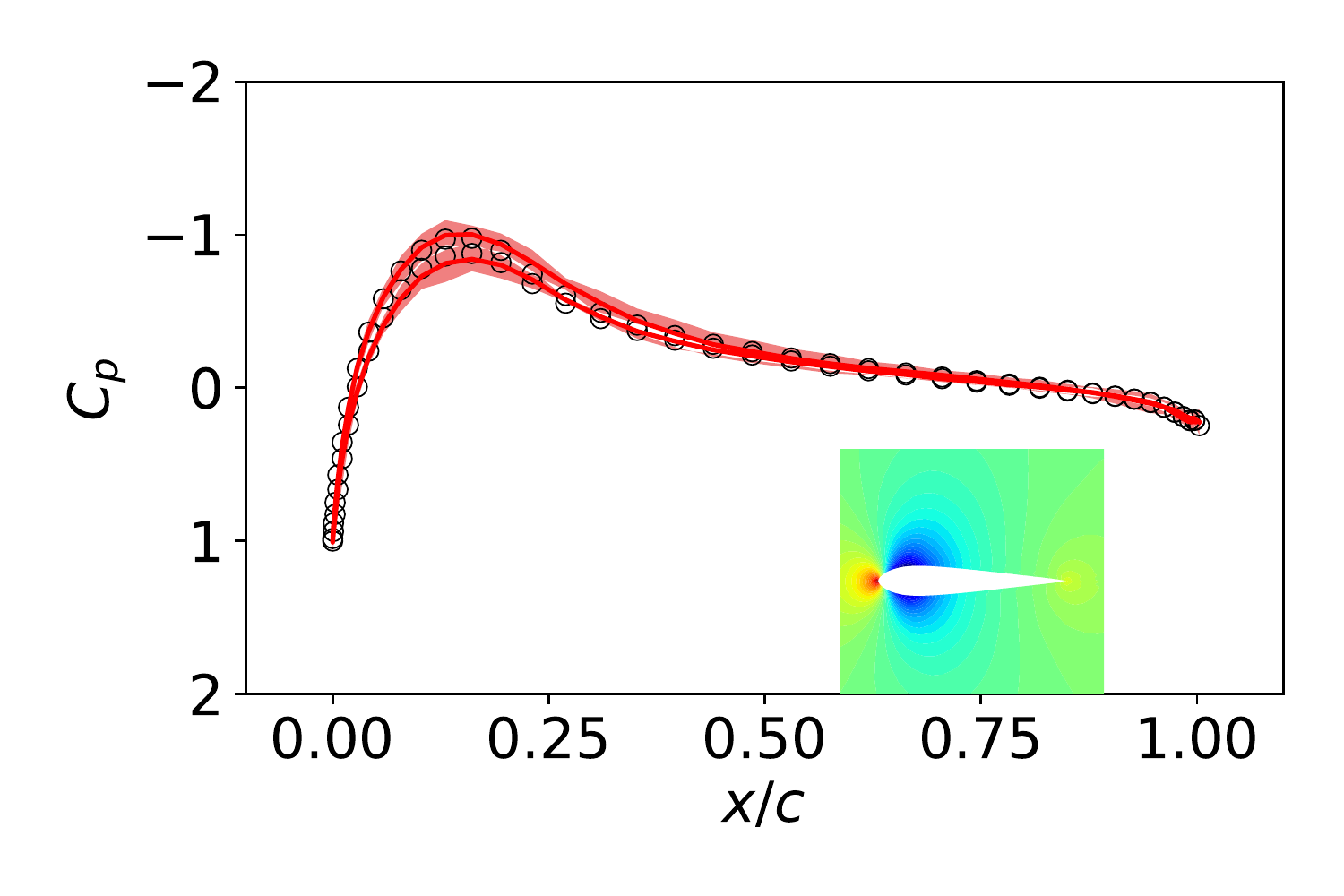}
\caption{e473}
\end{subfigure}

\begin{subfigure}{.45\textwidth}
\centering
\includegraphics[width=\linewidth]{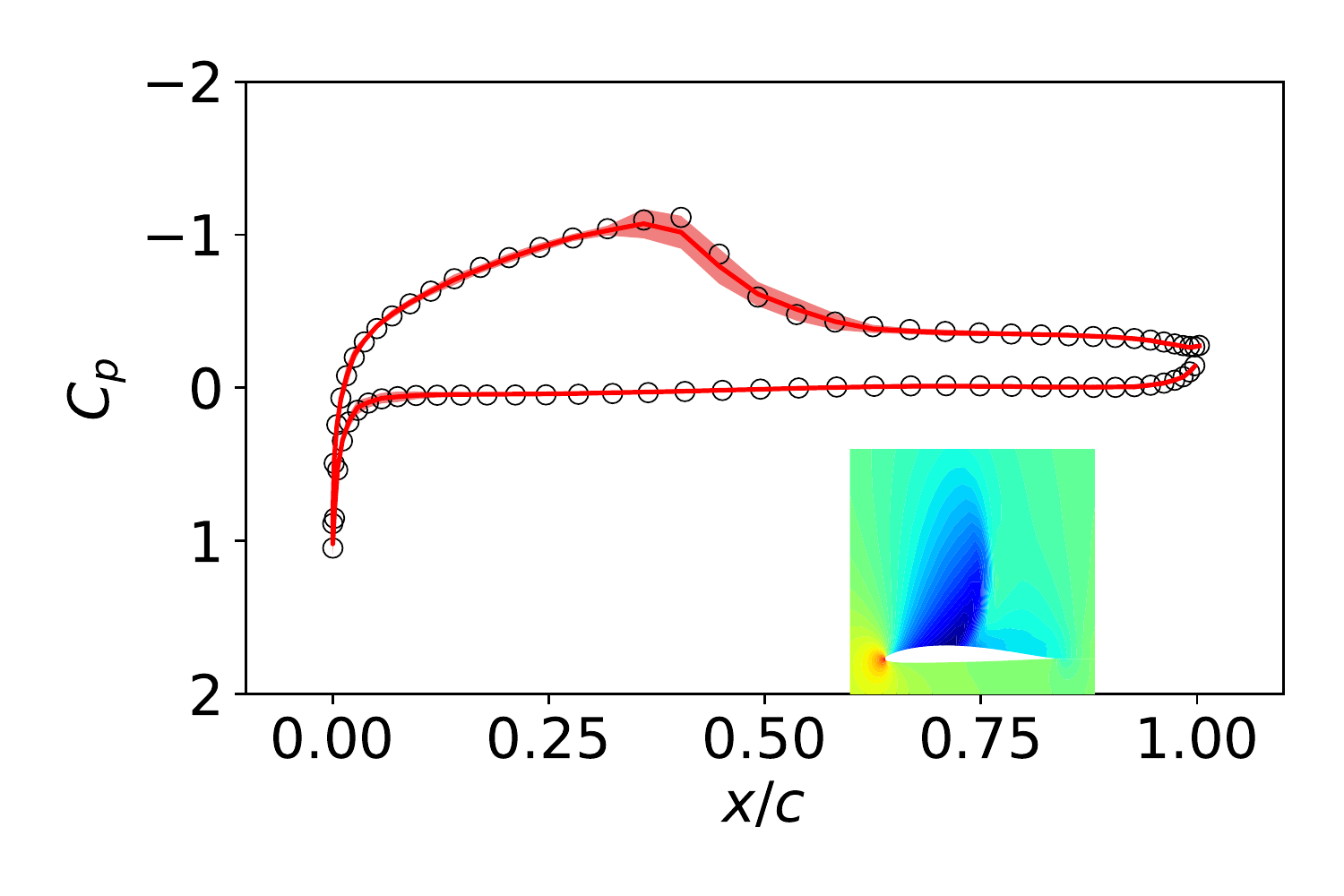}
\caption{fx84w097}
\end{subfigure}
\begin{subfigure}{.45\textwidth}
\centering
\includegraphics[width=\linewidth]{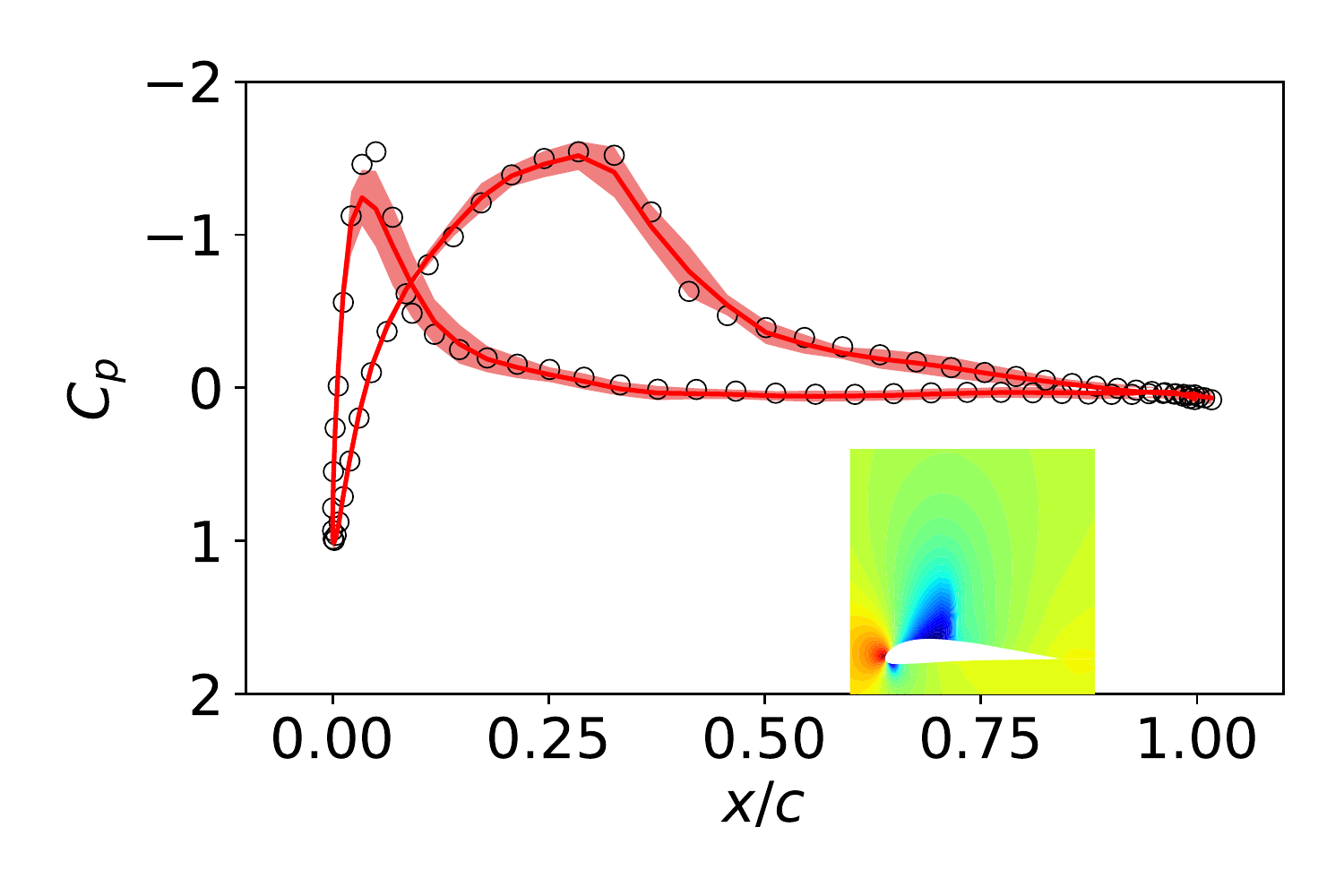}
\caption{mue139}
\end{subfigure}

\caption{Distributions of pressure coefficient in Case 4: a) e221, b) e473, c) fx84w097, and d) mue139. 
The red lines represent the DNN results with the shaded region visualizing $\pm3 SD$, and the black symbols represent the reference data. 
The embedded plots show the corresponding pressure fields. The free-stream conditions are listed in Table \ref{tab:exp_3_summary}.}
\label{fig:Exp3_Cp}
\end{figure}

\begin{figure}

\begin{subfigure}{.45\textwidth}
\centering
\includegraphics[width=\linewidth]{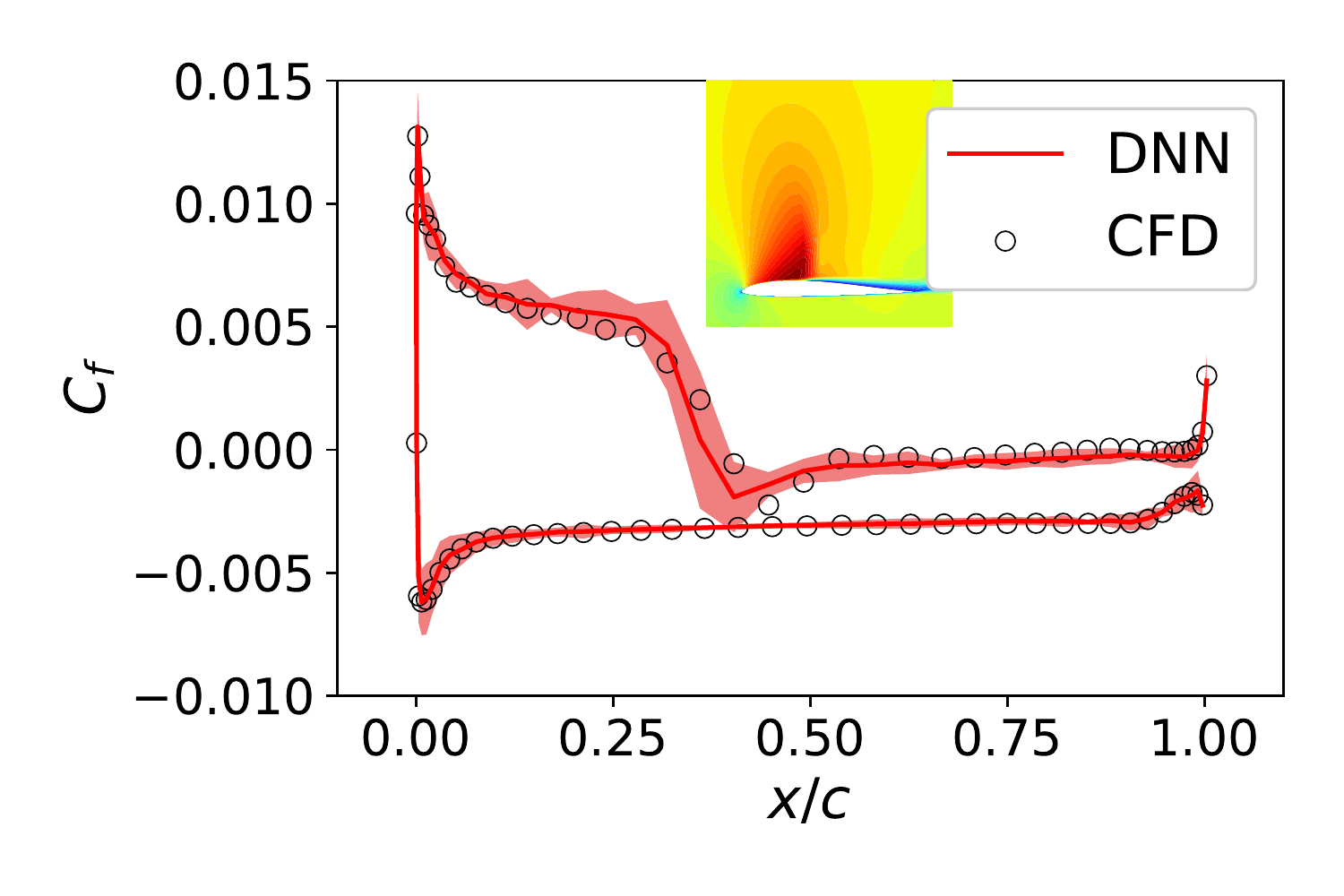}
\caption{e221}
\end{subfigure}
\begin{subfigure}{.45\textwidth}
\centering
\includegraphics[width=\linewidth]{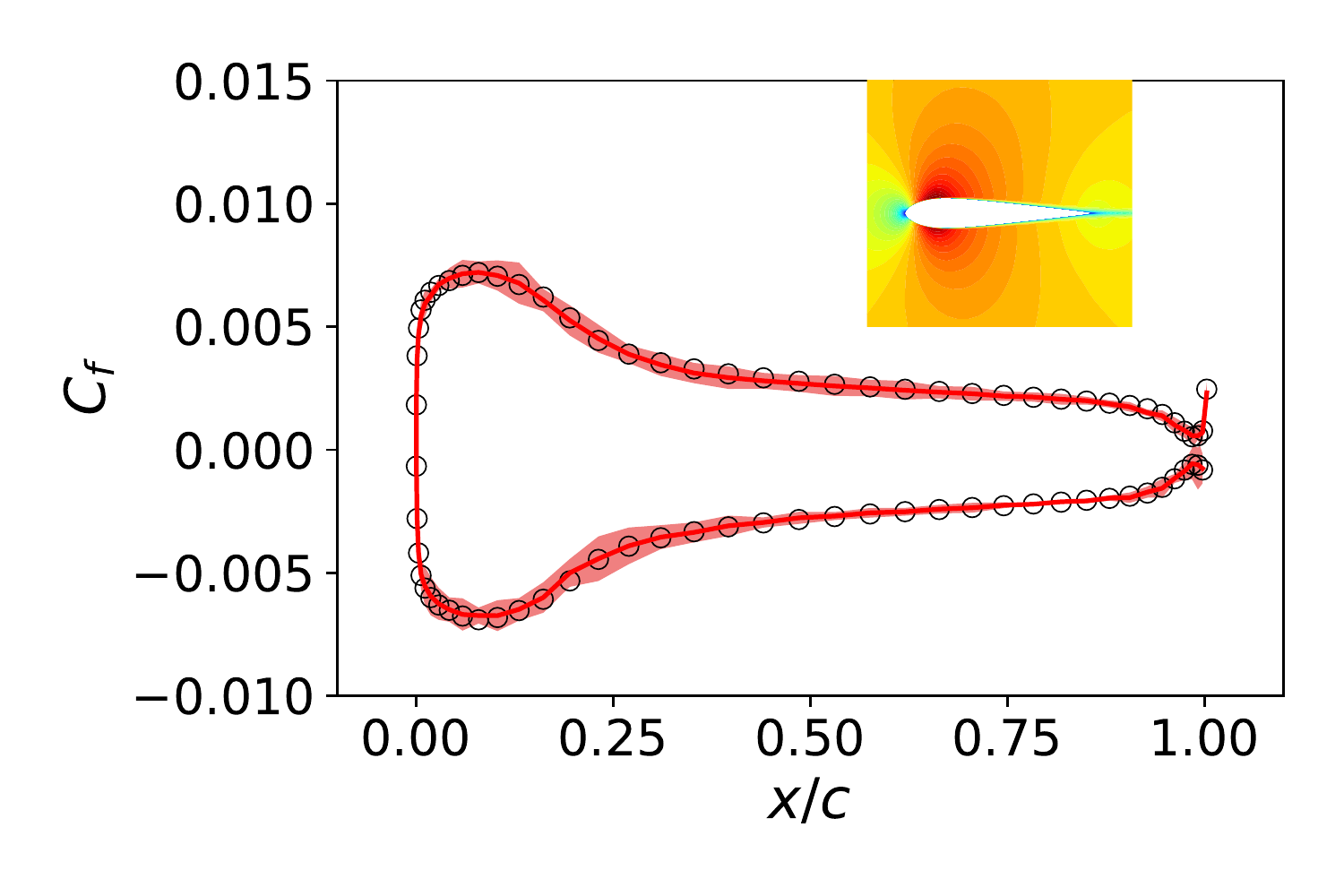}
\caption{e473}
\end{subfigure}

\begin{subfigure}{.45\textwidth}
\centering
\includegraphics[width=\linewidth]{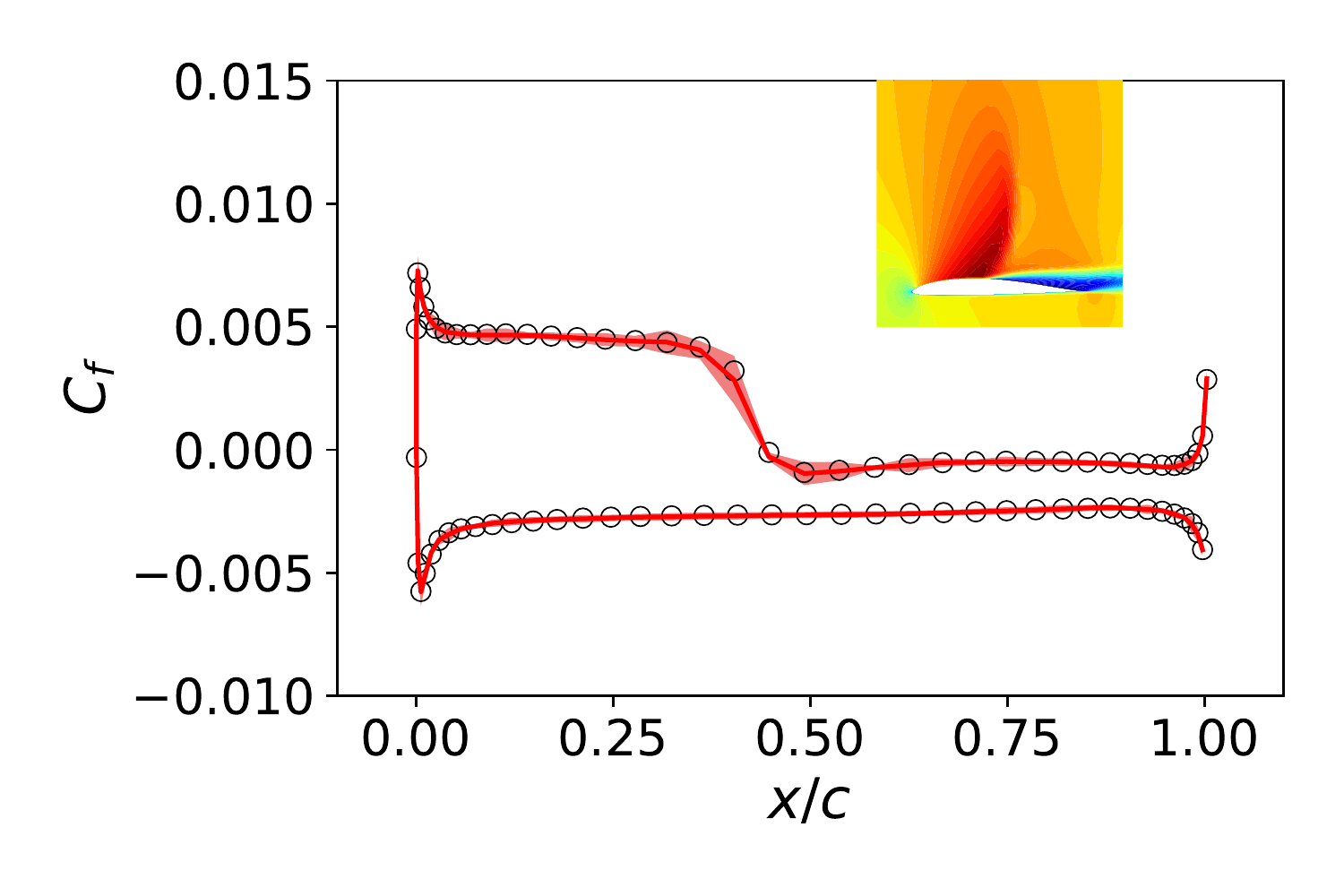}
\caption{fx84w097}
\end{subfigure}
\begin{subfigure}{.45\textwidth}
\centering
\includegraphics[width=\linewidth]{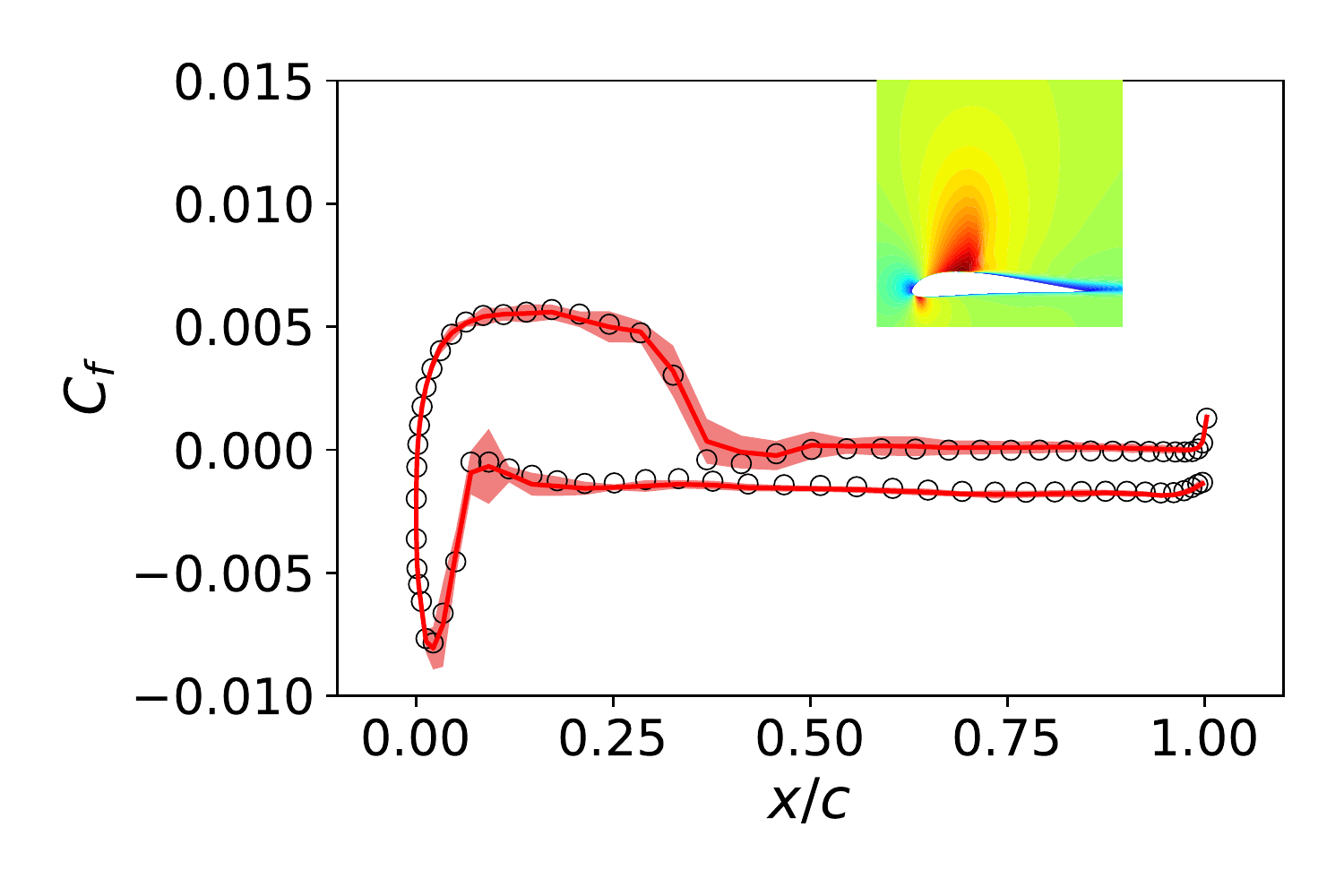}
\caption{mue139}
\end{subfigure}

\caption{Distributions of skin friction coefficient in Case 4. 
The red lines represent the DNN results with the shaded region visualizing $\pm3 SD$, and the black symbols represent the reference data.
The embedded plots show the corresponding \emph{x}-component velocity fields. The free-stream conditions are listed in Table \ref{tab:exp_3_summary}.}
\label{fig:Exp3_Cf}
\end{figure}

\begin{figure}

\begin{subfigure}{.3\textwidth}
\centering
\includegraphics[width=\linewidth]{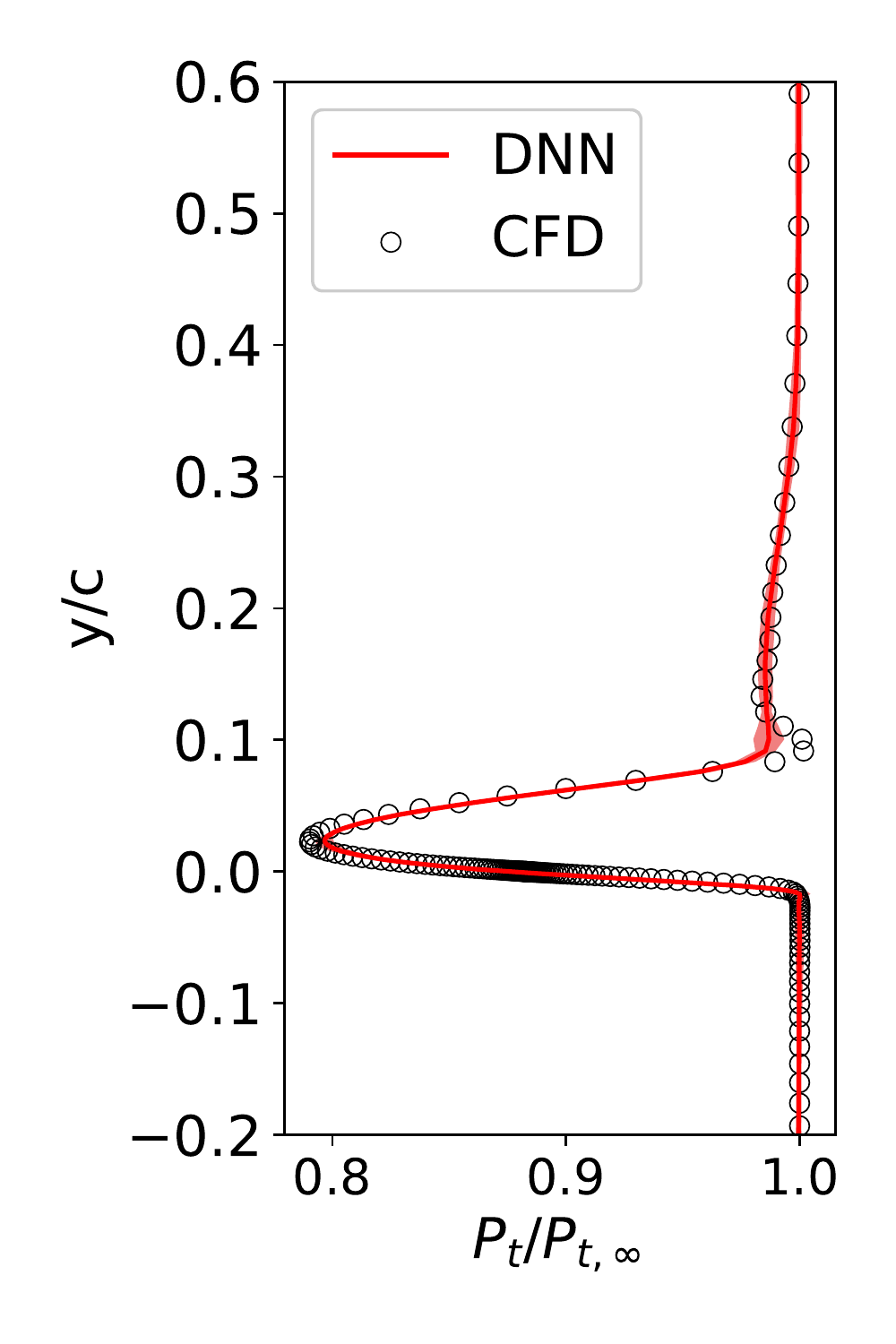}
\caption{e221}
\end{subfigure}
\begin{subfigure}{.3\textwidth}
\centering
\includegraphics[width=\linewidth]{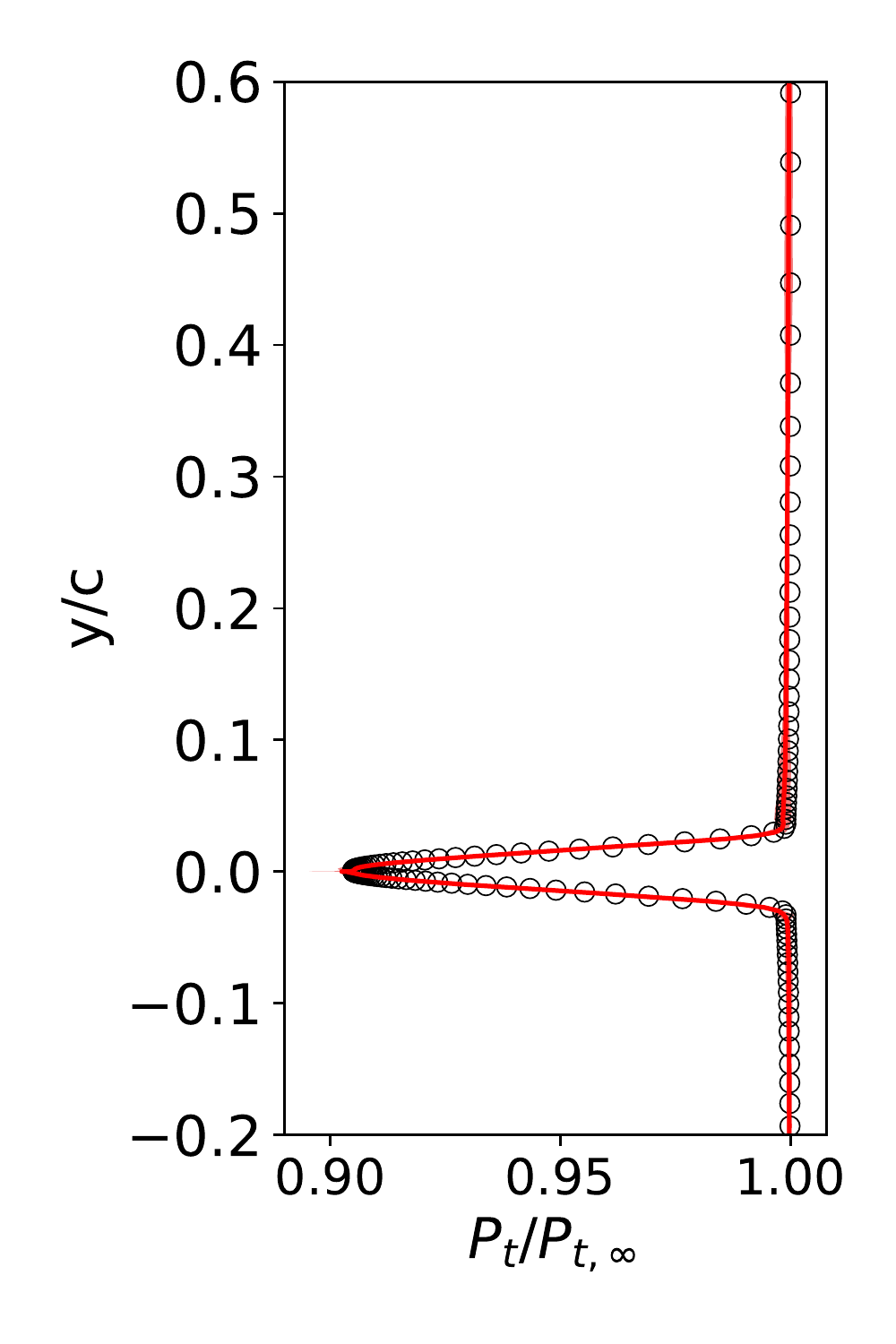}  
\caption{e473}
\end{subfigure}

\begin{subfigure}{.3\textwidth}
\centering
\includegraphics[width=\linewidth]{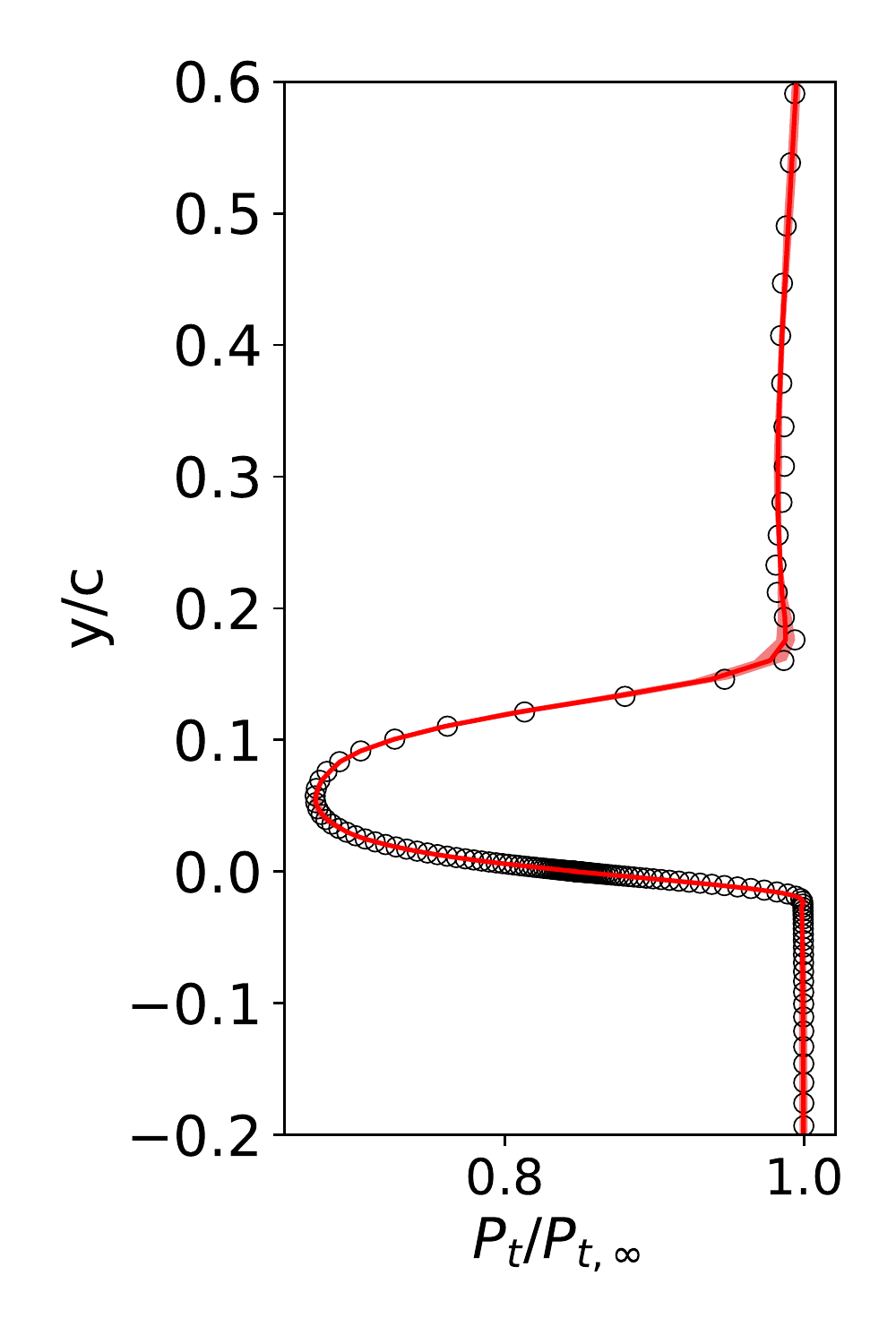}
\caption{fx84w097}
\end{subfigure}
\begin{subfigure}{.3\textwidth}
\centering
\includegraphics[width=\linewidth]{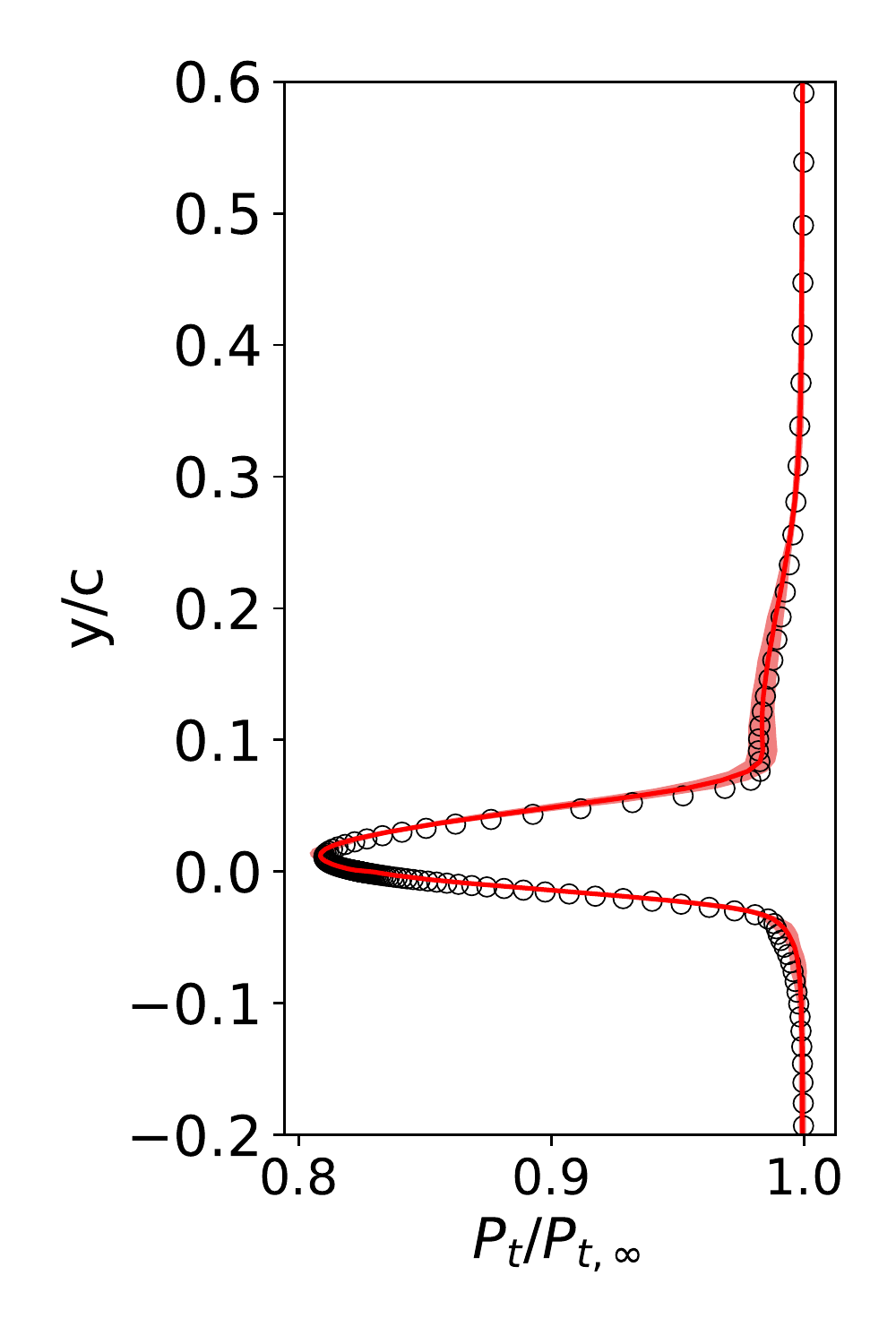}
\caption{mue139}
\end{subfigure}

\caption{Total pressure profiles at $x/c=1.2$ in Case 4.
The red lines represent the DNN results with the shaded region visualizing $\pm3 SD$, and the black symbols represent the reference data.}
\label{fig:Exp3_wake}
\end{figure}

We select four representative conditions/shapes from the test dataset for detailed discussions as shown in Table \ref{tab:exp_3_summary}. The ``worst'' case in terms of $L_1$ loss among the test results is ``e221'' at $\xmach_{\infty}=0.71$, $\alpha_{\infty}=3.43^{\circ}$ and $\Rey_{\infty}=0.661\times10^6$. ``e473'' is a thick aerofoil and at the given condition it turns out to be a shock-wave free case. In the case of ``fx84w097'' with the listed free-stream condition, the shock wave is very strong. On the other hand, ``mue139'' is chosen because of its negative angle of attack. 

The flow field of aerofoil ``e221'' at the test condition is characterised by a strong standing shock wave and shock-induced separation. 
Figure \ref{fig:Exp3_flowfield} shows the density, \emph{x} and \emph{y} components of velocity as well as the speed of sound predicted by DNN (top row) and the CFD solver (middle row). The shock wave is well reproduced by the DNN model and the flow fields in smooth regions are visually indistinguishable between the DNN and CFD predictions.
Despite that, noticeable relative errors (bottom row) are found in the vicinity of shock waves and separated regions induced by the strong adverse pressure gradient, 
the DNN model still provides reliable and accurate predictions in the smooth flow region.

Figure \ref{fig:Exp3_Cp} shows that the DNN model's predictions of pressure coefficients on the aerofoils compare favourably with the ground truth data. Key features are well captured by the DNN model, such as strong adverse pressure gradient due to the shocks on the upper surfaces (see Figs. \ref{fig:Exp3_Cp}a, \ref{fig:Exp3_Cp}c and \ref{fig:Exp3_Cp}d), and the ``cross-over'' of $C_p$ due to the negative angle of attack and the occurrence of a small shock wave near the leading edge shown in Fig. \ref{fig:Exp3_Cp}d. The shock-free attached flow case also agrees well with the CFD result as shown in Fig. \ref{fig:Exp3_Cp}b.

Figure \ref{fig:Exp3_Cf} gives a comparison of skin friction coefficients obtained by DNN and ground truth data. The distributions in Figs. \ref{fig:Exp3_Cf}a and \ref{fig:Exp3_Cf}c are typically characterised by leading edge peaks subjected to transonic fully turbulent inflow and abrupt drops to negative values at around $x/c=0.4$ due to the shock induced separation. Figure \ref{fig:Exp3_Cf}b exhibits smoother distribution as the flow field is shock free under a moderate Mach number. In Fig. \ref{fig:Exp3_Cf}d, there are two sudden changes: the one near the leading edge is due to the small shock wave on the lower surface and the other occurs at around $x/c=0.3$ is caused by the strong standing shock wave one the upper surface. It is clear from this comparison that the DNN predicts the characteristics of the boundary layer accurately.

In the wake region, total pressure profiles are measured vertically at $x/c=1.2$ given in Fig. \ref{fig:Exp3_wake}.
Some fine details caused by the complex flow behavior such as shock-induced separation on the upper surfaces are well reproduced, such as $y/c=0.1$ in Fig. \ref{fig:Exp3_wake}a and $y/c=0.2$ in Fig. \ref{fig:Exp3_wake}c. As aerofoil ``e473'' at the test condition is a shock-free and separation-free case, the wake profile is characterised by a smaller deficit and homogeneous distributions outside the wake. 
Figures \ref{fig:Exp3_wake}a, \ref{fig:Exp3_wake}b and \ref{fig:Exp3_wake}d indicate that the upper parts of the wake profiles exhibit $p_t/p_{t,\infty}<1.0$ in both DNN and CFD solutions likely due to the existence of shock waves, which lead to an increment in entropy and hence a total pressure loss. This phenomenon is well captured by the DNN model, which further demonstrates the high reliability of the DNN model. 
Therefore, the DNN model is clearly capable of reproducing physical details in the flow field, which lays the foundation for future applications such as loss-based optimization for turbine blades subjected to complex inflows \citep{MichelassiChen2015dns}.


It is found that the uncertainties are slightly more sensitive when shock waves occur. As shown in Fig. \ref{fig:Exp3_Cp}a, for example, the pressure coefficient distribution of the aerofoil ``e221'' exhibit wider regions (i.e. $\pm3 SD$) around the strong pressure gradient caused by the shock wave. As a result, the standard deviation of the pressure drag is $1.85\%$ of the mean total drag. Looking at Fig. \ref{fig:Exp3_Cp}d, the shaded regions with $\pm3 SD$ are pronounced in the vicinity of the two shocks over the aerofoil ``mue139'', and correspondingly the standard deviation of the pressure drag accounts for $4.47\%$ the mean total drag.
The predictions of skin friction distributions exhibit larger uncertainties in cases with strong shock, as depicted using $\pm3 SD$ shaded regions in Figs. \ref{fig:Exp3_Cf}a and \ref{fig:Exp3_Cf}d. For the aerofoil ``e221'', the standard deviation of the viscous drag accounts for $0.24\%$ of the mean total drag; for the ``mue139'', the standard deviation of the visous drag is $0.23\%$ of the mean total drag.
Compared to the quantities on the aerofoil surface, the predictions of wake profiles are generally robust for all the test aerofoils, as shown in Fig. \ref{fig:Exp3_wake}.

\subsection{Performance}\label{sec:performance}
\begin{table}
  \begin{center}
\def~{\hphantom{0}}
  \begin{tabular}{lcc}
      \toprule
{\textcolor{lCol}{Solver} }   & \textcolor{lCol}{Wall time} &\textcolor{lCol}{Platform} \\[3pt]
      \midrule 
      \textcolor{lCol}{CFL3D 8000 iterations}    & \textcolor{lCol}{296.96 s} & \textcolor{lCol}{CPU only, 1 core} \\
      \textcolor{lCol}{CFL3D 24000 iterations}   & \textcolor{lCol}{886.18 s} & \textcolor{lCol}{CPU only, 1 core} \\
      \textcolor{lCol}{Method-A batch size 1} & \textcolor{lCol}{2.286 ms} & \textcolor{lCol}{CPU, 1 core \& GPU} \\
      \textcolor{lCol}{Method-B batch size 1}  & \textcolor{lCol}{2.287 ms} & \textcolor{lCol}{CPU, 1 core \& GPU} \\
      \textcolor{lCol}{Method-C batch size 1}     & \textcolor{lCol}{2.287 ms} & \textcolor{lCol}{CPU, 1 core \& GPU} \\
      \textcolor{lCol}{Method-C batch size 5} & \textcolor{lCol}{2.610 ms}	 & \textcolor{lCol}{CPU, 1 core \& GPU} \\
      \textcolor{lCol}{Method-C batch size 20}     & \textcolor{lCol}{2.870 ms} &  \textcolor{lCol}{CPU, 1 core \& GPU} \\
      \textcolor{lCol}{Method-C batch size 40}     & \textcolor{lCol}{3.240 ms} &  \textcolor{lCol}{CPU, 1 core \& GPU} \\
      \bottomrule
  \end{tabular}
  \caption{\textcolor{lCol}{Run times at transonic speed at $\Rey_{\infty}=1\times10^6$}}.
  \label{tab:performance}
  \end{center}
\end{table}

\noindent The performance of trained DNN models is one of the central factors motivating their use, in particular, in aerodynamic shape optimizations. We evaluate our models using a regular workstation with 12 cores, i.e. Intel\textsuperscript{\textregistered} Xeon\textsuperscript{\textregistered} W-2133 CPU @ 3.60GHz, with an NVidia GeForce RTX 2060 GPU. Due to the strongly differing implementations, we compare the different solvers in terms of elapsed wall clock time, which is a pure computation time and is averaged over multiple runs without taking the start-up and initialization overheads into account.

\textcolor{lCol}{As listed in table \ref{tab:performance}, a typical RANS simulation with CFL3D solver at the current grid resolution $128\times128$ at $Re_{\infty}=1\times10^6$ requires 8000 iterations with a wall clock time of 296.96 seconds. As previously mentioned in \S\ref{sec:results_and_discussion}, some cases require an averaging of intermediate results to improve the convergence which takes 24000 iterations and 886.18 seconds, respectively.} Likewise, we evaluate the elapsed time for the DNN prediction,
taking about 2.29 ms with any of the three methods.
It worth mentioning that the run-time per solution can be reduced significantly when
evaluating multiple solutions at once. As an example with method C, for a batch size of
5, the evaluation time rises only slightly to 2.610 ms, and then to 3.240 ms for a batch size of 40.
\textcolor{lCol}{Therefore, relative to the CFD solver CFL3D, a speed-up factor between 130000 and 388000 is achieved.}
Even when considering a factor of ca. 10 in terms of GPU advantage due to improved on-chip memory bandwidth \textcolor{lCol}{and the fact that CFL3D simulation could be made roughly 10 times faster with the multigrid method \citep[p.~123]{Krist_Biedron_Rumsey1998manual},} these measurements indicate the significant reductions in terms of run time that can potentially be achieved by employing trained deep neural networks.
The time to train the DNN models varies with neural network size and the amount of training data. 
Taking Dataset-9700 (see \ref{appA:dataset_size}) as an example leads to training times of 
ca. 11 hours for 1200 epochs to obtain a large-scale model.
Given the potentially large one-time cost for training a model, learned approaches bear particular promise in settings where similar optimization problems are solved repeatedly.

To ensure reproducibility, the full code and training as well as test datasets for this work will be made available at: \href{https://github.com/tum-pbs/coord-trans-encoding}{https://github.com/tum-pbs/coord-trans-encoding} upon publication.

\section{Conclusion}\label{sec:concluding_remarks}
\noindent
In this paper, we presented a method to train a deep neural network (DNN) model that learns the Reynolds-averaged Navier-Stokes solutions based on widely-used structured grids for various angles of attack, Reynolds numbers and Mach numbers. 
Our approach yields networks that learn to generate precise flow fields for varying body-fitted structured grids by providing them with an encoding of the corresponding mapping to a canonical space for the solutions.
In the benchmark case of incompressible flow at randomly given angles of attack and Reynolds numbers 
the DNN model achieves an improvement of more than an order of magnitude compared to previous work.
For transonic flow cases, the DNN model accurately predicts complex flow behavior at high Reynolds numbers, such as shock wave/boundary layer interaction, and quantitative distributions like pressure coefficient, skin friction coefficient as well as wake total pressure profiles downstream of aerofoils. Furthermore the DNN model, being orders of magnitude faster than the conventional CFD solver, yields very significant computational speedups.

The proposed deep learning method offers several advantages. First, the method doesn't rely on auxiliary fields with information about the embedded aerofoil. Instead, it employs geometric information from structured grids that can be easily combined with common CFD applications to infer solutions with changing domain discretizations. Second, the model provides flexibility as it predicts a full flow field. Once trained, it can, for example, be used in different optimization tasks. Third, as the model is differentiable, it can be seamlessly integrated into gradient-based optimization algorithms.
Being trained with varying mesh deformations, the neural network architecture is able to deal with changing geometries, and can be easily extended to boundary value problems of PDEs discretized with structured body-fitted grids.

On the other hand, a limitation of the proposed method is that the DNN architecture assumes that the different discretizations share the same topology, e.g. a c-grid topology mesh. Meanwhile, the change of mesh resolution is not considered in the present work, though it is important. 
Nonetheless, the computational meshes of many of practical applications, such as design-space parameter studies, are usually generated with the same template based on best practices and existing knowledge and hence amenable to the approach.
In order to cope with different grid topologies, we believe it will be a interesting future direction to add varying boundary conditions and introduce physical inductive biases into the learning process.

\textcolor{lCol}{A prerequisite of the application of the DNN models is that the mapping to a monolithic structured grid exists. It is a requirement that the proposed method shares with many classic numerical approaches. 
We note that extensions such as the multi-block overset grid have been used with many successes in aerospace engineering, especially those with complex geometries \cite[]{Rumsey1997, Ali_Tucker2013multiblock}. Similarly, we believe the proposed model can be applied to those scenarios as well. In particular, if the flowfield solutions over a complex geometry can be represented on multiple monolithic structured grids, then the corresponding deep neural network architecture can be designed and trained based on the metrics of the given grids as well as free-stream features.}

\textcolor{lCol}{While the present work focuses on the steady RANS equations, we note that accurate predictions of transient turbulent flow physics have been achieved with direct numerical simulations (DNS), large-eddy simulations (LES), as well as hybrid RANS/LES methods \cite[]{SagautBookMMAT2013}, and the inference of high-fidelity data generated by those advanced approaches is a very interesting direction for future research.}

\section*{CRediT authorship contribution statement}
\textcolor{black}{Li-Wei Chen: Conceptualization, Methodology, Coding, Data generation, Original draft, Post-processing. 
Nils Thuerey: Resources, Reviewing and Editing, Supervision, Funding acquisition.}

\section*{Declaration of competing interest}
The authors declare that they have no known competing financial interests or personal relationships that could have appeared to
influence the work reported in this paper.

\section*{Acknowledgment} 
This work was supported by 
the ERC Consolidator Grant \textit{SpaTe} (CoG-2019-863850).

\appendix



\section{Network size}\label{appA:network_size}

\begin{figure}
\begin{subfigure}{.45\textwidth}
\centering
\includegraphics[width=\linewidth]{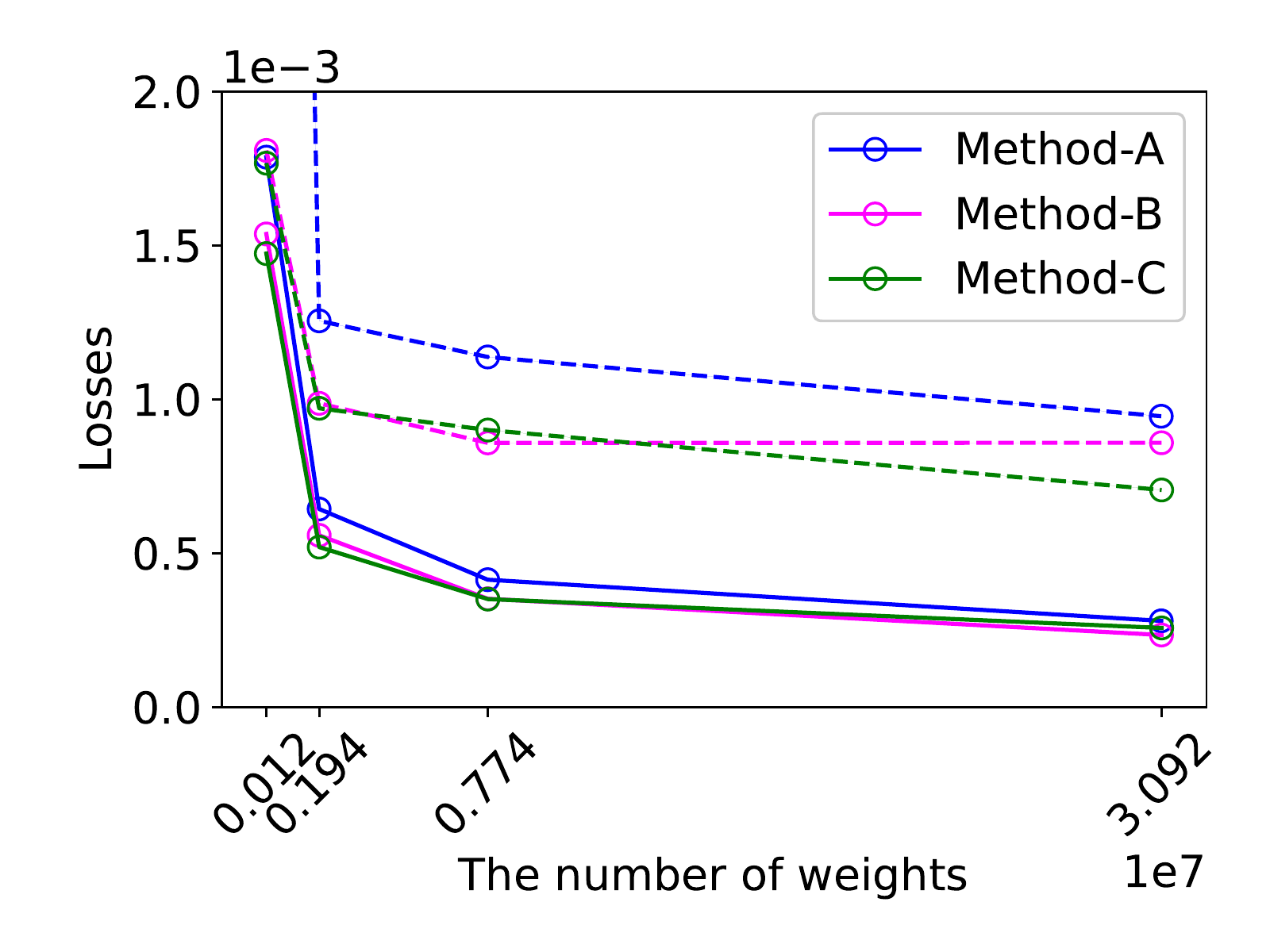}
\caption{Training and validation}
\end{subfigure}
\begin{subfigure}{.45\textwidth}
\centering
\includegraphics[width=\linewidth]{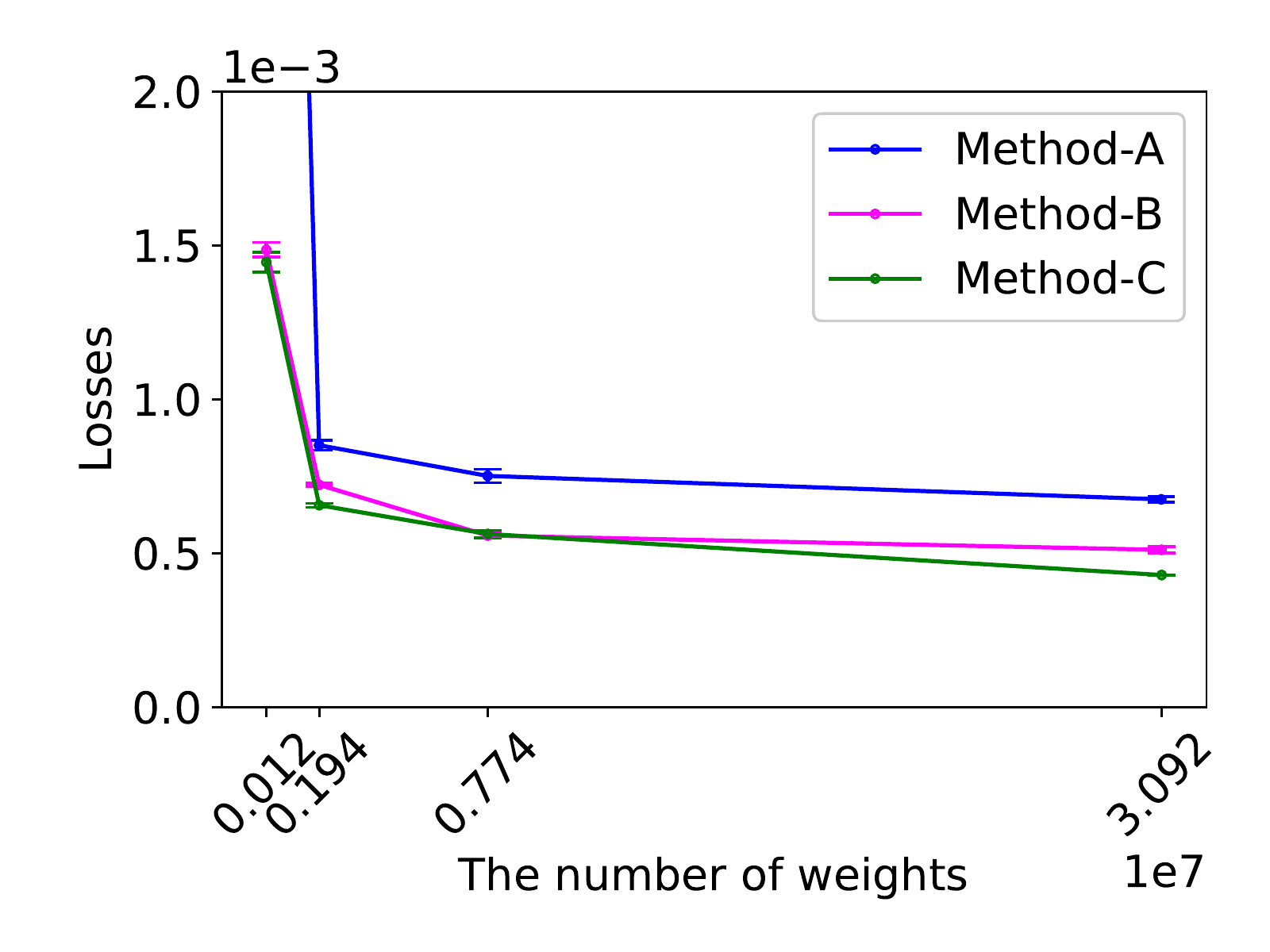}
\caption{Test}
\end{subfigure}
\caption{Training (solid lines), validation (dashed lines) and test L1 losses using different network sizes with the number of trainable parameters of $1.24\times10^5$, $1.94\times10^6$, $7.74\times10^6$, and $3.09\times10^7$.}
\label{fig-appA:Exp1_NN_size}
\end{figure}

\noindent
To investigate the impact of the neural network size on the accuracy, several training runs are conducted with four different network sizes for Case 1. 
\textcolor{black}{This can be achieved by varying the number of features in each layers.}
Figures \ref{fig-appA:Exp1_NN_size}a and \ref{fig-appA:Exp1_NN_size}b show the training, validation as well as test losses vary with the network size in terms of the number of trainable parameters, i.e. $1.24\times10^5$, $1.94\times10^6$, $7.74\times10^6$ and $3.09\times10^7$, respectively. The corresponding error bars indicate the standard deviations (i.e. $\pm SD$) of the test losses. Although volatility is still noticeable, the basic tendency is towards high accuracy and smaller losses with the increment of trainable parameters. There is no drastic change in loss curves when the number of trainable parameters is higher than $1.94\times10^6$. 

It worth noting that the DNN model with $1.24\times10^5$ parameters and using method A cannot give a reasonable prediction due to the high validation and test losses, i.e. 0.02688 and 0.01808, respectively. Note also that the largest scale DNN model with method B shows a high validation loss as shown in Fig. \ref{fig-appA:Exp1_NN_size}a, far from improving the prediction accuracy as expected, so we consider that method C outperforms the other two and will use the results obtained by DNN models with method C for detailed discussions.

\section{Dataset size}\label{appA:dataset_size}
\noindent
To assess the effects of the number of samples in the datasets on the training and validation losses, 
training runs for Cases 2 and 4 are conducted with various amounts of data. The results are listed in Table \ref{tab-appA:dataset_details_range}. 
Note that in this case any smaller dataset is a subset of a larger dataset and all follow the same probability distribution, i.e. the uniform distribution.
The size of the test set is kept the same, namely 20 samples. We found validation sets of several hundred samples to yield stable estimates, and hence use an upper limit of 400 as the maximal size of the validation dataset. The typical number of epochs for training is 1290. 

\begin{figure}
\begin{subfigure}{.45\textwidth}
\centering
\includegraphics[width=\linewidth]{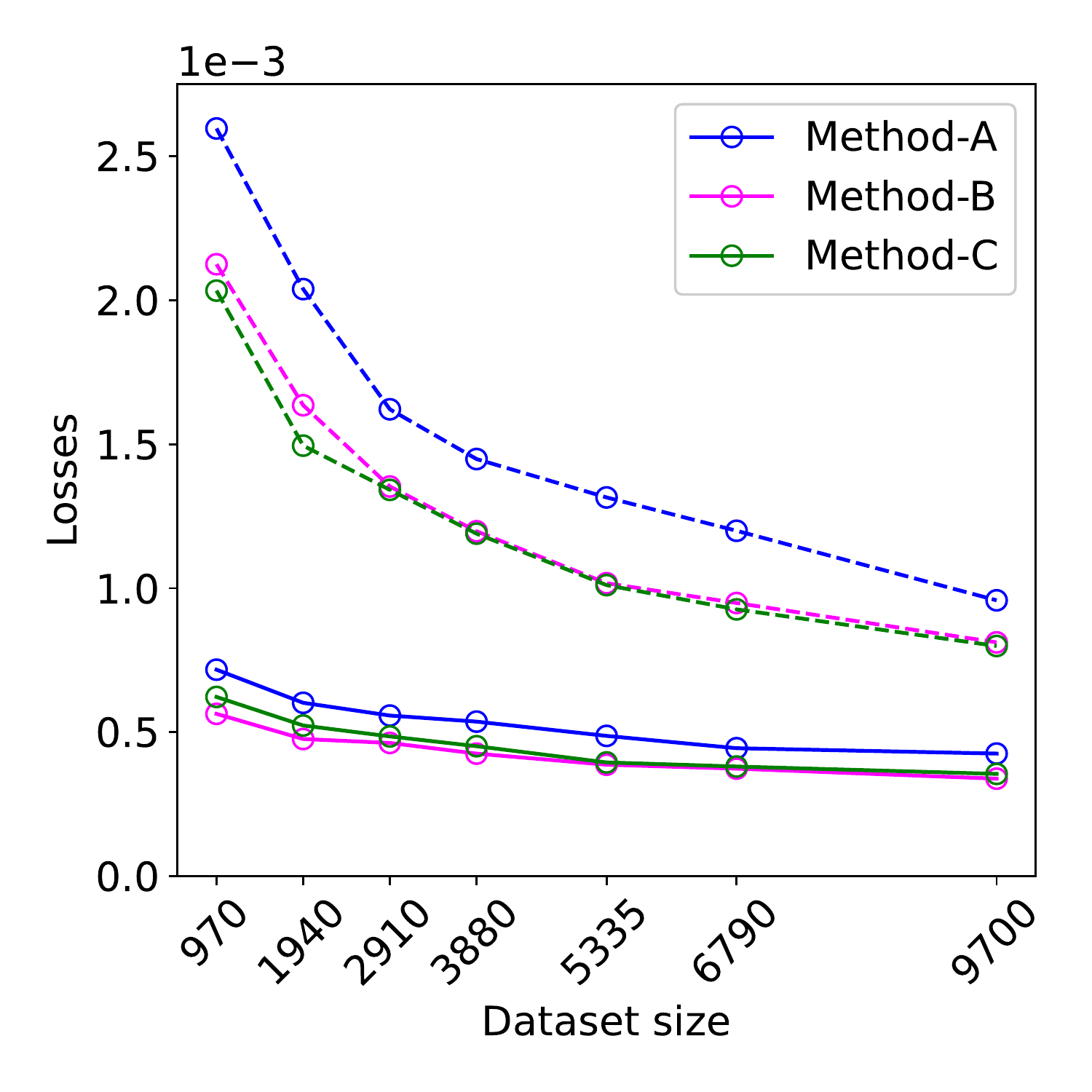}
\caption{Training and validation}
\end{subfigure}
\begin{subfigure}{.45\textwidth}
\centering
\includegraphics[width=\linewidth]{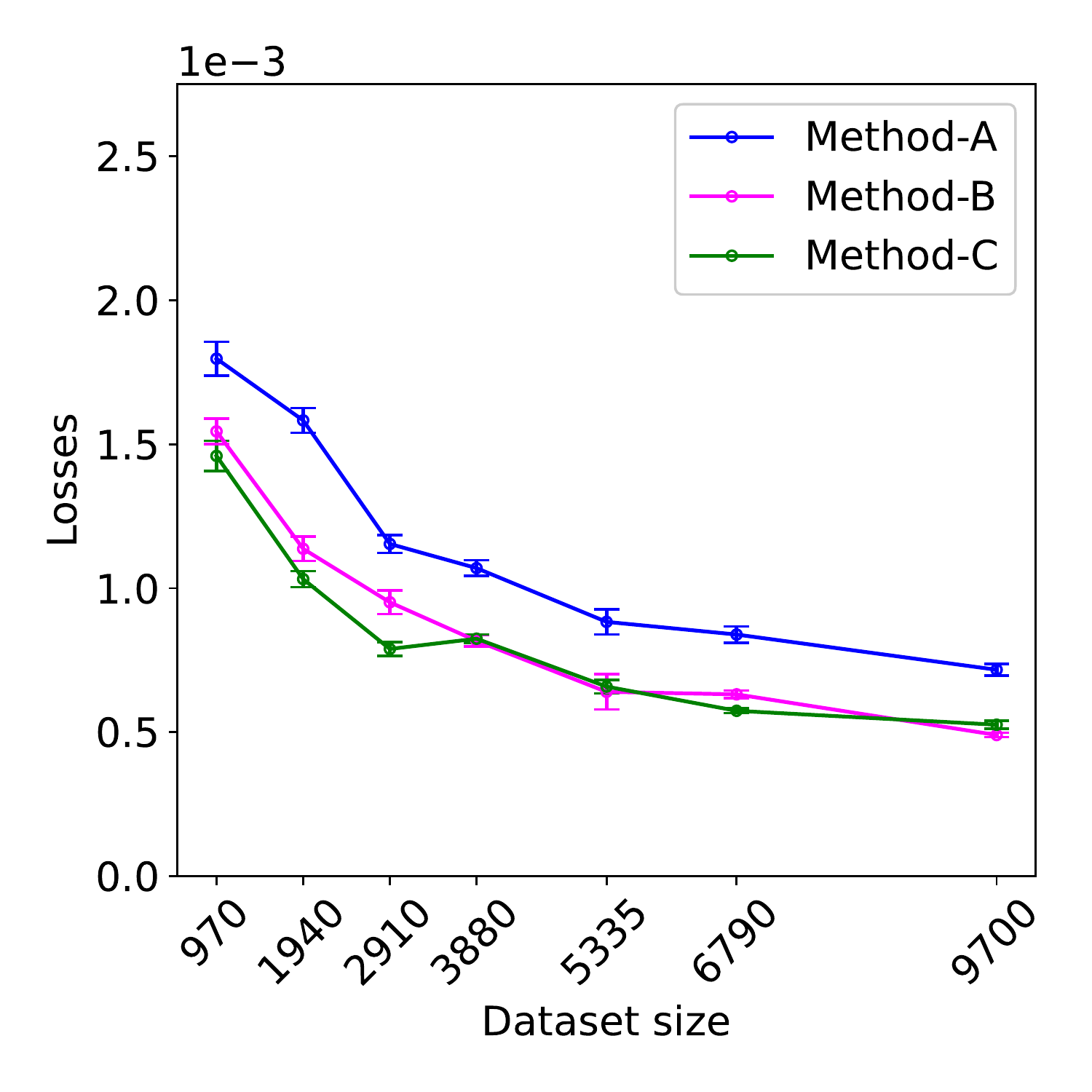}
\caption{Test}
\end{subfigure}
\caption{\textcolor{lCol}{Training (solid lines), validation (dashed lines) and test L1 losses using different methods and training data amount for Case 2.}}
\label{fig-appA:Exp2_loss_data_sens}
\end{figure}

Figures \ref{fig-appA:Exp2_loss_data_sens}a and \ref{fig-appA:Exp2_loss_data_sens}b show the values of training, validation and test losses for Case 2, and the error bars represent $\pm SD$ standard deviations of the test losses.
It can be observed that the models with small amounts of data exhibit larger losses and relatively larger standard deviations.
The behavior stabilizes with larger amounts of data being available for training, especially when the number of samples is greater than 6790.

\begin{figure}
\begin{subfigure}{.45\textwidth}
\centering
\includegraphics[width=\linewidth]{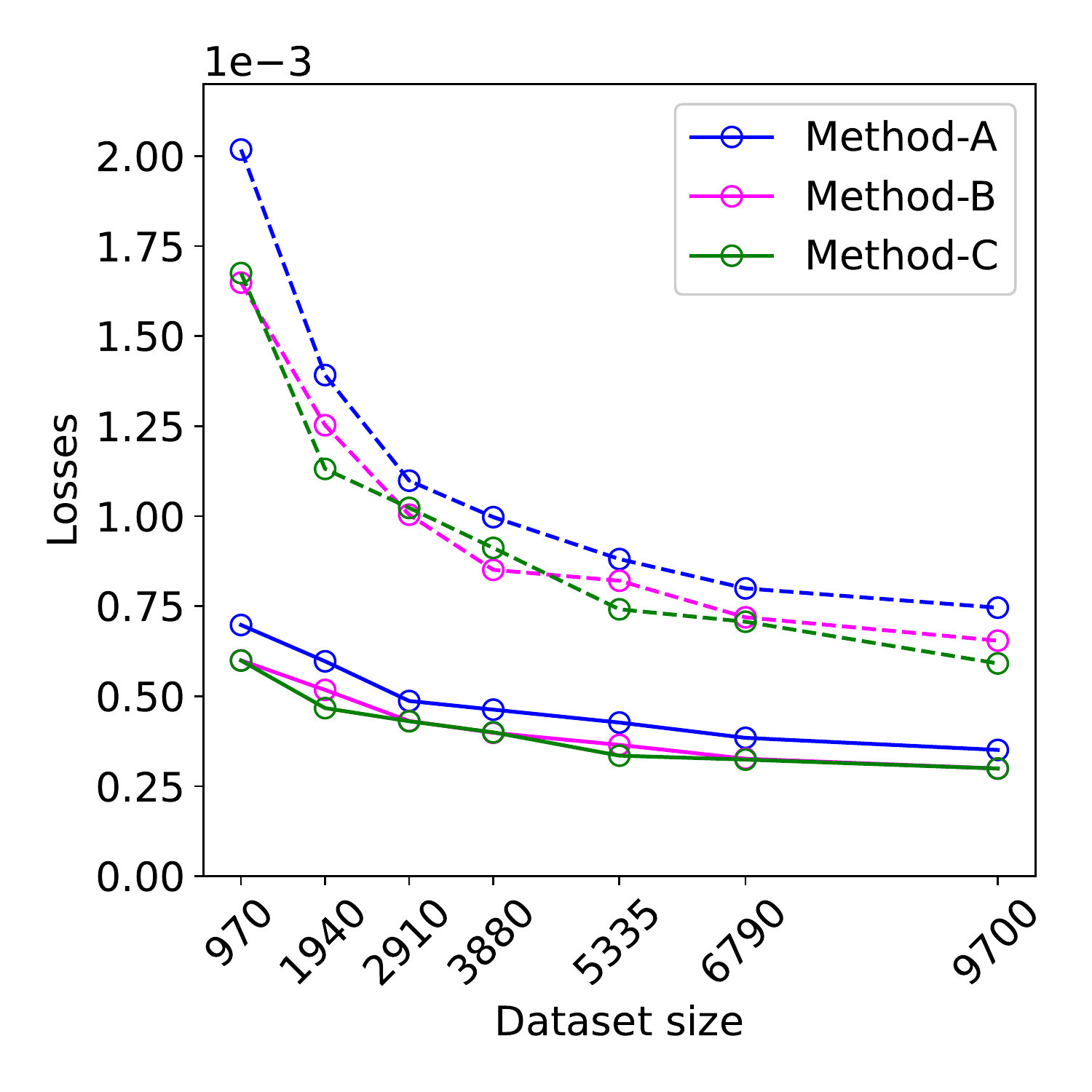}
\caption{Training and validation}
\end{subfigure}
\begin{subfigure}{.45\textwidth}
\centering
\includegraphics[width=\linewidth]{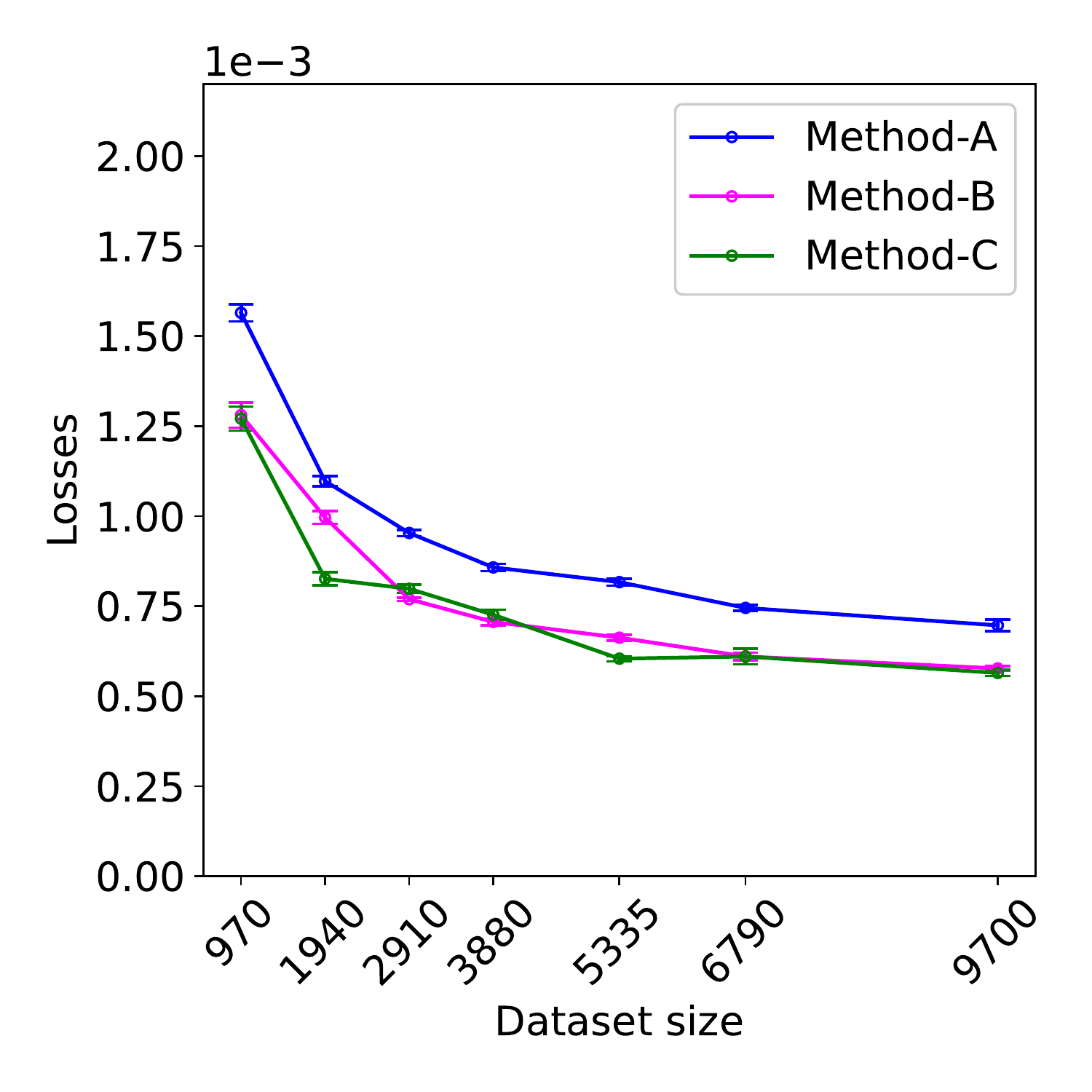}
\caption{Test}
\end{subfigure}
\caption{\textcolor{lCol}{Training (solid lines), validation (dashed lines) and test L1 losses using different methods and training data amount for Case 4.}}
\label{fig-appA:Exp3_loss_data_sens}
\end{figure}

From Figs. \ref{fig-appA:Exp3_loss_data_sens}a and \ref{fig-appA:Exp3_loss_data_sens}b for Case 4, it can be also seen that with an increased number of samples, the training, validation and test losses as well as the standard deviations follow an overall declining trend. 

To summarize, the above results in Cases 2 and 4 imply that further increasing the amount of data does not yield significant improvements in terms of inference accuracy. Therefore, the DNN model trained with ``Dataset-9700'' is chosen for the detailed study for Cases 2 and 4.
\textcolor{lCol}{As the loss of the model decreases only modestly for the larger datasets, it is likely that the variance in the dataset, especially that with 9700 samples, has exceeded the capacity of the chosen model configuration (i.e. 30.9M parameters). 
To further improve the model performance for the training set with 9700 samples, an even larger model could potentially be chosen.}
In most runs, the performance of methods B and C is very close. However, the validation loss curves of method B in Figs. \ref{fig-appA:Exp2_loss_data_sens}a and \ref{fig-appA:Exp3_loss_data_sens}a exhibit higher losses when the number of samples increase to 9700. Thus, method C is still considered to outperform the other two.

\begin{table}
  \begin{center}
\def~{\hphantom{0}}
  \begin{tabular}{lcccc}
      \toprule
      Name    & No. of flowfields & training  & validation & test \\
      \midrule 
      \textcolor{lCol}{Dataset-970} & \textcolor{lCol}{970} & \textcolor{lCol}{780} & \textcolor{lCol}{190} &\textcolor{lCol}{20} \\
      Dataset-1940    & 1940  & 1555 & 385 & 20 \\
      \textcolor{lCol}{Dataset-2910} & \textcolor{lCol}{2910} & \textcolor{lCol}{2510} & \textcolor{lCol}{400} &\textcolor{lCol}{20} \\      
      Dataset-3880   & 3880  & 3480 & 400 & 20 \\
      \textcolor{lCol}{Dataset-5335} & \textcolor{lCol}{5335} & \textcolor{lCol}{4935} & \textcolor{lCol}{400} &\textcolor{lCol}{20} \\         
      Dataset-6790    & 6790  & 6390 & 400 & 20 \\
      Dataset-9700    & 9700  & 9300 & 400 & 20 \\
      \bottomrule
  \end{tabular}
  \caption{Different dataset sizes used for training runs with corresponding splits into training and validation sets.}
  \label{tab-appA:dataset_details_range}
  \end{center}
\end{table}

\section{\textcolor{lCol}{CFD equations in generalized coordinates}}\label{appA:CFD_equations}
\noindent
\textcolor{lCol}{
By way of a concrete example, once the coordinate transformation is introduced, with the chain rule, let
\begin{equation}
\mathbf{F}=\frac{1}{J}\begin{bmatrix}
    \rho U\\
    \rho U u + p\ \partial_x\xi \\
    \rho U v + p\ \partial_y\xi \\
    (\rho E+p)U
\end{bmatrix}    \ \ 
\mathbf{F}_v=\frac{1}{J}\begin{bmatrix}
    0\\
    \partial_x\xi\ \tau_{xx} + \partial_y\xi\ \tau_{xy} \\
    \partial_x\xi\ \tau_{xy} + \partial_y\xi\ \tau_{yy} \\
    \partial_x\xi\ b_x + \partial_y\xi\ b_y
\end{bmatrix}    
\end{equation}
\begin{equation}
\mathbf{G}=\frac{1}{J}\begin{bmatrix}
    \rho V\\
    \rho V u + p\ \partial_x\eta \\
    \rho V v + p\ \partial_y\eta \\
    (\rho E+p)V
\end{bmatrix}    \ \ 
\mathbf{G}_v=\frac{1}{J}\begin{bmatrix}
    0\\
    \partial_x\eta\ \tau_{xx} + \partial_y\eta\ \tau_{xy} \\
    \partial_x\eta\ \tau_{xy} + \partial_y\eta\ \tau_{yy} \\
    \partial_x\eta\ b_x + \partial_y\eta\ b_y
\end{bmatrix}    
\end{equation}
where
\begin{equation}
\begin{aligned}
U=\partial_x\xi\ u + \partial_y\xi\ v \\
V=\partial_x\eta\ u + \partial_y\eta\ v
\end{aligned}
\end{equation}
\begin{equation}
\begin{aligned}
    b_x=u\tau_{xx}+v\tau_{xy}-q_x \\
    b_y=u\tau_{xy}+v\tau_{yy}-q_y
\end{aligned}
\end{equation}
and \begin{equation*}
    J=
    \begin{vmatrix}
    \partial_x\xi & \partial_y\xi\\
    \partial_x\eta & \partial_y\eta 
    \end{vmatrix}.
    \end{equation*}
Combining all the terms, the steady RANS equations \ref{eq:RANS-1}-\ref{eq:RANS-4} can be written
\begin{equation}
\frac{\partial(\mathbf{F}-\mathbf{F}_v)}{\partial \xi}+
\frac{\partial(\mathbf{G}-\mathbf{G}_v)}{\partial \eta}=0.
\end{equation}
Considering the aerofoil surface coordinates $(x_0, y_0)$ as reference points, the discretized implicit equation \ref{eq:discreteForm} can be expressed as:
\begin{equation}\label{eq:discreteForm2}
    \mathcal{R}(\mathbf{y}, \xmach_{\infty}, \alpha_{\infty}, \Rey{}_{\infty}, \partial_x\xi, \partial_y\xi, \partial_x\eta, \partial_y\eta, x_0, y_0) = 0
\end{equation}
or 
\begin{equation}\label{eq:discreteForm3}
\mathcal{R}(\mathbf{y}, \xmach_{\infty}, \alpha_{\infty}, \Rey{}_{\infty}, J^{-1}, \hat{\xi}_x, \hat{\xi}_y, |\nabla \xi|/J, \hat{\eta}_x, \hat{\eta}_y, |\nabla \eta|/J, x_0, y_0) = 0.
\end{equation}
}

\bibliographystyle{unsrt}
\bibliography{main}

\end{document}